\documentclass[fleqn,usenatbib]{mnras}

\usepackage[T1]{fontenc}

\DeclareRobustCommand{\VAN}[3]{#2}
\let\VANthebibliography\thebibliography
\def\thebibliography{\DeclareRobustCommand{\VAN}[3]{##3}\VANthebibliography}

\usepackage{graphicx}	
\usepackage{amsmath}	
\usepackage{comment}
\usepackage{amssymb}
\usepackage{xspace}
\usepackage{enumitem}
\usepackage{graphicx}
\usepackage{rotating}
\usepackage{color}

\usepackage[varvw]{newtxmath}

\DeclareMathOperator{\erf}{erfc}

\newcommand{\lya}        {Ly$\alpha$\xspace}
\newcommand{\hi}         {\ion{H}{I}\xspace}
\newcommand{\mgii}         {\ion{Mg}{II}\xspace}
\newcommand{\civ}         {\ion{C}{IV}\xspace}
\newcommand{\ovi}         {\ion{O}{VI}\xspace}
\newcommand{\nv}         {\ion{N}{V}\xspace}
\newcommand{\siiv}         {\ion{S}{IV}\xspace}

\newcommand{\RH}        {\relax\ifmmode{{\rm R}_{\rm H}\xspace} \else {${\rm R}_{\rm H}$}\expandafter\xspace\fi}
\newcommand{\rcl}        {\relax\ifmmode{{\rm r}_{\rm cl}\xspace} \else {${\rm r}_{\rm cl}$}\expandafter\xspace\fi}
\newcommand{\fc}        {\relax\ifmmode{f_{\rm c}\xspace} \else {${f}_{\rm c}$}\expandafter\xspace\fi}
\newcommand{\fccrit}        {\relax\ifmmode{{f}_{\rm c,crit}\xspace} \else {${f}_{\rm c,crit}$}\expandafter\xspace\fi}

\newcommand{\fmghi}        {\relax\ifmmode{{ f}_{\rm MgII/HI}\xspace} \else {${ f}_{\rm MgII/HI}$}\expandafter\xspace\fi}
\newcommand{\NHI}        {\relax\ifmmode{{ N}_{\rm HI}\xspace} \else {${ N}_{\rm HI}$}\expandafter\xspace\fi}
\newcommand{\Nmg}        {\relax\ifmmode{{ N}_{\rm MgII}\xspace} \else {${ N}_{\rm MgII}$}\expandafter\xspace\fi}
\newcommand{\tauK}        {\relax\ifmmode{{ \tau}_{\rm K}\xspace} \else {${ \tau}_{\rm K}$}\expandafter\xspace\fi}
\newcommand{\tauH}        {\relax\ifmmode{{ \tau}_{\rm H}\xspace} \else {${ \tau}_{\rm H}$}\expandafter\xspace\fi}

\newcommand{\lambdaK}        {\relax\ifmmode{{ \lambda}_{\rm K}\xspace} \else {${ \lambda}_{\rm K}$}\expandafter\xspace\fi}
\newcommand{\lambdaH}        {\relax\ifmmode{{ \lambda}_{\rm H}\xspace} \else {${ \lambda}_{\rm H}$}\expandafter\xspace\fi}

\newcommand{\Rmg}        {\relax\ifmmode{{ R}_{\rm MgII}\xspace} \else {${ R}_{\rm MgII}$}\expandafter\xspace\fi}
\newcommand{\Rmgemis}        {\relax\ifmmode{{R}^{\rm E}_{\rm MgII} \xspace} \else {${R}^{\rm E}_{\rm MgII}$}\expandafter\xspace\fi}
\newcommand{\Rmgcon}       {\relax\ifmmode{{R}_{\rm MgII}^{\rm C} \xspace} \else {${R}_{\rm MgII}^{\rm C}$}\expandafter\xspace\fi}
\newcommand{\Rmgtotal}        {\relax\ifmmode{{ R}_{\rm MgII, Total}\xspace} \else {${ R}_{\rm MgII, Total}$}\expandafter\xspace\fi}
\newcommand{\Rmghalo}        {\relax\ifmmode{{ R}_{\rm MgII, Halo}\xspace} \else {${ R}_{\rm MgII, Halo}$}\expandafter\xspace\fi}
\newcommand{\fesc}        {\relax\ifmmode{f_{\rm esc}\xspace} \else {$f_{\rm esc}$}\expandafter\xspace\fi}
\newcommand{\fesclya}        {\relax\ifmmode{f_{\rm esc, Ly\alpha}\xspace} \else {$f_{\rm esc, Ly\alpha}$}\expandafter\xspace\fi}
\newcommand{\fescmgii}        {\relax\ifmmode{f_{\rm esc, MgII}\xspace} \else {$f_{\rm esc, MgII}$}\expandafter\xspace\fi}
\newcommand{\fesclyc}        {\relax\ifmmode{f_{\rm esc, LyC}\xspace} \else {$f_{\rm esc, LyC}$}\expandafter\xspace\fi}
\newcommand{\lpath}        {\relax\ifmmode{l_{\rm path}\xspace} \else {$l_{\rm path}$}\expandafter\xspace\fi}
\newcommand{\taudabs}     {\relax\ifmmode {\tau_{\rm d,abs}} \else {$\tau_{\rm d,abs}$}\expandafter\xspace\fi}
\newcommand{\tauzero}     {\relax\ifmmode {\tau_{\rm 0}} \else {$\tau_{\rm 0}$}\expandafter\xspace\fi}
\newcommand{\NHIcl}      {\ifmmode{N_{\rm HI,cl}\xspace}\else{$N_{\rm HI,cl}\,$\xspace}\fi}
\newcommand{\Nmgcl}      {\ifmmode{N_{\rm MgII,cl}\xspace}\else{$N_{\rm MgII,cl}\,$\xspace}\fi}

\newcommand{\kms}        {\ifmmode{\rm \,km\,s^{-1}}\else\,km\,s$^{-1}$\xspace\fi}
\newcommand{\unitNHI}    {\ifmmode{\rm \,cm^{-2}}\else\,cm$^{-2}$\xspace\fi}  
\newcommand{\vexp}       {\relax\ifmmode {v_{\rm exp}} \else {$v_{\rm exp}$}\expandafter\xspace\fi}
\newcommand{\vran}       {\relax\ifmmode {v_{\rm ran}} \else {$v_{\rm ran}$}\expandafter\xspace\fi}
\newcommand{\vth}       {\relax\ifmmode {v_{\rm th}} \else {$v_{\rm th}$}\expandafter\xspace\fi}
\newcommand{\sigran}     {\relax\ifmmode {\sigma_{\rm Ran}} \else {$\sigma_{\rm Ran}$}\expandafter\xspace\fi}
\newcommand{\sigr}     {\relax\ifmmode {\sigma_{\rm R}} \else {$\sigma_{\rm R}$}\expandafter\xspace\fi}
\newcommand{\sigsrc}     {\relax\ifmmode {\sigma_{\rm src}} \else {$\sigma_{\rm src}$}\expandafter\xspace\fi}
\newcommand{\sigcl}     {\relax\ifmmode {\sigma_{\rm cl}} \else {$\sigma_{\rm cl}$}\expandafter\xspace\fi}
\newcommand{\dvpeak}     {$\Delta V_{\rm peak}$\xspace}
\newcommand{\Nscat}     {$N_{\rm scat}$\xspace}

\newcommand{\EWemis}        {\relax\ifmmode{|{ \rm EW}_{\rm emis}|\xspace} \else {$|{ \rm EW}_{\rm emis}|$}\expandafter\xspace\fi}
\newcommand{\EWabs}        {\relax\ifmmode{|{ \rm EW}_{\rm abs}|\xspace} \else {$|{ \rm EW}_{\rm abs}|$}\expandafter\xspace\fi}
\newcommand{\EWint}        {\relax\ifmmode{{ \rm EW}_{\rm int}\xspace} \else {${ \rm EW}_{\rm int}$}\expandafter\xspace\fi}

\title[Mg~II and Ly$\alpha$ Radiative Transfer]
{Probing cold gas with Mg~II and Ly$\alpha$ radiative transfer}

\author[Chang \& Gronke]{
Seok-Jun Chang$^{1}$\thanks{E-mail: sjchang@mpa-garching.mpg.de} and
Max Gronke$^{1}$ 
\\
$^{1}$ Max-Planck-Institut f\"{u}r Astrophysik, Karl-Schwarzschild-Stra$\beta$e 1, 85748 Garching b. M\"{u}nchen, Germany
}

\date{Accepted 2024 July 1. Received 2024 June 25; in original form 2023 July 14}

\pubyear{2024}

\begin{document}
\label{firstpage}
\pagerange{\pageref{firstpage}--\pageref{lastpage}}
\maketitle

\begin{abstract}
The Mg~II resonance doublet at 2796 \AA\ and 2803 \AA\ is an increasingly important tool to study cold, $T \sim 10^{4}\,$K, gas -- an observational driven development requiring theoretical support.
We develop a new Monte Carlo radiative transfer code to systematically study the joined Mg~II and Ly$\alpha$ escape through homogeneous and `clumpy' multiphase gas with dust in arbitrary 3D geometries. Our main findings are:
(i) The Mg~II spectrum differs from Ly$\alpha$ due to the large difference in column densities, even though the atomic physics of the two lines are similar.
(ii) the Mg~II escape fraction is generally higher than that of Ly$\alpha$ because of lower dust optical depths and path lengths -- but large variations due to differences in dust models and the clumpiness of the cold medium exist.
(iii) Clumpy media possess a `critical covering factor' above which Mg~II radiative transfer matches a homogeneous medium. The critical covering factors for Mg~II and Ly$\alpha$ differ, allowing constraints on the cold gas structure.
(iv) The Mg~II doublet ratio $R_{\rm MgII}$ varies for strong outflows/inflows ($\gtrsim 700 \mathrm{km\,s}^{-1}$), in particular, $R_{\rm MgII}<1$ being an unambiguous tracer for powerful galactic winds.
(v) Scattering of stellar continuum photons can decrease $R_{\rm MgII}$ from two to one, allowing constraints on the scattering medium. Notably, we introduce a novel probe of the cold gas column density -- the halo doublet ratio -- which we show to be a powerful indicator of ionizing photon escape. 
We discuss our results in the context of interpreting and modeling observations as well as their implications for other resonant doublets.
\end{abstract}




\section{Introduction}\label{sec:introduction}
The circumgalactic medium (CGM) plays a pivotal role in understanding galaxy formation and evolution. As an extended gaseous region surrounding galaxies, it serves as a bridge between the intergalactic medium and the galactic disk, facilitating the exchange of matter, energy, and feedback processes. Exploring the properties and dynamics of the CGM provides crucial insights into the mechanisms governing star formation, chemical enrichment, and the regulation of galactic outflows, ultimately shaping the growth and evolution of galaxies over cosmic time \citep[see extensive reviews][]{tumlinson17,faucher-giguere23}.

Traditionally, the CGM has been studied using absorption lines, which involve observing the light from a bright background source passing through the CGM and absorbing specific wavelengths \citep{prochaska13,crighton14,bouche16,chen20,weng23}. This usually constrains knowledge to singular sight lines and thus requires stacking to assemble a comprehensive picture of the CGM. However, recent advancements in observational techniques and instrumentation have opened up new avenues for studying the CGM using emission instead \citep{steidel10,steidel11,hennawi13,leclercq22,dutta23}.
One main advantage over the traditional absorption line approach is that the spatial information, such as the surface brightness profile and spatially varying spectral shapes, allows us to infer the physical properties of the CGM without a need for a background source.

In both emission as well as absorption line studies, metal resonance doublet lines play a crucial role.
The absorption features of the doublets in the quasar spectrum \citep[e.g.,][]{fox07,fox09,mas-ribas17,schroetter21,ranjan22} and the emission feature near the quasars and star-forming galaxies \citep[e.g.,][]{fabrizio15,hayes16,berg19,guo20,travascio20,leclercq22,dutta23}
have been utilized to study the kinematics and physical properties of CGM. 
Importantly, metal resonance doublets trace different temperature regions because of their different ionization energies.
For instance, the doublet \mgii at 2796 \AA\ and 2803 \AA\ traces the cold medium with $T\lesssim 10^4\, \rm K$, because the ionization energy of \mgii($\sim$ 15.0 eV) is comparable to that of \hi($\sim$ 13.6 eV).
The doublets of atoms with high ionization energy > 60~eV, C~IV $\lambda\lambda$1548, 1551, N~V $\lambda\lambda$1239, 1243, O~VI $\lambda\lambda$1032, 1038, and  \siiv$\lambda\lambda$ 1393, 1402 trace the warm medium with $T \gtrsim 10^5\, \rm K$. 
The absorption lines of these metal doublets coincide in the quasar spectrum with strong Lyman-$\alpha$ (\lya) absorption.
The coexistence of high and low ionized absorption lines shows the multiphase structure of CGM in one line of sight (\citealp{churchill99,schroetter21}, see also \citealp{tumlinson17}).\\


Apart from its fundamental role as the fuel of (future) star formation, studying the cold gas with $T = 10^4\ \rm K$ is also crucial in order to understand the epoch of reionization in the early universe. This is because cold \hi structures prevent the majority of ionizing photons to escape from galaxies and to reionize the intergalactic medium.

At high-$z$, \lya -- as the most prominent emission line -- serves as a valuable tracer of cold gas due to its resonant nature, which allows the imprinting of physical properties on the observed \lya line \citep[see reviews by][]{dijkstra14,dijkstra19,ouchi20}. 
One important example of such an imprint is the separation of \lya peaks correlating with the neutral hydrogen column density \citep{adams72,neufeld90}. It has, thus, been suggested that a narrow peak separation is an indication for Lyman continuum (LyC) escape \fesclyc \citep{verhamme15,2016ApJ...828...71D,Izotov18,flury22_b}. However, observing a clear \lya spectrum becomes challenging at higher $z \gtrsim 4$ due to extinction by the neutral intergalactic medium (IGM) \citep{laursen11,byrohl20,kulkarni18,hayes21,smith22}. 
Moreover, at $z < 2$, the observed wavelength of \lya is in Far-UV regime $< 3000$~\AA, necessitating the use of the Hubble Space Telescope (HST) for \lya observations.

To overcome the limitations of \lya, 
the \mgii doublet emission is spotlighted as a promising new tool for studying the cold CGM. Like \lya, the resonant nature of the \mgii doublet allows it to carry information about the cold medium \citep{prochaska11,henry18,katz22}. The \mgii emission undergoes less extinction by the neutral IGM than \lya due to its small \mgii fraction (approximately $10^{-4.5}$ in solar metallicity). Because the vacuum wavelength of \mgii doublet emission is near 2800 \AA, ground-based observations can capture \mgii at $0.1 <z < 2$, and it at $z > 6$ can be observable via the James Webb Space Telescope (JWST). Observations of the \mgii halo have already been made at $z < 2$ \citep[][]{rubin11,rickards19,zabl21,burchett21}. 

The advent of integrated field spectrographs (IFS) such as {MUSE} \citep[MUlti Unit Spectroscopic Explorer;][]{muse_obs}, {KCWI} \citep[Keck Cosmic Web Imager;][]{kcwi_obs}, and {HETDEX} \citep[Hobby-Eberly Telescope Dark Energy Experiment;][]{hetdex_obs} allows us the study extended regions spatially resolved in emission. Notably, many studies have been carried out using \lya \citep{steidel10,cantalupo14,wisotzki16,wisotzki18,cai19,fab19,li21,maja22a,maja22b,gronzalez23}
but more recently, also spatially resolved \mgii emission nebulae have been observed through IFS \citep{chisholm20,zabl21,burchett21,leclercq22,dutta23,guo23b}. 


Previous studies have explored the properties of \mgii in conjunction with \lya and LyC to leverage \mgii as an analog of \lya and a tracer of \fesclyc. 
\cite{chisholm21} proposed the doublet ratio of \mgii as a tracer of \fesclyc at $z > 6$ using the JWST. These previous studies highlight the potential of \mgii as an analog of \lya for probing the properties of the cold CGM.
\citet{henry18} conducted simultaneous observations of \lya and \mgii spectra in Green Pea galaxies at $z\sim 0.2-0.3$ to examine the relationship between the two lines. They discovered a linear correlation between the escape fractions of \lya and \mgii, accompanied by a broadening and redshifting of the \mgii spectrum at lower escape fractions. Furthermore, they established that the \mgii doublet ratio, represented by the ratio of \mgii $\lambda$2796 to $\lambda$2803, is contingent on \fesclyc \cite[also see][]{chisholm20,seive22,izotov22,xu23}. Galaxies with low \fesclyc exhibited a \mgii doublet ratio below 2, while the ratio approached 2 in cases with high \fesclyc. In a separate investigation, \citet{katz22} tested a method of measuring \fesclya using the doublet ratio post-processing a cosmological simulation. They highlighted the effectiveness of this method for optically thin \mgii lines in galaxies and proposed \mgii emission as an intriguing probe for studying galaxy kinematics at higher redshift in JWST observations.

Radiative transfer (RT) models that incorporate the scattering processes of metal resonance doublets are essential to interpret observational data correctly. 
While the study of \lya radiative transfer has a long history \citep[e.g.][]{ahn01,ahn02,zheng02,hansen06,verhamm06,duval14,gronke17,seon20,byrohl22,chang23}, less attention has been paid to other resonant lines. 
Previous work often employs Sobolev's approximation \citep{sobelev60,yoo01,kasen02}, which assumes large velocity gradients in the flow and generally is challenging to implement in complex geometries.
A full treatment of \mgii and Fe~II fluorescence lines was studied in \cite{prochaska11} through a Monte-Carlo RT method, focusing on a homogeneous, outflowing media. More recently, \citet{michel-dansac20} presented a versatile, 3D Monte Carlo radiative transfer of resonant lines concentrating on post-processing cosmological simulations.


In this paper, we want to focus on the resonant line transfer of \mgii and \lya in simplified, homogeneous, as well as multiphase geometries. This approach allows us to closely study the physics shaping observables, ultimately helping us to link them to physical quantities while at the same time circumventing convergence issues galactic and cosmological simulations usually suffer from \citep[e.g.,][]{faucher-giguere10,vandevoort19,hummels19}. For this purpose, we develop a new 3D radiative transfer simulation code -- dubbed \texttt{RT-scat} -- which can be used for \mgii and other resonance doublets. It is in parts based on the \lya RT simulation code presented in \citet{seon22,chang23}.

This paper is constructed as follows.
In \S~\ref{sec:RT}, we describe the physics of radiative transfer, the method of our Monte Carlo simulation, and the geometry of our model, which consists of a spherical scattering medium surrounding a point emission source.
\S~\ref{sec:smooth} and \S~\ref{sec:clumpy} present the simulated results of homogeneous and multiphase media, respectively, considering monochromatic light and in situ Gaussian emission. 
In these sections, we perform a comparative analysis of the radiative transfer of \mgii and \lya, focusing on their spectral profiles and escape fractions. 
In \S~\ref{sec:flat_continuum}, we study the scattering of the stellar continuum near \mgii doublet. Furthermore, we investigate the doublet ratio of \mgii $\lambda$2796 and $\lambda$2803 considering in situ \mgii emission and the continuum.
In \S~\ref{sec:discussion}, we discuss the implications of the results obtained from our \mgii and \lya radiative transfer simulations. 
Finally, \S~\ref{sec:conclusion} succinctly summarizes the main conclusions.

\section{Radiative Transfer for \mgii Doublet}\label{sec:RT}

Our simulation code for the \mgii resonance doublet \texttt{RT-scat} is based on the 3D \lya Monte-Carlo radiative transfer code used previously in \citet{chang23} and {\it LaRT} used, e.g., in \citet{seon20, seon22}. 
Since the atomic structure of metal resonance doublets is identical to the hydrogen atomic fine structure of \lya, we can easily adapt the atomic physics of \lya RT to \mgii RT.

The fine structure of \lya has two transition $(S_{1/2} - P_{1/2})$ and $(S_{1/2} - P_{3/2})$ commonly referred to {\it 'H`} and {\it 'K`} transitions, respectively. But, the energy difference between them is with $\sim 10^{-4}$ eV, corresponding to $\sim 1.5$\kms velocity separation of K and H transitions, too small to split the lines observationally.
However, this effect becomes observable in metal resonant doublets such as \mgii, where the doublet is located at 2796 \AA\ and 2803 \AA.
Hereafter, we will also refer to the \mgii doublet $\lambda\lambda$ 2796 \& 2803 as  {\it 'K`} \& {\it 'H`} line, respectively.

We use a 3D grid-based Monte Carlo simulation on a Cartesian coordinate system.
In a Monte-Carlo radiative transfer simulation, each photon package corresponds to a certain luminosity and is defined via a current wavelength $\lambda$ (frequency $\nu$), a position $\textbf{r}$, a direction ${\bf k}$ and a Stokes vector for polarization ${\bf S} = (I,Q,U,V)$. 
In the simulation, we consider two types of scattering media: homogeneous medium with uniform density and clumpy medium composed of small clumps.
Here, we use a uniform density and a constant effective temperature in our `smooth medium' models or for each clump in our `clumpy medium' models.
In this section, we briefly introduce the atomic physics of \mgii RT, the scattering geometry, and the method of our Monte Carlo simulation.



\subsection{Scattering Cross Section}\label{sec:cross_section}

\begin{figure*}
	\includegraphics[width=\textwidth]{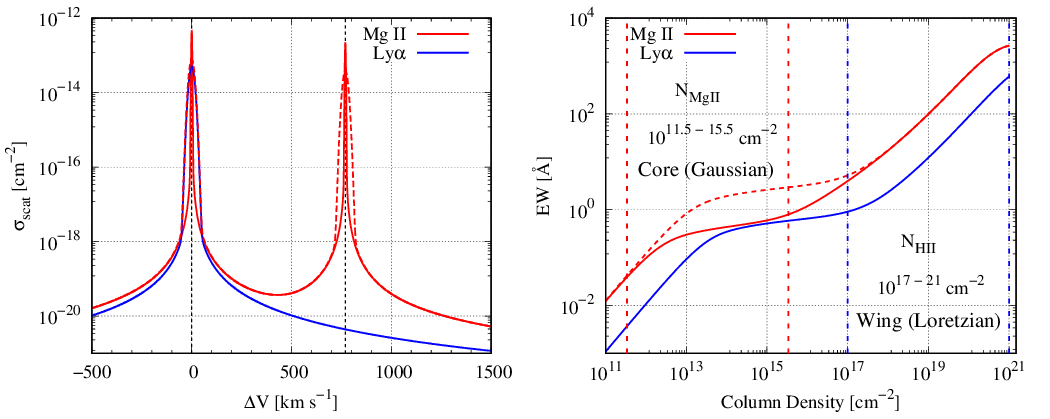}
    \caption{
    {\it Left panel:} Scattering cross section of \mgii doublet (red) and \lya (blue) as a function of the Doppler factor $\Delta V$ defined in Eq.~\eqref{eq:doppler_factor}.
    The solid and dashed red lines are for $\vth \sim 2.67 \kms$ and $15 \kms$, the thermal velocities of \mgii and \hi at temperature $T=10^4\, \rm K$, respectively. Note that to the oscillator strength, the scattering cross-section of \mgii is larger than that of \lya.
    The black vertical lines in the left panel indicate the line centers of the \mgii K and H lines.
    {\it Right panel:} Curve of growth of \mgii and \lya versus column density. The red and blue vertical lines represent reasonable ranges of \mgii and \hi column densities, respectively, assuming solar metallicity.
    The equivalent width of the \mgii doublet is larger than that of \lya at the same column density.
    }
    \label{fig:cross_section}
\end{figure*}

The scattering cross-section of each \mgii line is described by  the Voigt profile function, which yields a scattering cross-section for the \mgii doublet
\begin{equation}\label{eq:cross_section}
    \sigma_\nu = {\frac{\sqrt{\pi}e^2}{  m_e c } } \left[ \frac{f_{\rm K}}{\Delta \nu_{D,\rm K}} H(x_{\rm K},a) + \frac{f_{\rm H}}{\Delta \nu_{D,\rm H}} H(x_
{\rm H},a) \right],
\end{equation}
where $m_e$ is the electron mass and $c$ is the light speed. $f_{\rm K} = 0.608$ and $f_{\rm H} = 0.303$ being the oscillator strengths of the \mgii K and H transitions, respectively.
In this equation, $H(x,a)$ is the Voigt-Hjerting function given by
\begin{equation} \label{eq:voigt}
H(x,a) = \frac{a}{\pi} \int^{\infty}_{-\infty} \frac{e^{-y^2}}{{(x-y)^2 + a^2}}dy,
\end{equation}
where $a = \Gamma/(4\pi\Delta \nu_D)$ is the natural width parameter.
The damping constants of the K and H lines are $\Gamma_{\rm K} = 2.60 \times 10^{8} \rm\ s^{-1}$ and $\Gamma_{\rm H} = 2.57 \times 10^{8} \rm\ s^{-1}$, respectively.
The dimensionless frequency parameter $x$ is given by 
\begin{equation}
x=(\nu - \nu_0)/\Delta \nu_D
\label{eq:x}
\end{equation}
where $\Delta \nu_D = \nu_0 \vth/c$ is the thermal Doppler width with the thermal velocity $\vth = \sqrt{2k_BT/m_{\rm atom}}$, and $\nu_0$ are the line center frequencies -- which for the K and 
 H lines are $\nu_{0,\rm K} = 1.0724 \times 10^{15}\, \rm s^{-1}$ and  $\nu_{0,\rm H}= 1.0697 \times 10^{15}\, \rm s^{-1}$, respectively. Throughout the text, we will use $x_{\rm H}$ and $x_{\rm K}$ to specify the doublet lines and use $x = x_{\rm K}$ if not specified.
Note that if another random motion component is present,
\vth becomes \sigr $=\sqrt{\vth^2 + v_{\rm turbulent}^2}$.
Hereafter, note that \sigr is the random motion component instead of \vth.

Fig.~\ref{fig:cross_section} shows the scattering cross-sections and curves of growth of \mgii and \lya (in the left and right panel, respectively).
We used \vth $\sim 2.67 \kms$ and $15 \kms$ for \mgii and \lya, i.e., the thermal velocities at $T \sim 10^4 \, \rm K$,  respectively.

In the left panel, we plot the cross sections in velocity space where
\begin{equation}\label{eq:doppler_factor}
    \Delta V = \left(\frac{\nu_0}{\nu} - 1 \right)c \approx -v_{\rm th} x,
\end{equation}
where $\nu_0$ for \mgii is $\nu_{\rm 0, K}$.
The right panel of Fig.~\ref{fig:cross_section} shows that for a given column density, the equivalent width of \mgii is greater than that of \lya at the same column density because the cross section of \mgii is larger than that of \lya at the same thermal velocity since $f_{\rm K} = 608$ is greater than the \lya oscillator strength $f_{12} =0.4162$.
However, of course, generally, the \mgii column density \Nmg in a \hi region is much smaller than \NHI. This is because the typical Mg II fraction is $\sim 10^{-5.47}$ (i.e., $n_{\rm Mg II} \sim 10^{-5.47} n_{\rm HI}$), assuming solar relative abundances with an absolute metallicity of half solar metalicity \citep{prochaska11}; in the solar metalicity, the Mg fraction is $\sim 10^{-4.46}$ \citep{holweger01}.
Thus, \mgii photons scatter much less than \lya photons escaping the same medium.

In this work, we study the \mgii column densities from $10^{11.5}$ to $10^{15.5} \unitNHI$ corresponding to \NHI $=10^{17-21} \unitNHI$.
In this \Nmg range, an optical depth at the line center $\tauzero$ could be less than 1 at \Nmg$<10^{12} \unitNHI$, and wing (Rayleigh) scattering is negligible.
On the other hand, $\tauzero$ of \lya is always much larger than 1 ($\tauzero > 10^4$) in this \NHI range. 
Therefore, as we will show below, the behavior of \mgii RT is entirely different from that of \lya RT. 


\subsection{Geometry}\label{sec:geometry}

\begin{figure*}
	\includegraphics[width=160mm]{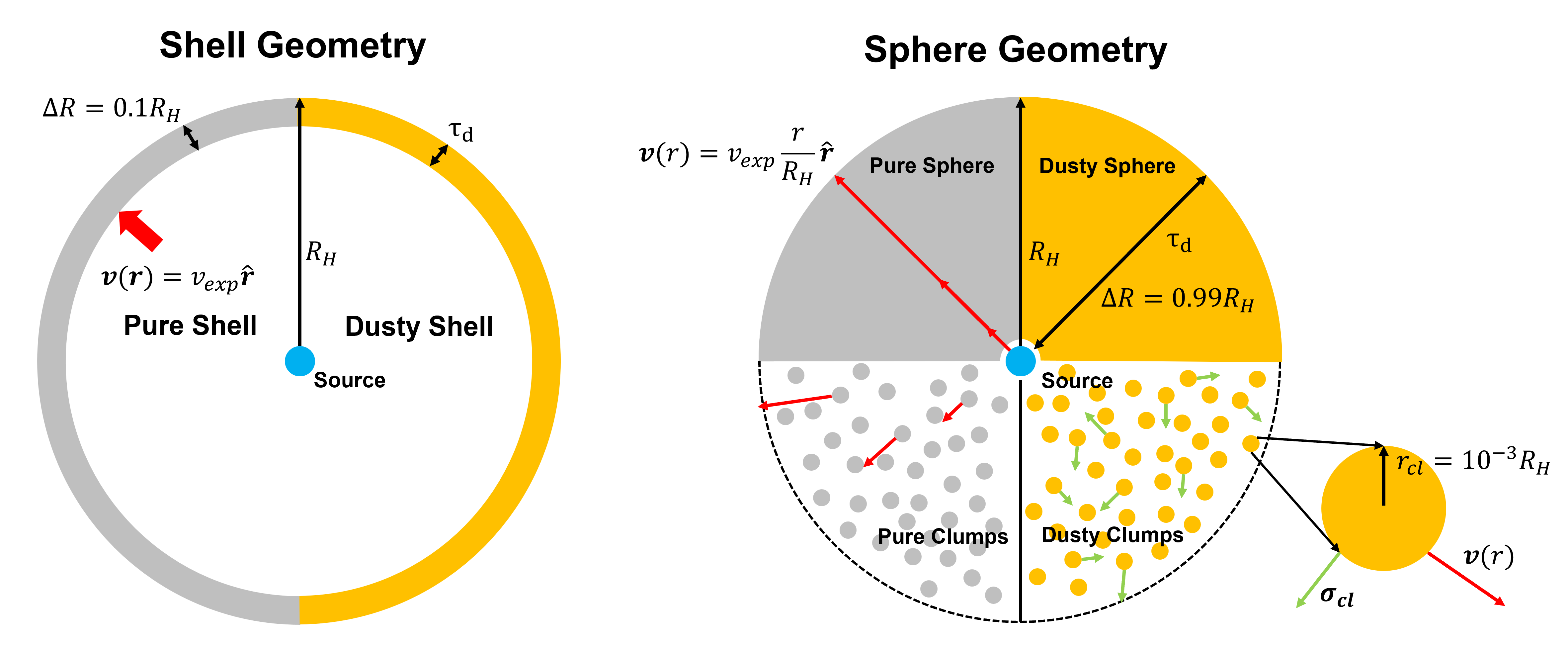}
    \caption{The schematic illustration of 'Shell` (left) and 'Sphere` (right) geometries.
    Both models possess a centrally located emission source.
In the `Shell' geometry, the \mgii and \hi filled region has a thickness of $\Delta R = 0.1 \RH$, where $\RH$ represents the total radius. The radial velocity $v(r)$ and the number density of \mgii \& \hi are constant.
In contrast, the `Sphere' geometry features a thicker \hi and \mgii filled region with $\Delta R = 0.99 \RH$. In this case, the radial velocity $v(r)$ is proportional to the distance from the source. The `Clumpy Sphere' (lower half of the right panel) mimics a multiphase system in which cold clumps with size of $10^{-3} \RH$ fill an otherwise empty spherical region with radius $\RH$. The clumps can exhibit two types of kinematics: an outflow characterized by $v(r)$ and a random motion $\sigcl$. In both models, since the dust-to-gas ratio is fixed, the dust distribution is characterized by a dust optical depth $\tau_{\rm d}$}.
    \label{fig:geometry}
\end{figure*}

In this study, we focus on three distinct geometries, `Shell', `Sphere', and 'Clumpy Sphere'.
Fig.~\ref{fig:geometry} shows schematic illustrations of these geometries.
In the 'Shell' and 'Sphere' geometries, \hi \& \mgii atoms are uniformly distributed, and the source is at the center of the halos.
The shell geometry consists of a thin \hi medium including \mgii with 
a constant radial speed and a thickness 10\% of a halo radius \RH. This geometry is widely used for \lya RT and was shown to reproduce a range of observed \lya spectral profiles \citep{ahn02,verhamm06,dijkstra08,gronke16_shell}.
We adopt the shell geometry with a monochromatic source to test our simulation in \S~\ref{sec:smooth_mono}.

In the sphere geometry, \hi \& \mgii are distributed in the radius from 0.01\RH to \RH, and
the radial velocity is proportional to the distance from the emission source $r$ ($v(r) = \vexp r\/\RH$).
The \hi \& \mgii halos in those geometries are characterized by a \hi column density \NHI, a \mgii column density \Nmg =  $10^{-5.5} \NHI$, a random motion speed $\sigr$, a maximum expansion velocity \vexp, and a dust absorption optical depth \taudabs. In Table~\ref{tab:par_sphere}, we summarize these parameters and their ranges.

We assume the range of \vexp in the sphere geometry from $-1000 \kms$ (inflow) to 1000 \kms (outflow).
When an expansion velocity \vexp $\gtrsim$ 750 \kms, which is the velocity separation of \mgii doublet, 
the \mgii transfer faces that the photons near the K line are scattered by the H transition of the outflowing medium.
Otherwise, H photons are scattered by the K transition in the inflowing medium.
In this case, \mgii doublet lines are mixed by scattering, and the ratio of the K and H lines varies.
Therefore, we assume an extensive inflow and outflow speed range from 0 to 1000 \kms, including a speed higher than the line separation of \mgii doublet.


The `Clumpy Sphere' geometry consists of spherical cold clumps of radius $\rcl$ embedded within a spherical region with radius $\RH$.
Apart from the total column density and bulk flow, this setup is characterized by two more parameters, namely a covering factor \fc and a clump's random motion \sigcl.
The covering factor \fc is the mean number of clumps in the line of sight from the source to a halo surface.
The clump's random motion \sigcl represents a random motion for each clump drawn from a Gaussian probability distribution.

Note that the size of clumps $\rcl$ does not affect the radiative transfer process when $\rcl \ll \RH$  -- the only important parameter is the mean number of clumps per sightline \fc \citep{hansen06,chang23}. We thus set the clump to halo radius ratio $\rcl/\RH=10^{-3}$. This value allows us to study and consider the range $\fc=[1,\,100]$ in a computationally feasible manner. \footnote{Since at fixed \fc, the total number of clumps and the volume filling fraction increase and decrease with decreasing $\rcl$, respectively, the clump radius cannot be too large, while too small values significantly increase the computational cost.
.}
 Since we assume spherical halo and \hi clump, the total \mgii column density of the clumpy medium is $\frac{4}{3} \fc \Nmgcl$ where \Nmgcl is a clump's column density.
The ranges of these parameters for the clumpy sphere are shown in Table~\ref{tab:par_sphere}.
We will discuss \mgii RT in the clumpy medium in \S~\ref{sec:clumpy}.


\begin{table*}
\centering
\caption{Parameters of the spherical geometry (cf. right panel of Fig.~\ref{fig:geometry}).}
\begin{tabular}{lll}
\hline
 Parameter   & Range                                                     & Note \\ \hline
 \Nmg        & $10^{11.5-15.5} \unitNHI$             & Total \mgii column density \\
 \NHI        & $10^{17-21} \unitNHI$             & Total \hi column density \\
 \vexp       & $-1000 \kms$ (inflow) $- \, +1000 \kms$ (outflow)    & Maximum expansion velocity \\
 \taudabs     & $0- 10$             & Dust absorption optical depth of \mgii  \\
 \sigr     & $2.7^{\rm \dagger}$  $- \ 30 \kms$  & Random velocity of scattering medium \\ 
 $\fc^{\rm *}$      &  $1 - 100$                                       & Covering factor \\
 $\sigcl^{\rm *}$      &  $0 - 100 \kms$                                       & Clumps' random velocity  \\
 $\rcl^{\rm *}$      &  $10^{-3} \RH$ (fixed)                         & Clump radius  \\
 $\sigsrc^{\rm **}$      &  $0 \kms$ (monochromatic) \& $100 \kms$ (Gaussian)                         & Width of intrinsic emission \\
 $\EWint^{\rm **}$      &  0 (flat continuum) $-\ 100$                        & Intrinsic equivalent width of \mgii emission \\
\hline
\end{tabular}
\\
\footnotesize{
$\dagger$: thermal speed of \mgii at $T = 10^4\, \rm K$, *:  used in the clumpy spherical geometry, **: parameters describing the emitting source}
\label{tab:par_sphere}
\end{table*}

\subsection{Emission Mechanism}\label{sec:source}

The formation of \mgii doublet emission feature is influenced by the intrinsic radiation and the physical properties of the cold gas. The cold gas properties have been studied via the absorption feature of \mgii doublet in QSO spectra \citep{prochaska13,bouche16,schroetter21,weng23}. However, the powering mechanism of the intrinsic \mgii emission is still unclear.
When \mgii emission mainly originates from collisional excitation of \mgii or recombination of \ion{Mg}{III}, the intrinsic profile is a Gaussian with width set by the random motion of the emission region.
Also, in this case, the line ratio of \mgii is intrinsically $\sim 2$
because the statistical weight of $P_{3/2}$ state for the K line is 2 times higher than that of $P_{1/2}$ state for the H line.
Alternatively, \mgii can be generated by scattering of stellar continuum near 2800 \AA. 
In this case, the doublet ratio of \mgii is not 2 even though the scattering cross section of the K line is 2 times higher than the H line \citep{prochaska11}.
The observable information of escaping \mgii photons depends on the density field, kinematics, and physical properties of cold CGM.

Here, we assume three types of the central source,
\begin{enumerate}
\item a monochromatic case: the source emits line center photons. While this is unrealistic, it helps us understand the basic principles of \mgii transfer. In this case, we use the canonical doublet ratio of $2$.
\item a Gaussian case: the intrinsic profile is Gaussian profile with width \sigsrc = 100 \kms. Also, in this case, we use an intrinsic doublet ratio of $2$.
\item a stellar continuum case: the incident radiation is a flat continuum near \mgii.
\end{enumerate}

In addition, we consider the synthesized source composed of the Gaussian and stellar continuum components, characterized by the equivalent width of the intrinsic Gaussian emission \EWint in Section~\ref{sec:ratio_emission_continuum}. The parameters for the source are shown in Table~\ref{tab:par_sphere}.

\subsection{Dust Extinction}\label{sec:dust}

UV radiation is affected by absorption and scattering by dust grains (\citealp{cardelli89,seon16}, also see reviews \citealp{mathis90,salim20}).
Our simulation considers dust absorption and scattering adopting the dust scattering phase function in \cite{henyey41} and albedo $\alpha_{\rm d}$(= 0.5718 at 2800 \AA\ and 0.3258 at 1216 \AA) adopting the fiducial dust model for the diffuse interstellar medium of the Milky Way described in \citet{draine03a,draine03b}. 

In the calculation, we also consider the contribution of free electrons in the optical constants of graphite, as described in \citet{draine84}.
We consider one dust property for \mgii doublet at 2796 and 2803 \AA\ because 
the absorption optical depth of the K and H lines are very similar, with a 0.2\% difference between them, and the difference of $\alpha_{\rm d}$ is 0.1 \%.
Note that in this Milky Way model -- and hence in our simulations -- the dust absorption optical depth of \lya is 2.53 larger than for \mgii.
The ratio of the dust absorption optical depth of \lya and \mgii is 3.79 and 9.26 in the LMC and SMC models, respectively. Therefore, a lower metallicity can cause a larger difference in the escape fraction of \lya and \mgii.
As illustrated in Fig.~\ref{fig:geometry},
we characterize the dust absorption optical depth for \mgii by \taudabs~=~$(1-\alpha_{\rm d}) \tau_{\rm d}$, where $\tau_{\rm d}$ is the dust extinction optical depth. Consequently, in the `clumpy' model, a clump's dust optical depth in the radius \rcl is $3\taudabs / {4\fc}$ as a clump's column density \Nmgcl =  $3\Nmg / 4\fc$.

Note that this simplified treatment of dust implies a fixed dust-to-gas ratio per simulation, i.e., perfectly mixed dust into the gas. While this does not represent realistic astrophysical conditions, we focus in this work on the radiative transfer effects only, thus leaving more complex dust structures -- which are supported by our simulation framework -- for future work. We specify the treatment of dust further in Appendix~\ref{sec:dust_model}.


\subsection{Monte Carlo Simulation}

Our Monte-Carlo radiative transfer simulation, {\tt RT-scat}, follows five steps,
\begin{enumerate}

\item
Initializing the scattering gas geometry and kinematics characterized by the parameters explained in \S~\ref{sec:geometry} on a 3D Cartesian grid with $200$ cells per dimension. For the smooth medium, the physical properties of each cell are determined based on the position of the cell's center. In the clumpy medium, clumps are randomly generated within each cell, and the physical properties are assigned based on the position of the clump's center. The medium between clumps is considered empty. The velocity of a clump is determined by random motion and radial velocity. In this simulation, a constant clump radius and clump column density are assumed, although different values can be considered.

\item
Generating an initial photon at the central source. The central source emits initial photons isotropically. The number of photons generated depends on the source type. For monochromatic sources, $10^5$ photons are generated at the line center. For Gaussian emission, the initial wavelength follows a Gaussian distribution with a width of 100 \kms. The simulation generates $10^5$ and $10^7$ photons for the Gaussian emission of \lya and \mgii doublet, respectively. In the Gaussian and monochromatic cases, the intrinsic \mgii doublet ratio is 2. In the case of a flat continuum for \mgii, $10^8$ photons are generated in the velocity range from $-2000\kms$ of the K line to $+2000\kms$  of the H line.

\item
Finding a scattering location by considering the cross sections for scattering with the atom (\hi or \mgii) and dust as a function of frequency. First, a dimensionless free path of the photon, $\tau_f = -\ln r_u$, is generated, where $r_u$ is a uniform random number between 0 and 1. The simulation then converts the free path to a physical distance to determine the location of the next scattering event. In the smooth medium, when the photon passes through a cell, $\nu$ is fixed at the frequency in the rest frame of the cell. The physical distance is computed based on the physical properties of each cell. Using the rest frame wavelength, the total scattering cross section $\sigma_{\rm tot}$ is the sum of the cross sections for scattering with \hi or \mgii, $\sigma_\nu$ (as described in Eq.~\eqref{sec:cross_section}), and the dust cross section, $\sigma_{\rm dust}$ (explained in \S~\ref{sec:dust}). In the case of the clumpy medium, the simulation identifies the closest clump in the direction of the photon and computes the free path assuming the wavelength in the rest frame of the clump and considering the sub-grid geometry.

\item
Computing the frequency $\nu$, direction $\bf{k}$, and Stokes vector ${\bf S}$ of the scattered photon. After determining the scattering position, the simulation selects which type of scattering the photon undergoes using a uniform random number. The computation of ${\bf k}$ and ${\bf S}$ for scattered \lya photons, considering the $S_{1/2} - P_{1/2}$ and $S_{1/2} - P_{3/2}$ transitions, is described in Section 2.2 of \cite{chang23}. Since the atomic structure of \lya and \mgii doublet are identical, the scattering phase function for \lya is used in \mgii scattering in our simulation. The computation of these vectors for dust-scattered photons is explained in Section 2.6 of \cite{seon22}.

\item
Repeatedly executing steps (iii) and (iv) until the photon breaks free from the 3D grid.
Once it escapes, $\nu$ is converted to the frequency in the observed frame. The simulation collects the escaping photon for further analysis.
After that, a new initial photon is continuously generated until the total number of photons generated so far is less than the total number specified in step (ii).

\end{enumerate}


\section{Results: Smooth Medium}\label{sec:smooth}

In this section, we present our numerical results in a smooth medium with two types of sources: a monochromatic light and an in situ Gaussian emission with the width $\sigsrc = 100 \kms$.
Regarding the geometries considered, in \S~\ref{sec:smooth_mono}, we investigate the monochromatic light case in a shell geometry to test our simulation and compare it with analytic solutions. Additionally, we consider a spherical geometry in \S~\ref{sec:smooth_mono} and \S~\ref{sec:smooth_gaussian}, for both monochromatic and Gaussian emission sources, respectively.
In all cases, the intrinsic doublet ratio of \mgii K and H lines is fixed to the fiducial value of $2$. 
We investigate the formation of the spectral profile of \mgii lines, the escape fraction, and the doublet ratio of escaping \mgii photons.
We also compare \mgii and \lya radiative transfer under the same physical conditions.

\subsection{Monochromatic Source}\label{sec:smooth_mono}

\begin{figure*}
	\includegraphics[width=\textwidth]{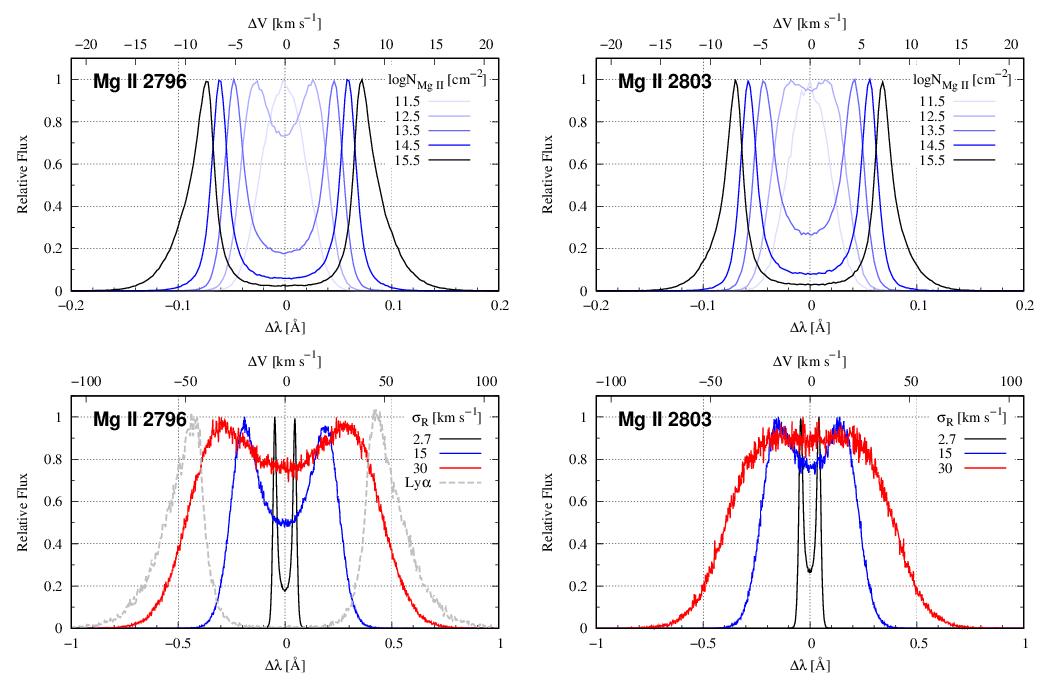}
    \caption{\mgii spectra of scattered photons in a static shell geometry with monochromatic emission. The left panel corresponds to the \mgii $\lambda$2796 (K line), while the right panel represents the \mgii $\lambda$2803 (H line).
    The top panel shows the spectrum for various \mgii column densities \Nmg\ at \sigran = 2.7 \kms. 
    The bottom panel displays the spectrum for various random speeds \sigran at \Nmg = $10^{13.5} \unitNHI$.
    The gray dashed line in the left bottom panel represents the \lya spectrum of the static medium with the monochromatic light at an \hi column density $\NHI = 10^{18} \unitNHI$.
    The \mgii spectrum broadens as \Nmg or \sigran increase.
    }
    \label{fig:spec}
\end{figure*}

\begin{figure}
\includegraphics[width=80mm]{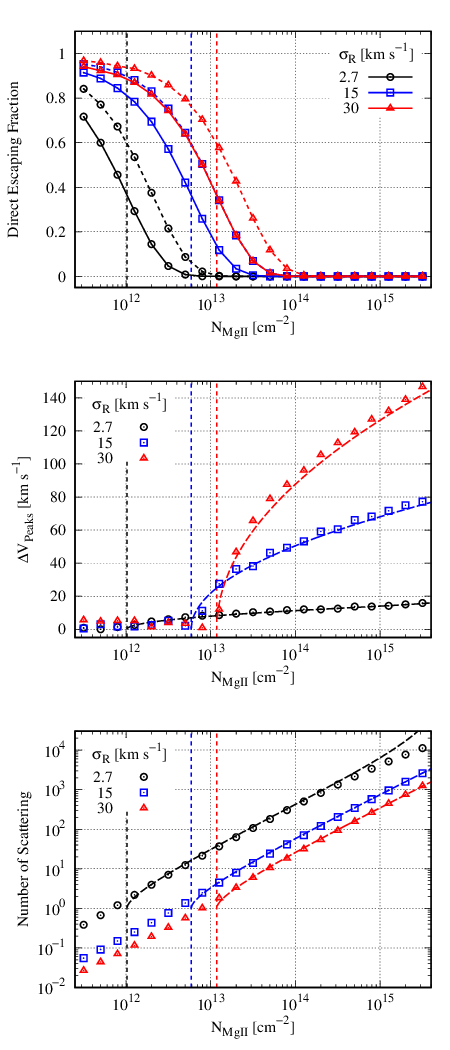}
    \caption{The fraction of directly escaping photons (top), velocity offset of the spectral peaks \dvpeak (middle), and the number of scattering \Nscat (bottom) of the \mgii K line as a function of \mgii column density \Nmg, in the shell model with the monochromatic source.
    The vertical dashed lines denote the K line center optical depths of unity.
    The solid and dashed lines in the top panel represent the direct escape fraction of the K and H lines, respectively.
    The colors of the lines represent the thermal motion of scattering medium $\sigran = $ 2 (black), 15 (blue), and 30 (red).
    The dashed lines in the middle and bottom panels are the analytic solutions of \dvpeak and \Nscat, respectively.
    The dashed lines in these panels represent the analytic solutions for \dvpeak and \Nscat, respectively.
    Overall, the simulated values closely follow the analytic solutions,
    }
    \label{fig:shell_result}
\end{figure}

In order to study \mgii radiative transfer, we first focus on the most simplified conditions and assume a shell geometry with a central point source emitting monochromatic light with frequencies at the line center of the \mgii K and H lines.
Fig.~\ref{fig:spec} shows \mgii spectra of the static shell for three random velocities of the scattering medium $\sigr = 2.7, 15$, and 30 \kms and various \Nmg=$ 10^{11.5-15.5} \unitNHI$.
Note that for illustration purposes, the spectra shown in Fig.~\ref{fig:spec} are composed of only scattered photons and normalized to the peak flux value.

The top panels of Fig.~\ref{fig:spec} show that the spectrum becomes broader with increasing \Nmg.
At $\Nmg \lesssim 10^{11.5} \unitNHI$, both profiles of \mgii doublet converge to a single Gaussian with width $\sigr$ because the optical depth at the line center $\tau_0$ is less than unity. 
In the bottom panels, the profiles become broadened with more substantial flux at the line center as \sigr increases.

As discussed in \S~\ref{sec:cross_section} (Fig.~\ref{fig:cross_section}), \mgii transfer differs from \lya transfer because of the different range of \mgii and \hi column densities and consequently very different optical depths.
In the case of \lya, the extremely high optical depth 
leads to a double peaked spectrum with zero flux at the line center (see the gray dashed line in the left bottom panel of Fig.~\ref{fig:spec}). 
On the other hand, \mgii spectra show non-zero flux at the line center due to much lower optical depth as compared to \lya. 

In the following sections, we show
(1) a comparison between the simulation results and analytic solutions,
(2) the escape fraction of \mgii dusty media, and
(3) the doublet ratio of \mgii.

\subsubsection{Simulation Comparing with Analytic Solution}

We compare the simulated results of the shell geometry without dust and analytic solutions from the literature \citep{osterbrock62,adams72,neufeld90}.
In Fig.~\ref{fig:shell_result}, we show three values of the \mgii K line (from top to bottom panel): the fraction of directly escaping photons (the fraction of escaping photons without scattering), the peak separation of double peaked spectra, and the mean number of scattering \Nscat as a function of \Nmg for \sigr = 2.7, 15, and 30 \kms.
In the top panel, the direct escape fraction decreases with increasing \Nmg because more photons experience scattering.
In the middle and bottom panels, the profiles of \dvpeak and \Nscat (solid lines) are comparable to analytic solutions (dashed lines).
Note that here the escaping frequency (here expressed as velocity) $v_{\rm esc}$ is simply given by $\tau_{\nu} = \Nmg \sigma_{v} = 1$ where $\sigma_v$ is the scattering cross section following a Voigt profile Eq.~\eqref{eq:cross_section}.

In the bottom panel of Fig.~\ref{fig:shell_result}, the numerical results of \Nscat follow the analytic solution, given by \Nscat $= 1/\erf(v_{\rm esc}/v_{\rm th})$ \citep{osterbrock62,neufeld90}. 
In summary, two statistical values from our simulation in the shell model, \dvpeak and \Nscat, closely match the analytic solutions.

Note that these solutions imply that the photon escapes via a single long flight and, hence, are also applicable to \lya radiative transfer but only for low/intermediate optical depths ($a \tau_0 \lesssim 1$, i.e., $\tau_0 \lesssim 2 \times 10^4$ or $\NHI \lesssim 10^{17} \unitNHI$ for $T\sim 10^4$~K). For more common optical depths, \lya photons escape via consecutive wing-scatterings, also known as `excursion' \citep{adams72}.

\subsubsection{Escape fraction \fesc}
\label{sec:fesc_mono_smooth}

\begin{figure*}
	\includegraphics[width=175mm]{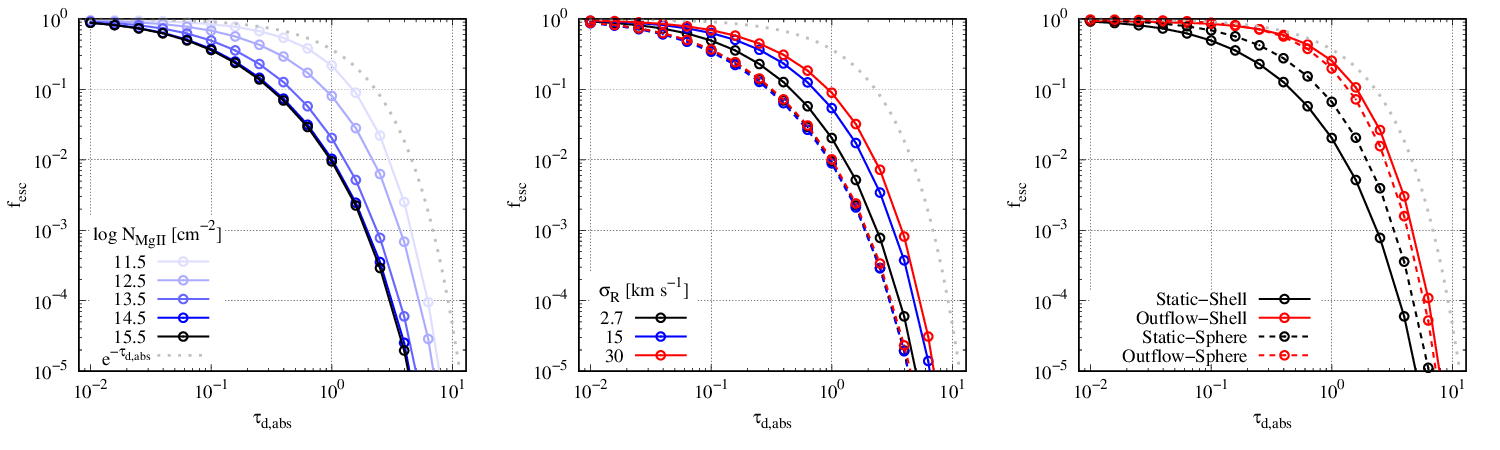}
    \caption{
    Escape fractions \fesc of \mgii as a function of dust absorption optical depth \taudabs for various \Nmg (left), random motion speed \sigr (center), kinematics-geometry (right panel).
    The gray dotted lines are a simple dust extinction curve $\exp (-\taudabs)$.
    {\it Left:} \fesc profiles of static shell geometry in the range of \NHI = $10^{11.5}$ to $10^{15.5} \unitNHI$. The shades of blue represent different \Nmg values.
    {\it Center:} \fesc profiles for three \sigr = 2.7 (black), 15 (blue), and 30 (red).
    The solid and dashed lines correspond to two distinct \Nmg values: $10^{13.5}$ and $10^{15.5} \unitNHI$, respectively.
    {\it Right:} \fesc profiles of static and outflowing media at \Nmg = $10^{13.5} \unitNHI$.
    `Static' and `Outflow' represent \vexp = 0 and 10 \kms, respectively.
    The solid and dashed lines are for the shell and sphere geometry, respectively.
    }
    \label{fig:fesc_smooth}
\end{figure*}

\begin{figure}
	\includegraphics[width=80mm]{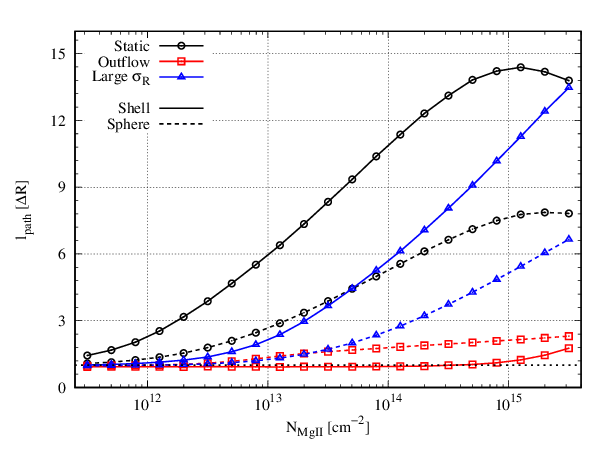}
    \caption{
    The path length \lpath of escaping \mgii K photons as a function of \Nmg in a unit $\Delta R$, which is 0.1\RH and 0.99\RH in the shell (solid) and sphere (dashed line) geometries, respectively.
    The black and red lines are identical to the `static' and `outflow' cases in the right panel of Fig.~\ref{fig:fesc_smooth}.
    The blue lines represent a static case with \sigr = 30 \kms.
    In the static cases, \lpath of the shell geometry is longer than that in spherical geometries (cf. \citealp{Adams75}).
    In outflowing cases, the sphere geometry has a longer path due to the different velocity distributions of the scattering medium.
    }
    \label{fig:path_smooth}
\end{figure}

The escape fraction of \mgii depends on how much dust the photons encounter on their escape path and, thus -- as in \lya -- primarily on the dust content but also on the \mgii column density \Nmg, the kinematics, and the structure of the gas.

Fig.~\ref{fig:fesc_smooth} shows the escape fraction \fesc of the K line as a function of dust absorption optical depth \taudabs in a homogeneous medium.
One obvious trend is that \fesc decreases with increasing \taudabs.
In the left panel, we show \fesc for various \Nmg=$10^{11.5-15.5} \unitNHI$.
If the incident photons experience only dust absorption, the profile of \fesc is equivalent to the simple dust extinction curve $\exp{(-\taudabs)}$.
However, one can note from Fig.~\ref{fig:fesc_smooth} that \fesc profiles are usually steeper than this dust extinction function (shown as the black dotted line).
This is because a larger \Nmg leads to more scatterings and a longer path length \lpath of each photon to escape from the scattering medium.

Fig.~\ref{fig:path_smooth} illustrates this point further and shows the average path length the photons travel in the scattering medium \lpath of the emergent \mgii K line in a dust-free medium in units of the physical thickness of scattering medium $\Delta R = R_o - R_i$.
The longer \lpath, the greater the dust extinction the photon experiences.
Hence, \fesc decreases with increasing \Nmg at the same \taudabs -- as seen in Fig.~\ref{fig:fesc_smooth}.

The left panel of Fig.~\ref{fig:fesc_smooth} shows that the profile of \fesc is similar to the simple dust extinction curve $\exp{(-\taudabs)}$ at low \Nmg ($\sim 10^{11.5} \unitNHI$) and \taudabs < 1.
However, even though most photons escape without \mgii scattering at this low \NHI, overall \fesc in \taudabs > 1 is smaller than $\exp{(-\taudabs)}$. 
Due to the fact that our simulation considers dust scattering and absorption together,
multiple dust scatterings occur in the medium with \taudabs > 1. 
These multiple dust scatterings cause a longer path length through the dusty medium, and thus $\fesc < \exp(-\tau_{\rm d})$.
In conclusion, an increased dust optical depth decreases the escape fraction in two ways: by directly destroying the \mgii photons and by increasing the path length of escaping photons.

Interestingly, for column densities $\Nmg > 10^{14} \unitNHI$, the escape fraction \fesc does not depend on \Nmg as shown in the left panel of Fig.~\ref{fig:fesc_smooth} where \fesc for \Nmg = $10^{14.5}$ and $10^{15.5}\unitNHI$ are identical.
In this high \Nmg regime ($\tauzero > 10$),
the initial monochromatic light must experience numerous scatterings to escape the scattering medium and has a longer \lpath.
However, Fig.~\ref{fig:path_smooth} shows 
the profile of \lpath for the static shell case (black solid line) becomes flattened near \Nmg$\sim 10^{15} \unitNHI$.
As \Nmg increases in this high \Nmg regime, a free path for each \mgii scattering gets shorter; the free path near the line center is $\sim \tauzero^{-1}$.
Hence, the total path length slightly increases due to the short free path.
We expect that the \fesc profile does not depend on \Nmg before \Nmg reaches the value ($> 10^{17} \unitNHI$), which is enough for the Lorentzian part of Voigt profile to be optically thick like \lya RT.

The central panel of Fig.~\ref{fig:fesc_smooth} shows \fesc for three random velocities \sigr= 2.7, 15, and 30 \kms.
We find that \fesc at \Nmg = $10^{13.5} \unitNHI$ increases with increasing \sigr at the same \taudabs-- but the dependence on \sigr is negligible at larger column densities \Nmg = $10^{15.5} \unitNHI$.
This is supported by Fig.~\ref{fig:path_smooth}, where the path lengths through the scattering medium \lpath for \sigr = 2.7 \kms (black curves) is longer than those at \sigr = 30 \kms (blue curves) at $\Nmg < 10^{15} \unitNHI$.
\lpath for \sigr = 2.7 \kms and = 30 \kms are similar at \Nmg $10^{15.5} \unitNHI$.

In the right panel of Fig.~\ref{fig:fesc_smooth}, 
we show that \fesc in and outflowing medium is greater than that of the static medium.
This is simply because, for the considered monochromatic source, most atoms are out of resonance and do not interact with the medium; in other words, the path length \lpath is $\sim 1$ (cf. Fig.~\ref{fig:path_smooth}.
This short \lpath due to an outflow medium leads to less interactions with dust and aids the escape of \mgii photons.
Specifically, Fig.~\ref{fig:path_smooth} shows that overall \lpath for outflowing cases is $\sim 3\Delta R$ which is 5-7 times smaller than that for static cases at \Nmg = $10^{15.5} \unitNHI$.

In addition to the effects described above, the geometry of the scattering medium also affects the escape fraction \fesc. 
The right panel of Fig.~\ref{fig:fesc_smooth} shows that in the static case, the escape fraction from a shell geometry is smaller than that of a filled sphere. 
This is because the width of the shell is much smaller than its radius. Hence, the escape through the shell model is similar to a semi-infinite slab where (straight) crossing sightlines can, in principle, intercept the slab for an infinite distance. This is different in the spherical model, where the maximum crossing path length from the central source to the surface is equal to the sphere's radius. This effect was studied in \citep{Adams75} in detail and numerically again, e.g., in \citep[][; see their Appendices B \& C]{seon20}.

However, for the outflowing cases, this order is reversed, and \fesc in the sphere is slightly smaller than in the shell. 
This is because the radial velocity $v(r)$ is constant in the shell and proportional to the distance from the source in the sphere. 
Thus, in the shell, the monochromatic light becomes quickly optically thin because the shell has only one radial velocity, whereas, in the sphere, several photons of monochromatic light can be scattered by the slowly moving inner halo.
Fig.~\ref{fig:path_smooth} also shows this effect through the small difference of \lpath of the outflowing shell (red solid) and sphere (red dashed).

\subsubsection{\mgii doublet ratio}\label{sec:ratio_mono}

\begin{figure*}
	\includegraphics[width=\textwidth]{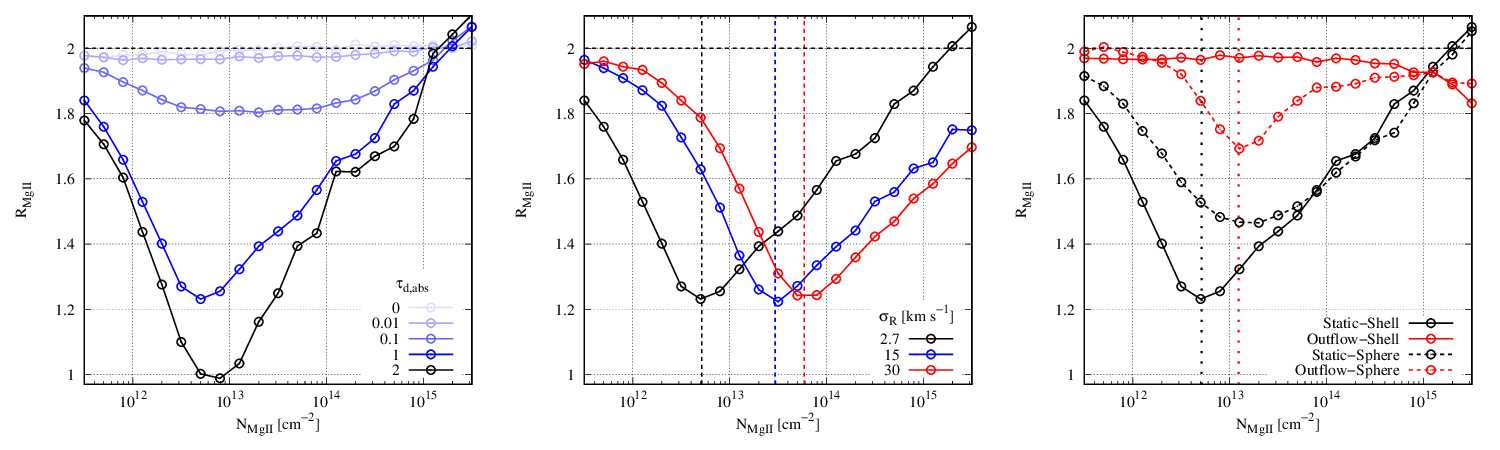}
    \caption{ 
    The \mgii doublet ratio \Rmg of the monochromatic light case as a function of \Nmg for various \taudabs (left), \sigr (center), and kinematics-geometry (right). The left panel displays \Rmg for the static shell with \taudabs = 0 (dust free) -- 2. (since \fesc is too small to be observed for \taudabs > 2; cf. Fig.~\ref{fig:fesc_smooth}). 
    The central panel shows \Rmg for three different \sigr while \taudabs$=1$ is fixed. The vertical dashed lines in the corresponding color represent $\tauzero = 5$. 
    The right panel shows \Rmg of static and outflowing media and highlights the differences between the shell and the sphere geometries while \sigr = 2.7 \kms is fixed. The vertical dotted lines indicate $\tauzero=5$ (black) and 12 (red). For \taudabs $\gtrsim 1$, \Rmg depends on \Nmg. As \Nmg increases, \Rmg decreases up to a specific optical depth ($\tauzero \approx 5$). For larger column densities, \Rmg increases again. 
    }
    \label{fig:ratio_smooth}
\end{figure*}

\begin{figure}
	\includegraphics[width=80mm]{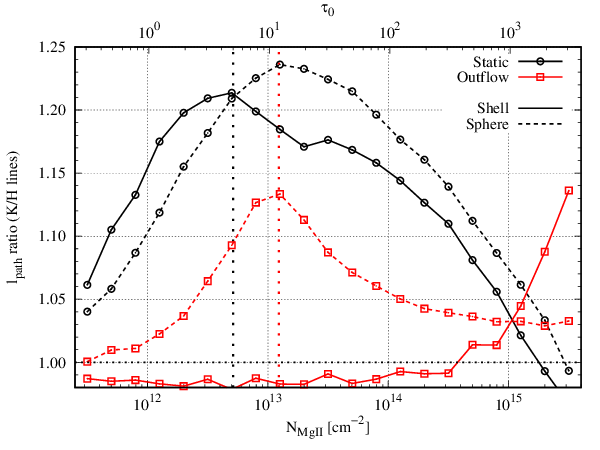}
    \caption{
    The ratio of \lpath of \mgii K and H photons for a static/outflowing medium in the shell/sphere geometry.
    The vertical dashed lines represent $\tauzero= 5$ (black) and 12 (red) as the dashed lines in the right panel of Fig.~\ref{fig:ratio_smooth}.
    }
    \label{fig:path_ratio_smooth}
\end{figure}

The previous section shows that the escape fraction \fesc depends on a \mgii column density \Nmg at the same dust absorption optical depth \taudabs due to variations in the path length through the dusty medium \lpath.
A longer \lpath leads to a decreased \fesc because the escaping photons get more chance to be absorbed by dust.
Since the scattering cross section of \mgii K is twice that of the H line,
\lpath of the K line can be larger than \lpath for the H line.
Thus, the doublet ratio of \mgii can decrease in a dusty medium.
To understand this effect, we clearly define the doublet ratio of \mgii K and H lines \Rmg.
Generally, the \mgii doublet ratio of the \mgii emission case
\begin{equation}\label{eq:doublet_ratio_gaussian}
    \Rmgemis = \frac{\int^{\lambdaK + \Delta \lambda_{300}}_{\lambdaK - \Delta \lambda_{300}} F(\lambda) d \lambda  }
    {\int^{\lambdaH + \Delta \lambda_{300}}_{\lambdaH - \Delta \lambda_{300}} F(\lambda) d \lambda  }
\end{equation}
where $F(\lambda)$ is the \mgii spectrum as a function of a wavelength, $\lambda_{\rm K,H}$ is the line center wavelength of \mgii K,H line, and $\Delta \lambda_{300} \sim 2.8$ \AA\ is the wavelength shift corresponding to 300 \kms.
Indeed, if the \mgii K and H lines can not be mixed via scattering processes, the doublet ratio can become the ratio of the escaping fraction.
Hence, as we set the intrinsic doublet ratio of the K and H line emission to its canonical value of 2, the doublet ratio of the monochromatic light case is given by
\begin{equation}\label{eq:doublet_ratio_mono}
    \Rmgemis = \frac{2 f_{\rm esc,K}}{f_{\rm esc,H}},
\end{equation}
where $f_{\rm esc,K}$ and $f_{\rm esc,H}$ are escape fractions of \mgii K and H lines, respectively.

Fig.~\ref{fig:ratio_smooth} shows examples of this behavior. It displays \Rmg profiles as a function of \Nmg in the same setups as shown in Fig.~\ref{fig:fesc_smooth}.
As expected, \Rmg decreases with increasing \taudabs at $\Nmg < 10^{13} \unitNHI$ since  K line photons experience more dust absorption than H line photons.
Also, expected for the same reason is the decreasing trend of \Rmg with increasing column density for $\Nmg > 10^{13}\,{\rm cm}^{-2}$.
However, for larger column densities, this trend is reversed, and there is a characteristic minimum in the \mgii doublet ratio at the line center optical depth of the K line $\tauzero \approx 5$ (corresponding to $\Nmg\approx 10^{13}\,\mathrm{cm}^-2$).

The central panel of Fig.~\ref{fig:ratio_smooth} shows \Rmg for three \sigr in the static medium versus the column density.
The minimum \Rmg exists at the vertical dashed lines representing the value of \Nmg for \tauzero = 5 of different \sigr.
The right panel shows \Rmg for the static and outflow cases of the shell and sphere.
Also, the dip of \Rmg profiles of the static shell lies at the vertical lines for $\tauzero = 5$.

To understand \Rmg profile near \tauzero = 5, Fig.~\ref{fig:path_ratio_smooth} shows the ratio of the path lengths of \mgii K and H lines as a function of \Nmg.
The peak of the ratio of \lpath of the static shell appears at the vertical black dotted line.
For the static sphere case, this peak is slightly shifted to \tauzero = 12 (marked with the red vertical dotted line in Fig.~\ref{fig:path_ratio_smooth}). This is because the escape path \lpath in the sphere is generally shorter than that of the shell -- as seen in the left panel of Fig.~\ref{fig:path_smooth} and discussed in \S\ref{sec:fesc_mono_smooth}.

For $\tauzero < 1$ corresponding to $\Nmg < 10^{12} \unitNHI$, both of \mgii lines are optically thin, and most photons directly escape without scattering; \lpath $\sim 1$ and the ratio of \lpath $\sim 1$.
Thus, the difference of \lpath between two lines is too small to affect the interaction with dust.

For $\tauzero \gg 1$ (\Nmg $> 10^{15} \unitNHI$), both of \mgii lines are optically thick, and all photons must experience multiple scatterings before escaping; \lpath $> 10$.
But, in this high \Nmg regime, the profile of \lpath versus \Nmg flattens as seen in Fig.~\ref{fig:path_smooth}.
Hence, in this case, the factor of 2 difference between the cross section of the K and H lines is negligible. Consequently, \lpath of the two lines are similar (cf. Fig.~\ref{fig:path_ratio_smooth}).

Only in the intermediate regime, $\tauzero \sim 5$, the two times larger scattering cross section makes a notable difference in the \lpath ratio. It is also noteworthy that in this regime, the doublet ratio of \mgii lines \Rmg strongly depends on \taudabs (cf. left panel in Fig.~\ref{fig:ratio_smooth}).

For an outflowing medium, we find that the doublet ratio \Rmg is very different.
The right panel of Fig.~\ref{fig:ratio_smooth} shows the profiles of \Rmg for static and outflow cases.
The profiles \Rmg of the outflow case are flat in the shell and have a minor dip in the spherical case.
As shown in Fig.~\ref{fig:path_smooth}, in the outflowing shell case, \lpath $\sim 1$ regardless of \Nmg. Because of constant radial velocity, all monochromatic light directly escapes.
In the outflowing sphere case, the inner sphere with a slowly outflowing medium causes the monochromatic light to scatter.
However, the optical depth of this inner halo is smaller than the total optical depth \tauzero.
If the source is not monochromatic, the variation of \Rmg in the outflowing medium becomes significant.
We will discuss it in the following sections using Gaussian emission.

\subsection{Gaussian Source}\label{sec:smooth_gaussian}

In this section, we simultaneously run the simulation for \lya and \mgii in the smooth spherical geometry.
We adopt the Gaussian profiles as incident radiation for \lya and \mgii doublet with the same intrinsic width \sigsrc = 100 \kms.
The random speed of the scattering medium \sigr is fixed at the thermal speed with $T=10^4\, \rm K$, $\sim 2.7 \kms$, and 14 \kms for \mgii and \hi media, respectively.
The ratio of \mgii and \hi column densities \Nmg/\NHI is fixed at $10^{-5.5}$ following solar metalicity.
Even though the atomic physics affecting the radiative transfer for \mgii and \lya are alike,
we will see that the results of the two lines are not similar due to the very different column densities in the neutral medium.
The following sections compare spectra and escape fractions of \mgii and \lya. 
Furthermore, we show the doublet ratio of \mgii for various \Nmg and expansion velocities \vexp.

These different ranges of the \mgii and \hi column densities lead to very different scattering effects.
As \NHI $> 10^{19} \unitNHI$, 
the Lorentzian part of the scattering cross section ($\Delta V > 3 \vth$), starts becoming optically thick ( see the left panel of Fig.~\ref{fig:cross_section}).
This leads to multiple \lya scattering even though the wavelength of the incident photon is far from the line center.

On the other hand, in the range of \Nmg = $10^{11-16} \unitNHI$, the Lorentzian part is always optically thin for \mgii photons.
This means that only photons in the Gaussian part ($\Delta V \lesssim 3 \vth$) of the rest frame of an \mgii atom undergo scattering.
Because of these two very different scattering behaviors and thus different escape paths, comparing emergent \mgii and \lya observables can lead to additional insights into the physical parameters of the scattering medium (see \S~\ref{sec:discussion} where we discuss this further).


\subsubsection{\mgii \& \lya spectra}

\begin{figure*}
	\includegraphics[width=170mm]{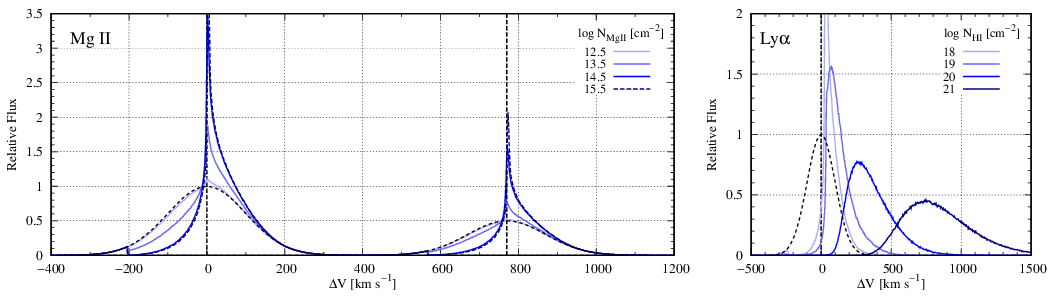}
    \caption{ The spectra of \mgii (left) and \lya (right) for various \mgii \& \hi column densities at \vexp  = 200 \kms.
    The dashed black lines represent the intrinsic Gaussian profile.
    Note that in the left panel, the \mgii spectra for $\Nmg = 10^{14.5}$ (blue) and $10^{15.5} \unitNHI$ (dashed dark blue) are nearly identical.
    }
    \label{fig:smooth_spec_NH}
\end{figure*}

\begin{figure*}
	\includegraphics[width=170mm]{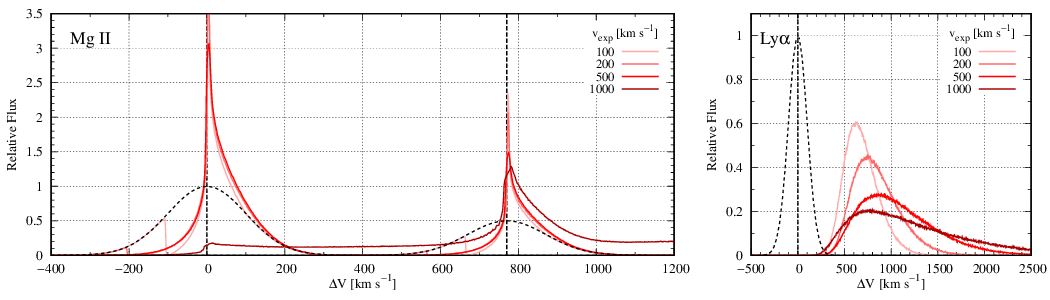}
    \caption{ The spectra of \mgii (left) and \lya (right) for various expansion velocities \vexp at \Nmg = $10^{15.5} \unitNHI$ and \NHI = $10^{21} \unitNHI$, respectively.
    }
    \label{fig:smooth_spec_vexp}
\end{figure*}

Fig.~\ref{fig:smooth_spec_NH} shows the emergent spectra of \mgii and \lya for various column densities, \Nmg and \NHI, at an expansion velocity \vexp = 200 \kms.
The right panel shows the total integrated \lya spectra for various \NHI = $10^{18-21} \unitNHI$.
The \lya spectrum becomes more extended and redshifted with increasing \NHI; the spectral peaks move to the redward.
However, \mgii spectra for \Nmg = $10^{12.5-15.5} \unitNHI$ (shown in the left panel of Fig.~\ref{fig:smooth_spec_NH}) possess absorption-like features in the blueward part of the spectrum and the steep asymmetric profile redwards. Also -- as opposed to the \lya spectra -- the spectral peak is always at the line center.

It is noteworthy that for \Nmg $\le 10^{13.5} \unitNHI$, the H and K lines of the emergent \mgii spectra differ. Specifically, the blue absorption features of the K line are roughly two times stronger than that of the H line due to the larger cross-section of the K line. Also, some blueward photons can directly escape because of the low optical depth at the line center \tauzero $\lesssim 1$.
But, at in the high \Nmg $\ge 10^{14.5} \unitNHI$ regime, the spectra of K and H lines are nearly identical because \tauzero of both lines $\gg 1$ akin to \lya transfer. Consequently, those spectra are very similar to the \lya spectrum with $\NHI = 10^{18} \unitNHI$.

In summary, \mgii transfer for large \Nmg is analogous to \lya transfer for the low \NHI, leading to similar spectral shapes in \mgii H and K lines. On the other hand, for lower \Nmg,  the absorption features of the K and H lines differ. \\

Fig.~\ref{fig:smooth_spec_vexp} shows the \mgii (\lya) spectra for four expansion velocities \vexp = 100, 200, 500, and 1000 \kms at \Nmg = $10^{15.5}$ (\NHI = $10^{21} \unitNHI$).
The left panel shows that \mgii spectra for \vexp = 100-500 \kms are very similar on the redward side because here, most photons directly escape from the source.
The blueward side, however, shows absorption features from -\vexp to 0 \kms.

In the right panel of Fig.~\ref{fig:smooth_spec_vexp}, the \lya spectra show a very different trend with varying \vexp. 
Here, the red wing is more extended and red-shifted with increasing \vexp. This trend continues even to very large outflow velocities.
For the \mgii spectrum, this is very different. For the largest outflow velocity shown ($\vexp = 1000 \kms$), the spectrum is composed of a weak K line and a strong H line.
This is because if the outflow velocity is stronger than the peak separation of the \mgii doublet $\sim 750 \kms$, the K line photons can be scattered by the H transition, leading to a redistribution near the K line.

Thus, the scattering of the H transition in the strong outflow can suppress the flux of the K line.
For that reason, the doublet ratio of \mgii decreases with increasing the outflow velocity, especially \vexp $> 750 \kms$.
Otherwise, in the case of an inflowing medium, the doublet ratio increases with increasing inflow velocity.
We will measure the doublet ratio and discuss it in \S~\ref{sec:ratio_smooth}.\\


\begin{figure*}
	\includegraphics[width=170mm]{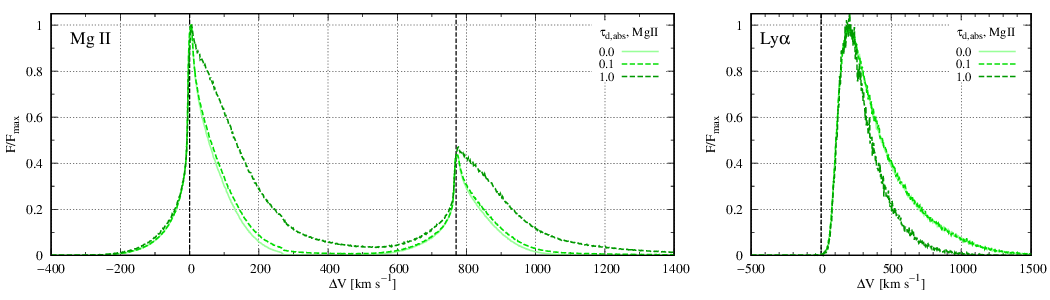}
    \caption{
    The normalized spectra of \mgii (left) and \lya (right) for various dust optical depth \taudabs at \Nmg = $10^{14.5}\unitNHI$ (\NHI = $10^{20} \unitNHI$) and \vexp = 500 \kms. 
    The solid line is the spectrum of the dust free medium (\taudabs = 0).
    The green dashed lines represent the spectra from the dust medium at \taudabs 0.1 and 1.0.
    The spectrum is the total integrated spectra divided by the peak value.
    The spectra of \mgii (\lya) at \taudabs = 1 show a stronger (weaker) red wing compared to the dust-free case.
    }
    \label{fig:smooth_spec_tauD}
\end{figure*}

Dust affects \mgii spectra in a different way compared to \lya.
Fig.~\ref{fig:smooth_spec_tauD} shows the spectra of \mgii (\lya) for three dust optical depths \taudabs = 0, 0.1, and 1.0 at \Nmg = $10^{14.5} \unitNHI$ (\NHI = $10^{20} \unitNHI$).
The spectra at \taudabs = 0.1 are similar to those from the dust free medium (\taudabs = 0).
At \taudabs = 1, the spectral profile of \lya has a less extended red wing.
This is because the photons in the red wing of the spectrum, on average, undergo more scatterings with atomic hydrogen, thus have a longer path length, and are more likely to be absorbed by dust.
Therefore, the strength of the red extended wing decreases with increasing \taudabs.

This picture is reversed for \mgii (shown in the left panel of Fig.~\ref{fig:smooth_spec_tauD}). Here, the red wing of at \taudabs = 1 is stronger than at \taudabs = 0.1.
This is because, without dust, the red part of the \mgii spectrum is mainly composed of directly escaping photons.
As the dust scattering does not strongly depend on the wavelength in the range of several tens of \AA, all the escaping photons can be scattered by outflowing dust forming a red wing\footnote{For very low column densities \NHI $< 10^{17} \unitNHI$, \lya spectra show a similar trend (not shown here).}
Importantly, note that this effect cannot be observed if a radiative transfer simulation does not consider dust scattering.

\subsubsection{Escape fractions of \mgii and \lya}

\begin{figure}
	\includegraphics[width=80mm]{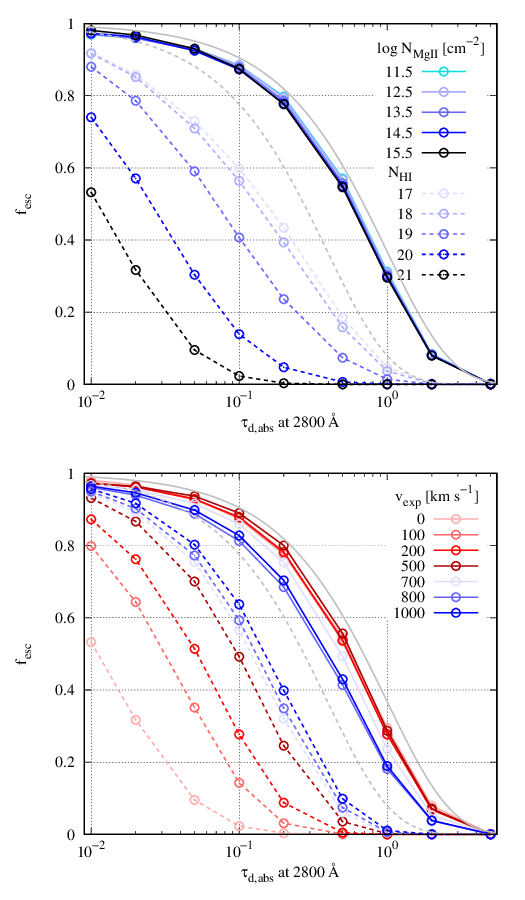}
    \caption{
    Escape fractions \fesc of \mgii (solid) and \lya (dashed lines), \fescmgii and \fesclya, in the smooth medium versus the dust absorption optical depth at 2800\AA, \taudabs.
    In the top panel, each line color represents various column densities, and \vexp is fixed at 0 \kms.
    In the bottom panel, the color corresponds to various \vexp while \Nmg (\NHI) is fixed at $10^{15.5}\unitNHI$ ($10^{21} \unitNHI$).
    The solid and dashed gray lines are the simple dust extinction function $\exp(-\taudabs)$ for \mgii and \lya, respectively.
    Overall, the dashed line is smaller than the solid line since the dust optical depth of \lya is 2.53 times higher than that of \mgii. 
    The profile of \fesclya strongly depends on \vexp and \NHI. 
    The profile of \fescmgii only slightly varies, with changing \vexp and \Nmg.
    }
    \label{fig:smooth_f_esc}
\end{figure}

The \mgii escape fraction of monochromatic light depends on both the column density and the kinematics because of the \mgii scattering effects discussed in the previous section.
But with a width of the intrinsic profile \sigsrc $\gg$ the random speed of scattering medium \sigr in a static medium, most initial photons escape without any \mgii scattering, and the path length of these escaping photons is $\sim R_{\mathrm{H}}$.

Fig.~\ref{fig:smooth_f_esc} shows \fescmgii and \fesclya, escape fractions of \mgii doublet and \lya, respectively.
The top panel shows \fescmgii in a static medium which follows $\exp(-\taudabs)$ since
only a small fraction of photons near the line center experience scatterings with \mgii atoms.
Consequently, the \fescmgii profile does not depend on the \mgii column density \Nmg.

In a smooth medium, the escape fraction of \lya is always smaller than that of \mgii for two reasons.
Firstly, \taudabs of \lya is $\sim 2.53$ times higher than that of \mgii.
Secondly, \NHI $\gg$ \Nmg causes more scatterings and a longer path length.
The top panel of Fig.~\ref{fig:smooth_f_esc} shows that \fesclya decreases with increasing \NHI.
In the high column density regime ($\NHI \ge 10^{18} \unitNHI$), \lya photons experience multiple scatterings -- even if the photon's intrinsic wavelength is in the Lorentzian part of the cross section. 
This scattering process increases the path length of \lya photons significantly.
Thus, unlike for \mgii, the dependence on the column density is not negligible for \lya radiative transfer due to the different range of column densities between \NHI and \Nmg in astrophysical systems.

Even without radiative transfer effects (i.e., for extremely low column densities $\NHI < 10^{17} \unitNHI$), the different dust cross sections and corresponding \taudabs values cause a large difference between \lya and \mgii escape fractions as illustrated in Fig.~\ref{fig:smooth_f_esc}.
In this figure, we consider the Milky Way dust model where the dust cross section for \lya is $2.53$ times larger than the one for \mgii. However, it is noteworthy that this ratio can be significantly larger in other dust models. In the dust model of the Small and Large Magellanic Clouds (SMC and LMC, respectively) the ratios are  3.79 and  9.26, respectively  \citep{draine03a,draine03b}. 
In these models, the \lya escape fraction becomes significantly smaller. We explore the different dust models in Appendix~\ref{sec:dust_model}.\\

The bottom panel of Fig.~\ref{fig:smooth_f_esc} shows \fescmgii  and \fesclya for various expansion velocities \vexp at fixed column densities \Nmg = $10^{15.5} \unitNHI$  and \NHI = $10^{21} \unitNHI$.
We find that \fescmgii in \vexp $> 700 \kms$ becomes $\sim 10\%$ smaller compared to lower outflow velocities and that the dependence on \vexp is negligible in the low \vexp regime.
In this high \vexp regime, the outflowing medium induces scatterings of the K photons in both the blueward and redward by H transition, thus increasing their path length in the exposure to dust.
Note that for an inflowing medium with similar absolute velocities $\vexp < -700 \kms$, this difference decreases to $\sim 5\%$ due to the intrinsic doublet ratio of \mgii of 2.
This dependence on \vexp becomes imperceptible as \Nmg $< 10^{13} \unitNHI$.\\

In the bottom panel of Fig.~\ref{fig:smooth_f_esc}, we confirm that \fesclya increases with increasing expansion velocity \vexp.\footnote{Here, we consider only an outflowing medium (\vexp > 0 \kms) but note that \fesclya of an inflowing medium with the same absolute velocity is identical.}
This is because the rapidly moving medium induces a large Doppler shift to the scattering photons, which escape easily with a short path length.
Note, however, that even for the highest outflow velocities $\sim 1000 \kms$, \fesclya is still lower than the simple dust extinction function $\exp(-\taudabs)$  because the large column density \NHI $= 10^{21} \unitNHI$ makes the \hi region optically thick over $\Delta V \pm 1000 \kms$ (see the left panel of Fig.~\ref{fig:cross_section}).
This dependence on \vexp becomes weaker as \NHI decreases and for $\NHI < 10^{19} \unitNHI$ and large outflow velocities |\vexp| $> 700 \kms$ \fesclya follows $\exp(-\taudabs)$.

\subsubsection{\mgii doublet ratio}\label{sec:ratio_smooth}

\begin{figure}
	\includegraphics[width=80mm]{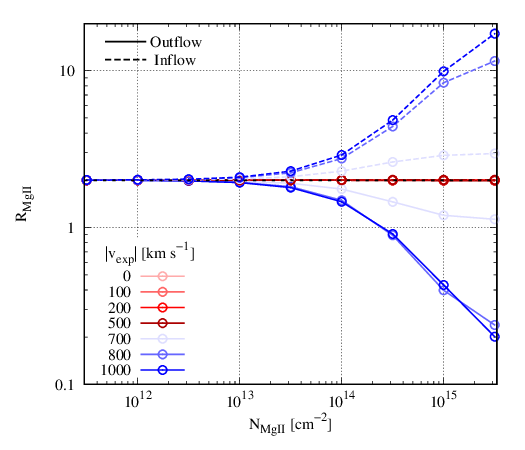}
    \caption{The \mgii doublet ratio \Rmg of the Gaussian emission case for various \vexp. 
    The solid and dashed lines represent outflow and inflow cases, respectively.
    }
    \label{fig:smooth_ratio}
\end{figure}

As shown in Fig.~\ref{fig:smooth_spec_vexp}, \mgii K and H photons can be mixed by the strong outflow or inflow medium ($|\vexp|$ > 700 \kms).
To investigate the behavior of these mixed photons by scattering processes, we measure the doublet ratio \Rmg of the \mgii spectra adopting Eq.~\ref{eq:doublet_ratio_gaussian}.
Fig.~\ref{fig:smooth_ratio} shows the doublet ratio of the Gaussian emission case as a function of \Nmg for various \vexp.
The doublet ratio \Rmg is invariant for bulk velocities $|\vexp| < 700 \kms$ regardless of other parameters.
At \vexp > 700 \kms and \Nmg > $10^{13.5} \unitNHI$,
\Rmg decreases with increasing \vexp and \Nmg because the strong outflow allows the K line photons to be scattered by the H line transition.
In the strong inflow case, $\vexp < -700 \kms$, \Rmg increases with decreasing \vexp and \Nmg; H line photons move to the blueward of the K line as shown in Fig.~\ref{fig:smooth_spec_vexp}. 
As a result, \Rmg can vary by resonance scattering at $\vexp > 700 \kms$.

\section{Clumpy Medium}
\label{sec:clumpy}

\subsection{Scattering processes in clumpy medium}\label{sec:surface_scattering}

Resonant line transfer through a multi-phase clumpy medium is fundamentally different than through a single-phase smooth medium. The main reason for this is that while in a homogeneous medium, photons can escape through `excursion' or `single-flight' (as discussed above), in a clumpy medium, a third form of escape via `random walk' between the dense clumps exists \citep{neufeld91,gronke17}. The main additional parameter controlling radiative transfer effects in a clumpy medium is the mean number of clumps per line of sight \fc \citep{hansen06}.

This implies, for instance, that a clumpy medium can enhance the \lya escape fraction due to the so-called surface scattering effect (\citealp{neufeld91,hansen06} but see, e.g., \citealp{laursen13,duval14}). This effect occurs in the multi-phase clumpy medium when the hot inter-clump medium is optically thin or empty, the cold clumps are optically thick ($\tau_{\rm cl}\gg 1$), and relatively few clumps exist (see below for an exploration of the transition in a clumpier medium). If these conditions are fulfilled, the photons cannot penetrate the clumps. Instead, they undergo several scatterings merely on the surface of the clump.
In that case, the \lya photons escape from the clumpy medium through a shorter path length within the cold gas. If the clumps are dusty, the escaping photons have less exposure to dust \citep{neufeld91}.
Therefore, the escape fraction of \lya in a clumpy medium increases with increasing clump's \hi column density \NHIcl.

If the multiphase medium becomes very clumpy, that is, if the mean number of clumps per line of sight \fc is greater than some critical value, it is advantageous for the  \lya photons to escape as in a homogeneous medium and the spectra \citep{gronke16,gronke17}, as well as the surface brightness profiles and polarization signal \citep{chang23}, converge to the homogeneous one.

Here, we adopt the algorithm to describe the scattering process in a clumpy medium described in \citet{chang23} to investigate the emergent spectra and escape fractions of \mgii and \lya in a clumpy medium. 
The right panel of Fig.~\ref{fig:geometry} shows the scattering geometry composed of a central source and a clumpy spherical medium, additionally regarding the covering factor \fc. 
The total \hi column density \NHI in the clumpy medium is $4/3\fc\NHIcl$ where \NHIcl is clump's \hi column density; in the case of \mgii, \Nmg = 4/3\fc\Nmgcl.
In this section, we investigate \mgii \& \lya radiative transfer and study the emergent spectra, escape fractions, and the doublet ratio \Rmg.
Also, our simulation considers two types of sources, a monochromatic light and Gaussian emission in \S~\ref{sec:clumpy_mono} and \ref{sec:clumpy_gaussian}, respectively.

\subsection{Monochromatic Source}\label{sec:clumpy_mono}

\begin{figure}
	\includegraphics[width=\columnwidth]{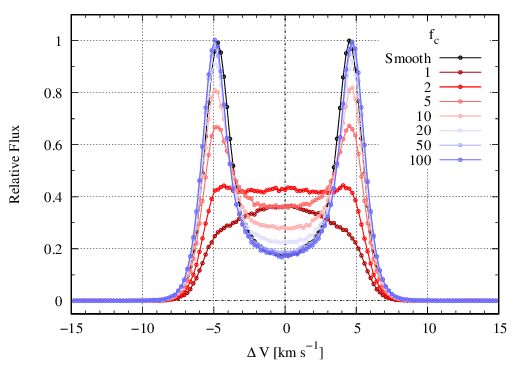}
    \caption{\mgii K line spectra of the clumpy sphere for various covering factors \fc = $1 - 100$. 
    \Nmg is fixed at $10^{13.5} \unitNHI$. 
    The spectra from the clumpy sphere become similar to the spectrum of the smooth sphere with increasing \fc.
    }
    \label{fig:clumpy_spec}
\end{figure}

\subsubsection{\mgii spectrum}

Fig.~\ref{fig:clumpy_spec} shows the spectra of the \mgii K line for various covering factors \fc at $\Nmg = 10^{13.5} \unitNHI$ adopting a monochromatic intrinsic emission.
This figure clearly shows that \mgii spectrum from a clumpy medium becomes similar to that from a smooth medium as \fc increases like \lya in a clumpy medium as shown in the previous work \citep{gronke16,gronke17,chang23}.

The black solid line in Fig.~~\ref{fig:clumpy_spec} is the spectrum of a smooth medium with the same column density \Nmg.
While the spectra of a clumpy medium with large \fc resemble that of a homogeneous one, the peak structure is washed out at  \fc $\leq 2$.
In this low \fc case, all clumps are optically thick for the monochromatic light from the source as $\Nmgcl \approx \Nmg$.
Thus, the clumps are hard to penetrate for the photons, which go through several scatterings merely on the surface of the clumps.
This implies that the photons `random walk' between the clumps experience less overall column density (as opposed to a homogeneous medium) and less line broadening.

Fig.~\ref{fig:clumpy_spec} also shows that the \mgii spectrum with \fc$ \gtrsim 50$ is a clear double-peaked profile, which is identical to that of the smooth medium.
We discuss the critical covering factor for this behavior and compare it with that of \lya in \S~\ref{sec:f_crit}.

\subsubsection{Escape fraction \fesc}\label{sec:fesc_mono_clumpy}

\begin{figure*}
	\includegraphics[width=175mm]{  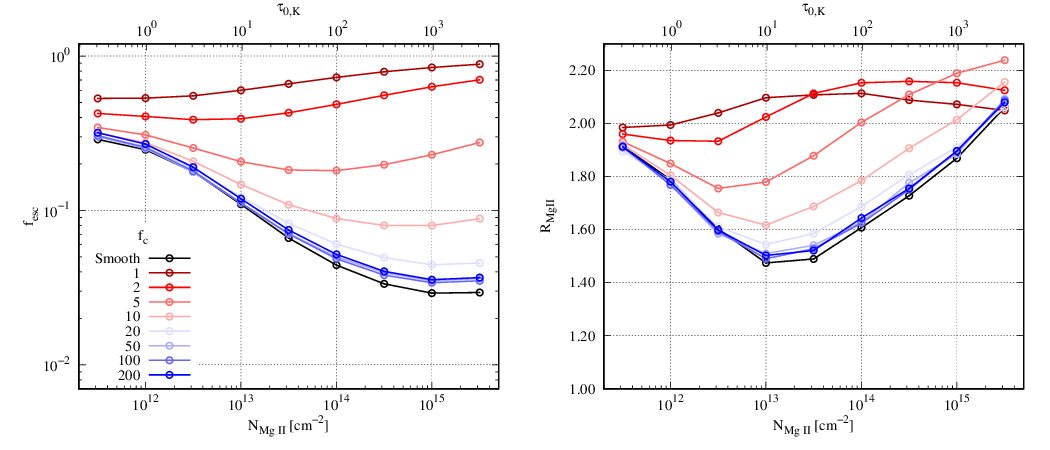}
    \caption{
    The escaping fraction \fesc (left) and doublet ratio \Rmg (right) of the monochromatic light case as a function of $\Nmg=4/3 \fc\Nmgcl$ comparing results of the clumpy medium for various covering factor \fc to the smooth medium.
    \sigr and \taudabs are fixed at 2.7 \kms and 1, respectively.
    The colors of lines represent \fc = 1 (red) - 200 (blue) of a clumpy medium and static smooth medium (black).
    As \fc increases, the profiles of \fesc and \Rmg in the clumpy medium become similar to those in the smooth medium.
    }
    \label{fig:fesc_clumpy}
\end{figure*}

Fig.~\ref{fig:fesc_clumpy} shows the escape fraction \fesc of the K line and \mgii doublet ratio \Rmg as a function of \Nmg for various covering factors \fc.
The left panel shows that \fesc increases with decreasing \fc.
Note that since the total column density is kept constant, a smaller \fc means a higher \Nmgcl. 
The increase of \Nmgcl leads to a shorter path length \lpath within the scattering medium of the escaping photon due to the surface scattering effect described in \S~\ref{sec:surface_scattering}.

In a smooth medium, \fesc decreases with increasing \Nmg because of more scatterings causing a longer \lpath (as discussed in \S~\ref{sec:fesc_mono_smooth}).
Similarly, for $\fc > 10$, the escape fraction is close to the one of a homogeneous medium.
However, for \fc $\leq 2$, the trend is clearly reversed: higher column densities imply a higher escape fraction.
This can be explained through the enhanced the surface scattering effect as discussed in \S~\ref{sec:surface_scattering}.

\subsubsection{\mgii doublet ratio }
In the right panel of Fig.~\ref{fig:fesc_clumpy}, we show the doublet ratio of the monochromatic light case, defined in Eq.~\ref{eq:doublet_ratio_mono}, emergent from a clumpy medium with total column density \Nmg. 
As before, in the escape fraction, the \Rmg for \fc $>10$ is similar to the smooth medium. 
This means there is also a characteristic minimum of \Rmg at $\tau_0\sim 12$ in the spherical geometry (cf. \S~\ref{sec:fesc_mono_smooth} for a discussion of this effect).

The trends of \Rmg versus \fc are more complicated at $\fc < 10$.
Generally \Rmg increases with decreasing \fc, however for $\Nmg\gtrsim 10^{14}\,{\rm cm}^{-2}$ this trend is non-monotonous and intermediate $\fc \sim 5$ leads to the highest doublet ratios (even larger than the fiducial value of $2$).
To understand this effect, it is important to recall that the cross section of the K line is a factor of 2 larger than that of the H line. Hence, the surface scattering effect is stronger for the K line than the H line, leading to \Rmg > 2.

\subsection{Gaussian Source}\label{sec:clumpy_gaussian}

\begin{figure*}
	\includegraphics[width=\textwidth]{  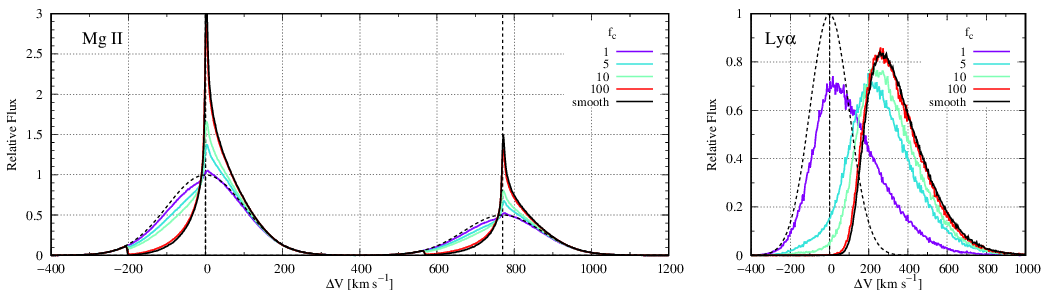}
    \caption{Integrated spectra of \mgii (left) \& \lya (right) for various \fc at $\Nmg = 10^{14.5} \unitNHI$ \& $\NHI = 10^{20} \unitNHI$. The width of the intrinsic Gaussian profile \sigsrc and \vexp are fixed at 100 \kms and 200 \kms, respectively. The color and black lines represent the spectra of the clumpy medium and smooth medium, respectively, at the same total column density \Nmg = 4/3 \fc\Nmgcl. The black dashed line is the intrinsic spectrum.
    The vertical dashed lines represent the line center of \mgii K \& H lines and \lya.
    }
    \label{fig:spec_fc}
\end{figure*}

\begin{figure*}
	\includegraphics[width=\textwidth]{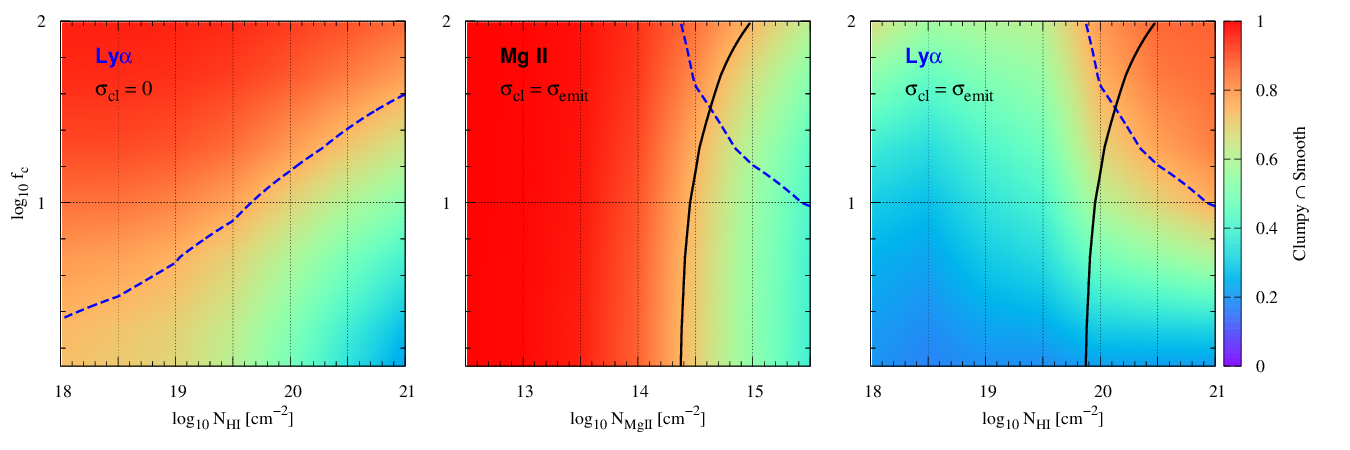}
    \caption{
    The 2D map of the similarity between \mgii (\lya) spectra from a clumpy and smooth medium as a function of \Nmg (\NHI) and covering factor \fc. 
    The similarity is the fraction of the common area of the two types of spectra as defined in Eq.~\eqref{eq:similarity}.
    The left panel shows the similarity maps of \lya in the static medium with the clumps' random velocity \sigcl = 0 \kms and \vexp = 0 \kms.
    The center and right panels show the similarity maps of \mgii and \lya, respectively, with the random motion (\sigcl and \sigr) being identical to the intrinsic emission width \sigsrc = 100 \kms. 
    The black solid and blue dashed contours represent Clumpy $\cap$ Smooth = 0.8 of \mgii and \lya, respectively.
    Here, the dashed lines represent the critical covering factor \fccrit, which is high enough \fc for the spectra from a clumpy medium similar to that from a smooth medium.
    The map of \mgii in the static medium (\sigr = \vth $\sim 2.7 \kms$) is not plotted here and is completely red (Clumpy $\cap$ Smooth $\sim$ 1) because most initial photons directly escape without scattering due to \sigsrc $\gg \vth$.
    }
    \label{fig:fc_map}
\end{figure*}

\begin{figure}
	\includegraphics[width=\columnwidth]{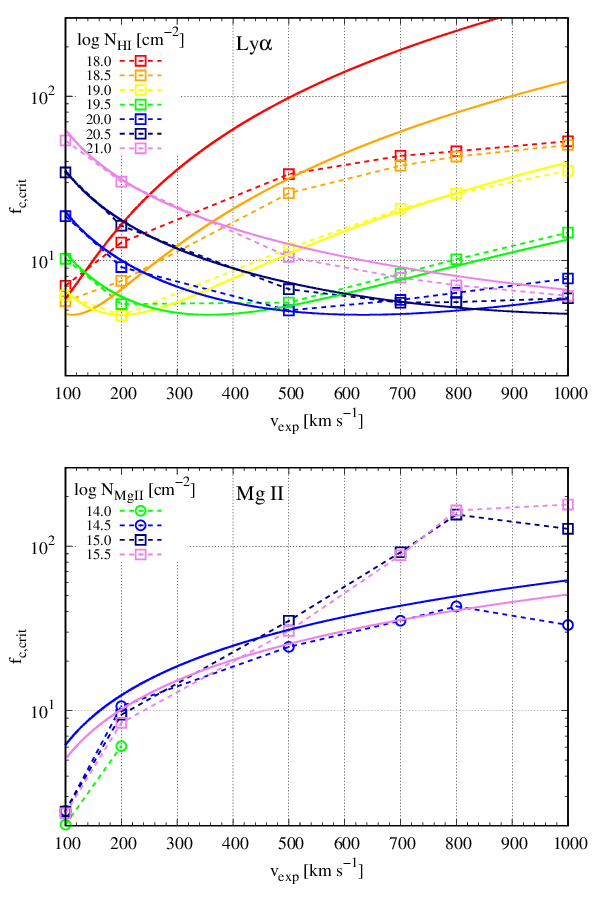}
    \caption{The critical covering factor \fccrit of \lya (top) and \mgii (bottom panel) as a function of expansion velocity \vexp when the fraction of directly escaping photons is less than 80\%.
    \fccrit represents the value of \fc when C$\cap$S = 0.8.
    The dashed lines are \fccrit from the simulation, and the colors represent the various column densities.
    The solid lines in the top and bottom panels show Eq.~\eqref{eq:fit_lya} and \eqref{eq:fit_mgii}, respectively.    
    }
    \label{fig:fc_crit}
\end{figure}

This section shows the results of \mgii \& \lya transfer in a clumpy medium with Gaussian emission with intrinsic width $\sigsrc= 100 \kms$.
We study the dependence of the \mgii radiative transfer process on the covering factor \fc and compare it to \lya. We will mainly focus on the spectral profile and the escape fraction \fesc.

\subsubsection{Spectrum and critical covering factor \fccrit}\label{sec:f_crit}

As discussed in \S~\ref{sec:clumpy_mono} and Fig~\ref{fig:clumpy_spec} for large enough \fc resonant line transfer through a multiphase, clumpy medium resembles that through a homogeneous medium.
\cite{gronke17} investigated this high \fc regime for \lya radiative transfer and found a critical covering factor \fccrit above which the photons escape through a homogeneous medium.
This prior investigation provides analytic solutions for \fccrit in various situations: static media, outflowing media, and media composed of randomly moving clumps.
In this section, we expand upon this work and find similar critical covering factors for \mgii.

Fig.~\ref{fig:spec_fc} shows the \mgii \& \lya spectra of a clumpy medium for various \fc and a smooth medium at the same total column densities, \Nmg = $10^{14.5}\unitNHI$ \& \NHI = $10^{20} \unitNHI$ and \vexp  = 200 \kms.
The spectra at \fc = 1 are different from those of a smooth medium (black solid line) and comparable to the intrinsic profile (black dashed line). This is because while, on average, each line-of-sight intercepts one clump, this is not sufficient to fully sample the kinematics of the clumps.
On the other hand, for \fc = 100, the \mgii and \lya spectra are identical to those of a smooth medium.
We confirm that both spectra of \mgii and \lya in the clumpy medium become similar to those of the smooth medium at high enough \fc.

To investigate this dependence on \fc of the spectrum further,
we measure the similarity of the spectra from the clumpy and smooth media.
Fig.~\ref{fig:fc_map} shows maps of the similarity C$\cap$S, which is a fraction of the common area overplotting the spectra of the clumpy and smooth media at the same total column density; the range of C$\cap$S is from 0 to 1.
Specifically, we define this similarity parameter as
\begin{equation}\label{eq:similarity}
    {\rm C}\cap{\rm S} = {{\int \min(F_{\rm Clumpy}, F_{\rm Smooth} ) d (\Delta V)} \over {\int F_{\rm Smooth} d (\Delta V)}}
\end{equation}
where $F_{\rm Clumpy}$ ($F_{\rm Smooth}$) is the spectrum as a function of the velocity $\Delta V$ of the clumpy (smooth) medium, which corresponds to the colored (black) solid line shown in Fig.~\ref{fig:spec_fc}.

At C$\cap$S = 1, the spectrum of the clumpy medium is identical to that of the smooth medium.
Each panel shows the map of C$\cap$S as a function of a column density \Nmg (\NHI) -- \fc.
The first--third and second panels of Fig.~\ref{fig:fc_map} are maps for \lya and \mgii, respectively.
The black (blue) contour means C$\cap$S = 0.8 for \mgii (\lya). Hereafter, we define that these contours represent the critical covering factor \fccrit because 80\% similarity is enough for the spectrum of the clumpy medium to be similar to that of the smooth medium.

In the left panel of Fig.~\ref{fig:fc_map},
we compute the map C$\cap$S for \lya in the static clumpy medium with \vexp = 0 \kms and the clump's random motion \sigcl = 0 \kms.
The critical covering factor \fccrit of \lya (blue dashed line) increases with increasing \NHI.
\cite{gronke17} already noted this behavior of \fccrit.
Equation~7 in their work suggested the analytic solution for \fccrit of \lya in the static medium that \fccrit is proportional to $\sqrt{\NHI}$.

In the case of \mgii, we do not plot the map because  C$\cap$S of \mgii is always $\sim 1$ regardless of \fc and \Nmg.
When \Nmg $=10^{11.5-15.5} \unitNHI$, only the Gaussian part of the Voigt profile \textbf{($\Delta V \lesssim 3 \vth$)} becomes optically thick (see the left panel of Fig.~\ref{fig:cross_section}). In both clumpy and smooth media, most intrinsic \mgii photons directly escape without scattering since the width of the intrinsic profile \sigsrc = 100 \kms is much broader than \vth of \mgii at $10^4$~K ($\sim 2.7 \kms$).
Therefore, the spectral profiles from smooth and clumpy media are similar to the profile of incident radiation; hence, C$\cap$S $\sim 1$. \\

In the randomly moving clumpy medium,
the critical covering factor \fccrit of \mgii and \lya show different behavior 
The center and right panels of Fig.~\ref{fig:fc_map} exhibit C$\cap$S of \mgii and \lya in the clumpy medium with the random motion of clumps \sigcl = 100 \kms, corresponding to the width of the intrinsic Gaussian profile \sigsrc.
The right panel shows that \fccrit of \lya (blue dashed line) decreases with increasing \NHI.
This behavior is similar to the analytic solution in Equation~12 of \cite{gronke17} for \lya RT in the randomly moving clumps $\fccrit \propto 1/\sqrt{\ln(4/3 \NHIcl)}$.
However, the center panel of Fig.~\ref{fig:fc_map} shows that \fccrit for \mgii is steeply increasing with increasing \Nmg ($\gtrsim 10^{14.5} \unitNHI$) -- i.e., a very different behavior as predicted.

To understand this inverse relation, it is important to recall that for \Nmg < $10^{14} \unitNHI$, C$\cap$S is always > 0.8 because most photons (> 80 \%) directly escape without scattering.  
Due to the different trends of \fccrit, \fccrit of \lya (blue contour) is smaller than that of \mgii (black contour) in the randomly moving clumpy medium at $\Nmg < 10^{14} \unitNHI$ (\NHI < $10^{19.5} \unitNHI$).
Consequently, the different trends and values of \fccrit for \lya \& \mgii allow us to constrain the clumpiness of cold gas. We will discuss these prospects further in \S~\ref{sec:f_crit_clumpy}.\\

In an outflowing clumpy medium, \fccrit behaves very differently \citep{gronke17}. In Equation~13 of their work, when \vexp is higher and smaller than the particular velocity ${\hat v_{\rm max}}$, \fccrit is proportional to $\vexp^2$ and $1/\vexp$, respectively. Here, ${\hat v_{\rm max}}$ is proportional to $\NHI^{1/3}$ and $\sim 100 \kms$ at \NHI = $10^{19} \unitNHI$.
Fig.~\ref{fig:fc_crit} shows \fccrit (dashed lines) defined as the similarity $    {\rm C}\cap{\rm S} = 0.8$ as a function of \vexp for \mgii and \lya for various column densities.

In the top panel of Fig.~\ref{fig:fc_crit}, we show the \lya results and confirm that \fccrit depends on \NHI. As \vexp increases, \fccrit increases and decreases at \NHI $\le 10^{19} \unitNHI$ and $\ge 10^{20} \unitNHI$, respectively.
In the bottom panel, \fccrit of \mgii increases with increasing \vexp.
This means it follows the trend of \lya in the low \NHI regime (< $10^{19}\unitNHI$), which is sensible because generally \Nmg $\ll \NHI$.

To compare the trend of \fccrit from our simulation and in the analytic solutions,
we find modified analytic solutions of \fccrit as a function of \vexp, which is necessary since the previous solutions do not capture the full trend observed.
The modified analytic solution of \fccrit is given by
\begin{equation} \label{eq:fit_lya}
    \fccrit(\vexp) = \sqrt{\NHI \over 10^{17} \unitNHI} {{x_* \vth} \over \vexp} + {{\sqrt{\pi}\vexp^2} \over {a_v \NHI \vth^2 \sigma_0}},
\end{equation}
where $x_*\approx 3.26$, $a_v\approx 4.7 \times 10^{-4}$, and $\sigma_0\approx 5.9 \times 10^{-14} {\rm cm}^2$ (at $T=10^4\,$K).
The modified solution is composed of the equations for \fccrit in \citet{gronke17}.
The first term $\propto \sqrt{\NHI}$ is from their equation 9 for the static medium. 
The second term $\propto 1/\vexp$ and the third term $\propto \vexp^2$ are from their equation 13.
In the top panel of Fig.~\ref{fig:fc_crit}, the modified analytic solutions (solid lines) are comparable to the simulated \fccrit at \NHI > $10^{19} \unitNHI$.
However, \fccrit at \NHI $< 10^{19} \unitNHI$ and \vexp > 500 \kms shows the difference between simulated values and analytic solutions because of the approximation of the scattering cross section.

The analytic solution of \fccrit for \lya cannot be readily adopted to \mgii because \cite{gronke17} assumes that the scattering cross section is the approximation of the Lorentzian function due to the high optical depth at the line center ($\tauzero \gg 1$) to derive the analytic solution.
In other words, they consider the Voigt-Hjerting function $H(x,a)$ in Eq.~\eqref{eq:voigt} is $\sim a/\sqrt{\pi} x^2$.
However, in \mgii RT, when \Nmg $< 10^{16} \unitNHI$, the optical depth in the Lorentzian part of the \mgii cross section is much less than unity ($\tau_\nu = \Nmg \sigma_{\nu} \ll 1$), as shown in the left panel of Fig.~\ref{fig:cross_section}.
Thus, to derive the analytic solution of \fccrit for \mgii, we should instead approximate the cross section as purely Gaussian.

For a clumpy medium with bulk velocity to behave as a homogeneous medium, the entire velocity range has to be covered by clumps. This means, in our case, that the velocity range from 0 to \vexp has to be covered by clumps along each line of sight.
For each clump, the velocity width of the optically thick region ($\tau > 1$) $\Delta V_{\rm clump}$ is given by
\begin{equation}
    \Delta V_{\rm clump} = 2 \vth \sqrt{ 2 \log (\Nmg \sigma_0)},
\end{equation}
since $H(x,a) \approx e^{-x^2}$ in the approximation of the Gaussian function. 
In this equation, \vth and $\sigma_0$ are the thermal speed of \mgii atom at $10^4$~K and the scattering cross section at the line center of the \mgii K line, respectively. Since on average \fc clumps can be found per line-of-sight the velocity range covered is $\fc \Delta V_{\mathrm{clump}}$ (i.e., \vexp $= \fccrit \Delta V_{\rm clump}$), and hence
\begin{equation} \label{eq:fit_mgii}
\fccrit(\vexp) = {\vexp \over{ 2 \vth \sqrt{2 \log (\Nmg\sigma_0})  }}.
\end{equation}

In the bottom panel of Fig.~\ref{fig:fc_crit}, we compare the analytic solutions of \fccrit for \mgii (solid lines) and numerical \fccrit profiles as a function of \vexp (dashed lines).
At \Nmg < $10^{15}$, the analytic solution is comparable to \fccrit from the simulation.
But, when \Nmg > $10^{15}$ and \vexp $\ge 700 \kms$, the numerical \fccrit is higher than the analytic value.
This gap originates from the physical properties of \mgii doublet.
As discussed \S~\ref{sec:ratio_smooth}, the K line is suppressed by scattering of the H transition in the strong outflow.
This induces the mixing of K and H line photons, which is not considered in our simple analytical estimate.
Consequently, since our analytic solution is for the K line transition, in the strong outflow/inflow regime (|\vexp| $\ge 700 \kms$), \fccrit shows the value higher than the analytic solution.

These different trends of \fccrit for \mgii and \lya open the possibility of studying the structure of a multiphase gas; we will discuss this further in\S~\ref{sec:f_crit_clumpy}.

\subsubsection{Escape fraction \fesc}\label{sec:fesc_clumpy}

\begin{figure}
	\includegraphics[width=80mm]{  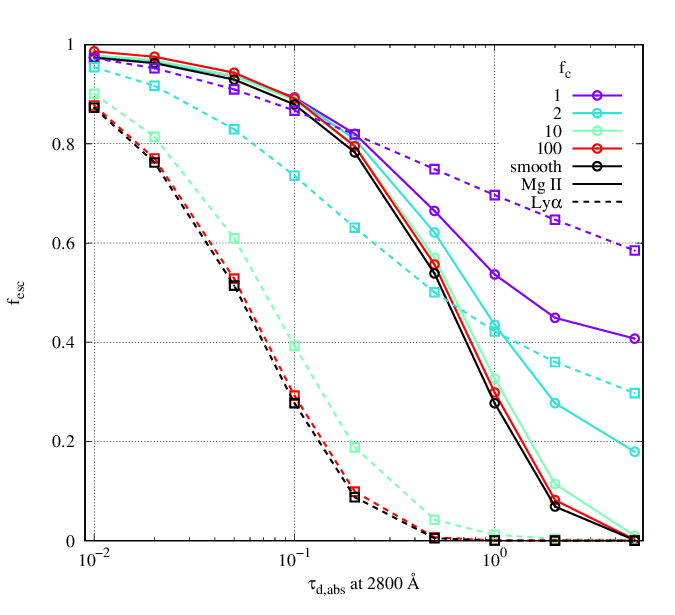}
    \caption{Escape fractions \fesc of \lya (dashed) and \mgii (solid) versus the optical depth of dust absorption at 2800 \AA\ (\taudabs) for various covering factor \fc in the static medium (\vexp = 0 \kms).
    The column densities \Nmg and \NHI is fixed at $10^{15.5} \unitNHI$ and $10^{21} \unitNHI$. 
    The colors represent different \fc. The black solid and dashed lines are the profiles of \fesc of \mgii and \lya at the same \Nmg and \NHI, respectively, in Fig.~\ref{fig:fesc_smooth}.
    The dust optical depth of \lya is 2.53 times higher than of \mgii. 
    \fesc of \lya is higher than that of \mgii when \taudabs $\geq 1$ and $\fc \leq 10$.
    }
    \label{fig:fesc_fc}
\end{figure}

Fig.~\ref{fig:fesc_fc} shows the \mgii and \lya escape fractions as a function of \taudabs for various \fc at \Nmg = $10^{15.5} \unitNHI$ and \NHI =$10^{21} \unitNHI$.
We confirm that both escaping fractions of \mgii and \lya increase with decreasing \fc due to the `surface scattering' effect (cf. \S~\ref{sec:surface_scattering}).
  
Furthermore, we find that the \lya escape fraction is higher than that of \mgii lines at \fc$< 10$ and \taudabs$> 1$.
In the smooth medium, the escape fraction of \lya is always smaller than that of \mgii because of the dust optical depth of \lya $\sim 2.5$ times higher than of \mgii (see Fig.~\ref{fig:smooth_f_esc} and discussion in \S~\ref{sec:dust}). In addition, the path length of \lya is longer than that of \mgii because of the much higher column density.

However, in a clumpy medium, this can be reversed, i.e., the \lya path length through the scattering medium is shorter, and thus, the escape fraction is higher compared to \mgii.
Surface scatterings induce a short path length when a clump is optically thick for incident photons; it causes less interaction with dust in a clump.
Thus, the escape fraction increases with decreasing \fc (increasing \NHIcl).
This surface scattering effect is much stronger for \lya than for \mgii photons because \NHI $\gg$ \Nmg. We will discuss its details in \S~\ref{sec:fesc_mgii_lya}.

\subsubsection{\mgii doublet ratio }\label{sec:ratio_clumpy}

\begin{figure}
	\includegraphics[width=\columnwidth]{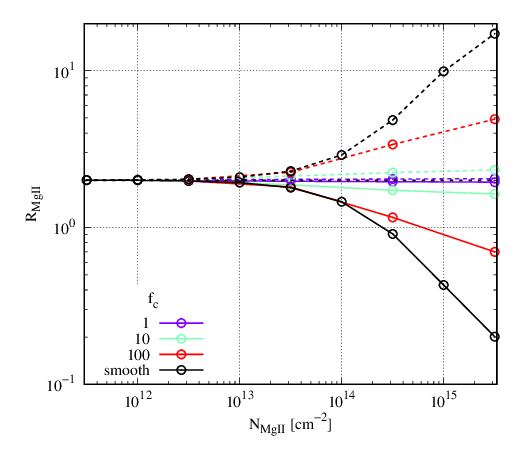}
    \caption{ \mgii doublet ratio of the Gaussian emission case in a clumpy medium for various \fc at \vexp = 1000 \kms. The solid and dashed lines represent outflow and inflow, respectively. The colors of lines represent various \fc. The solid black line is the result of the smooth medium corresponding to the dark blue line for \vexp = 1000 \kms in Fig.~\ref{fig:ratio_smooth}. 
    \Rmg approaches the intrinsic line ratio 2 with decreasing \fc.}
    \label{fig:ratio_clumpy}
\end{figure}

In the smooth medium, we confirmed that the doublet ratio \Rmg varies at \NHI  = $10^{14} \unitNHI$ and |\vexp| > 700 \kms (see \S~\ref{sec:ratio_smooth}).
However, we find that this trend becomes weaker as \fc decreases. 
Fig.~\ref{fig:ratio_clumpy} shows the doublet ratio (for the Gaussian emission case defined in Eq.~\ref{eq:doublet_ratio_gaussian}) in the clumpy medium for various \fc at \vexp = 1000\kms.
The black solid line is the result of the smooth medium in Fig.~\ref{fig:ratio_smooth}.
When \fc decreases, overall \Rmg approaches the intrinsic ratio 2.
At \fc = 1, one finds, on average, only a single clump per line of sight with a distinct expanding velocity. 
If this expanding velocity is less than 700 \kms, the line ratio is invariant. 
In summary, we note that the variation of \Rmg by strong inflow and outflow is sensitive to \fc.

\section{\mgii scattering of stellar continuum}\label{sec:flat_continuum}

\begin{figure*}
\includegraphics[width=175mm]{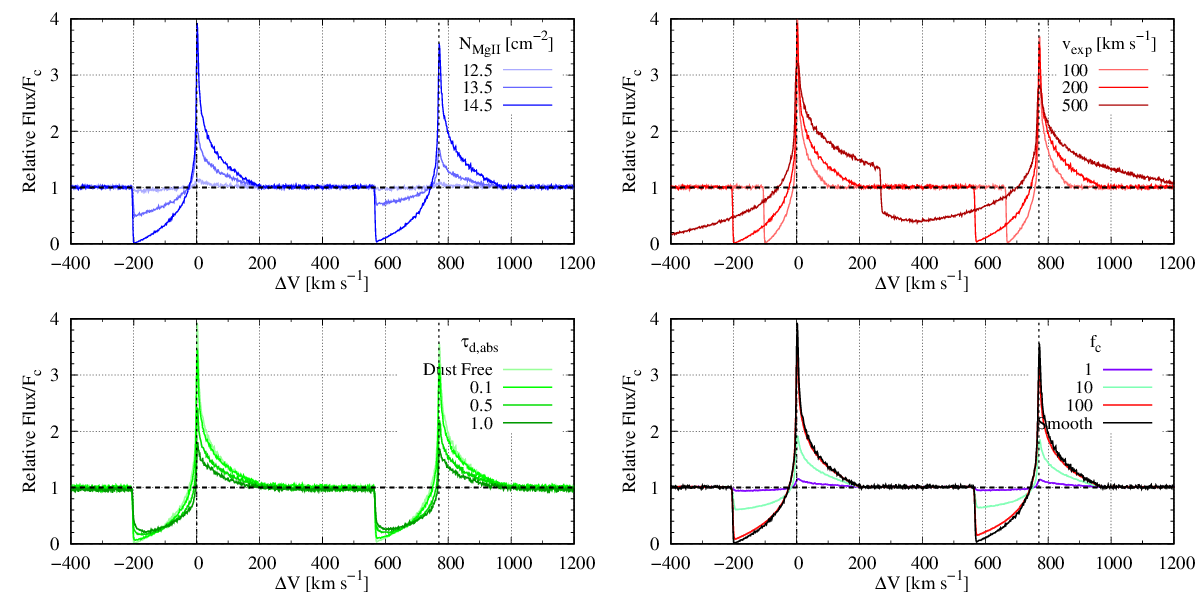}
    \caption{\mgii spectra assuming an flat continuum for various \Nmg (left top), \vexp (right top), \taudabs (left bottom), and \fc (right bottom). 
    The left top panel shows the blue spectra for \Nmg = $10^{12.5-14.5}$ when \vexp is fixed at 200 \kms.
    The red spectra in the right top panel are for three \vexp = 100, 200, and 500 \kms at \Nmg $= 10^{14.5} \unitNHI$.
    In the bottom panels, \NHI and \vexp are fixed at $10^{14.5} \unitNHI$ and 200 \kms.
   Most spectra are from the dust free medium (\taudabs = 0) except the spectra with \taudabs > 0 in the left bottom panel.
   In the right bottom panel, the black line is the spectrum of the smooth medium, which is identical to the spectra at \Nmg = $10^{14.5} \unitNHI$ in the left top panel. The blue, green, and red colors represent \fc = 1, 10, and 100 at the same \Nmg.
    The spectra are shown in units of normalized flux, i.e., the relative flux divided by the continuum level $\rm F_c$. 
    }
    \label{fig:spec_smooth_flat}
\end{figure*}

\begin{figure*}
	\includegraphics[width=175mm]{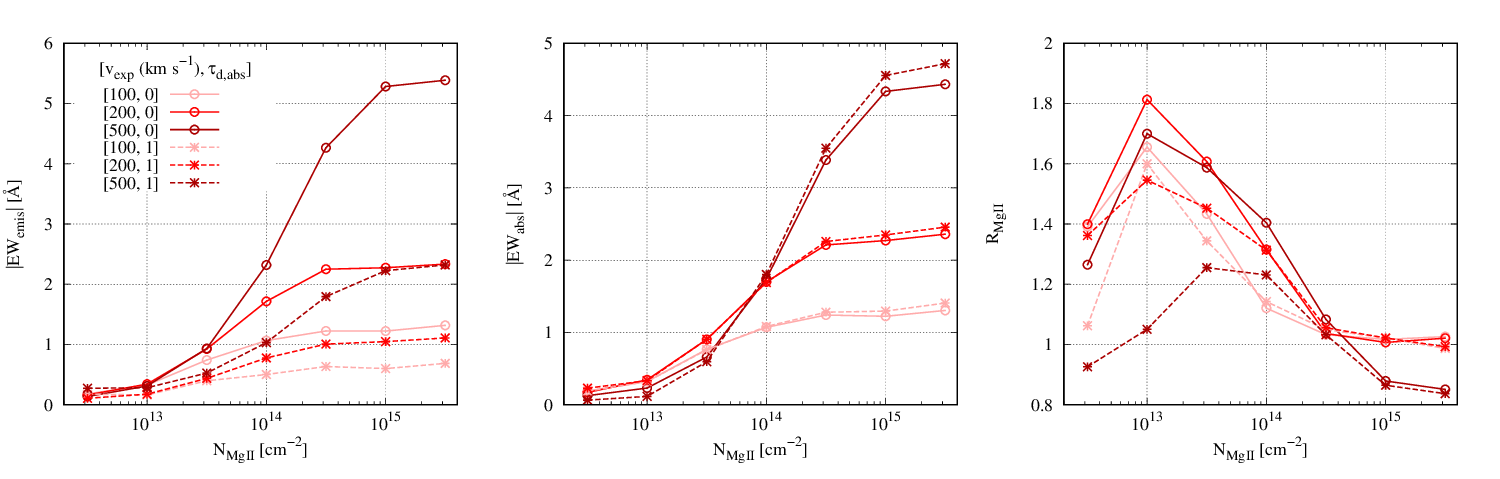}
    \caption{Equivalent width of \mgii emission (left) and absorption (center) features as a function of \Nmg. The right panel shows the \mgii doublet ratio of the spectrum with the continuum defined in Eq.~\eqref{eq:doublet_ratio_continuum}, which is the ratio of the equivalent widths of \mgii K and H emission lines.
    The line colors represent various \vexp. The solid and dashed lines are for the dust-free and dusty medium at \taudabs = 1, respectively. 
    }
    \label{fig:EW_smooth}
\end{figure*}

\subsection{\mgii emission from flat continuum}\label{sec:spec_flat}

\mgii emission features can be generated by stellar continuum radiation near 2800 \AA\ (i.e., continuum pumping).
For instance, \citet{xu23} measured the equivalent width of \mgii emission to be around $2 - 10$ \AA, which means the contribution of the stellar continuum is not negligible.
When the continuum radiation is scattered by \mgii in an outflowing medium, the escaping spectrum exhibits a P-Cygni profile \citep{prochaska11}.
Interestingly, as we will show, in this case, the \mgii doublet ratio is not necessarily two.
In this section, we investigate the emergent spectrum and the doublet ratio of \mgii, considering a flat continuum as incident radiation.

Fig.~\ref{fig:spec_smooth_flat} displays spectra assuming an intrinsically flat continuum.
Each panel shows the dependence on \Nmg, \vexp, \taudabs, and \fc. The spectra are normalized by dividing by the continuum level. A P-Cygni profile appears in most spectra due to the outflow.

  Furthermore, to study the trend varying those parameters, in Fig.~\ref{fig:EW_smooth}, we measure the equivalent widths of emission \EWemis and absorption \EWabs of the spectra from the flat continuum in the range of the wavelength from $\lambdaK - \Delta \lambda_{300}$ to $\lambdaH + \Delta \lambda_{300}$ (corresponding the velocity range from $-300 \kms$ of the K line to $+300 \kms$ of the H line). Thus, the equivalent widths are given by
\begin{eqnarray}\label{eq:EW_continuum}
         \EWemis = \frac{\int^{\lambdaH + \Delta \lambda_{300}}_{\lambdaK - \Delta \lambda_{300}} (F(\lambda) - F_{\rm c}) d \lambda  }{F_{\rm c}}, \quad {\rm when\,} F(\lambda) > F_{\rm c}
          \nonumber \\
         \EWabs = - \frac{\int^{\lambdaH + \Delta \lambda_{300}}_{\lambdaK - \Delta \lambda_{300}} (F(\lambda) - F_{\rm c}) d \lambda  }{F_{\rm c}}, \quad {\rm when\,} F(\lambda) < F_{\rm c}
\end{eqnarray}
where the continuum flux $F_{\rm c}$ is the flux level near 2880 \AA\ ($-1500 \kms$ from the K line).

To investigate the doublet ratio of the spectrum with a continuum, we measure the ratio of \EWemis of K and H lines in the right panel of Fig.~\ref{fig:EW_smooth}.
The ratio of \EWemis represents the doublet ratio of \mgii emission features in the emergent spectrum with a continuum resembling the doublet ratio of the intrinsic emission case in Eq~\eqref{eq:doublet_ratio_gaussian}.
To compute the \EWemis ratio, we calculate the \EWemis of both the K and H lines within a velocity range of $\pm 300 \kms$ from their respective centers.
Consequently, the \EWemis of the K line is determined by the equation
\begin{eqnarray}\label{eq:EW_K}
         {|{\rm EW}_{\rm emis, K}|} = \frac{\int^{\lambdaK + \Delta \lambda_{300}}_{\lambdaK - \Delta \lambda_{300}} (F(\lambda) - F_{\rm c}) d \lambda  }{F_{\rm c}}, \quad {\rm when\,} F(\lambda) > F_{\rm c}.
\end{eqnarray}
Similarly, ${|{\rm EW}_{\rm emis, H}|}$ follows the same equation within the identical velocity range $\pm 300 \kms$ from the line center of the H line. 
Hence, we define the \mgii doublet ratio of the spectrum with a continuum,
\begin{equation}\label{eq:doublet_ratio_continuum}
\Rmgcon = \frac{|{\rm EW}_{\rm emis, K}|}{|{\rm EW}_{\rm emis, H}|}.
\end{equation}
In this section, we specify that the doublet ratio \Rmg is \Rmgcon since the \mgii spectrum has a continuum.

\subsubsection{Dependence on \Nmg}

The left top panel of Fig.~\ref{fig:spec_smooth_flat} shows \mgii spectra for three \Nmg = $10^{12.5}$, $10^{13.5}$, and $10^{14.5}$ \unitNHI at \vexp = 200 \kms.
The red emission and blue absorption features become stronger with increasing \Nmg.
In the left and center panels of Fig.~\ref{fig:EW_smooth}, \EWemis and \EWabs increase with increasing \Nmg. 
In the dust free cases (solid lines), the profile of \EWemis is akin to that of \EWabs because the photons from the blueward become the emission features in the redward via scattering in an outflowing medium.

In the left top panel of Fig.~\ref{fig:spec_smooth_flat}, 
at $\Nmg=10^{13.5} \unitNHI$, the absorption feature in the blueward of the K line is two times stronger than that of the H line.
At $\Nmg = 10^{14.5}\unitNHI$, the two blue absorption features are similar
since this high \Nmg regime is enough to cause nearly all continuum photons in the blueward of K and H lines to undergo scattering.
This effect is shown more quantitatively in the right panel of Fig.~\ref{fig:EW_smooth} showing \Rmg as a function of \Nmg. 
The trend from the intrinsic $\Rmg\sim 2$ to $\Rmg\sim 1$ at $\Nmg\gtrsim 10^{15}\unitNHI$ is apparent (but to various strengths) for a variety of dust optical depths and outflow velocities.

In the right panel of Fig.~\ref{fig:EW_smooth}, \Rmg decreases with decreasing \Nmg in the low \Nmg regime (\Nmg < $10^{13} \unitNHI$) because of the noisy spectrum.
Measuring the doublet ratio of the noisy spectrum at $\Nmg = 10^{12.5} \unitNHI$ in the left top panel of Fig.~\ref{fig:spec_smooth_flat}, some noise can be counted as emission features and contribute equally increasing \EWemis of K and H lines.
Hence, the doublet ratio from the noise fluxes is $\sim 1$. The doublet ratio is underestimated in this low \Nmg regime.
However, since the emission is hard to be observable, we focus on the trend of \Rmg at \Nmg $> 10^{13} \unitNHI$.

In short, if the absorption is saturated, the emission doublet ratio will be -- per construction -- $\sim 1$, whereas for lower optical depths, the different values of the cross section play a role and lead to ratios closer to the fiducial ratio of $\sim 2$. 
We will discuss this effect further and argue that it might explain observations of \citet{chisholm20} in \S~\ref{sec:ratio_emission_continuum}.

\subsubsection{Dependence on \taudabs}
The left bottom panel of Fig.~\ref{fig:spec_smooth_flat} shows the spectra normalized by the continuum level for four values of dust extinction \taudabs = 0, 0.1, 0.5, and 1.
Even though the absorption features do not strongly depend on \taudabs,
the emission features weaken with increasing \taudabs. This is due to the same effect found for the Gaussian emission case discussed in Fig.~\ref{fig:smooth_spec_tauD}.

How this variation in dust content affects \Rmg is shown with dashed lines for \taudabs = 1 in the right panel of Fig.~\ref{fig:EW_smooth}. As discussed, the absorption feature is unaffected by dust, and the profile of \EWabs at \taudabs = 1 is identical to that of a dust-free medium. 
However, when the continuum photon undergoes scattering to become the red emission features seen, the path length of the scattered photon is longer than the direct escaping photons.
Thus, \EWemis decreases with increasing \taudabs.
In other words, dust extinction decreases the strength of the \mgii emission features from the continuum pumping.

Because the increase in path length causing the decrease in the escape probability depends on the different cross sections of the two lines, the line ratio is slightly decreased with increasing dust content, as shown by the dashed line in the right panel of Fig.~\ref{fig:EW_smooth}.
This effect is negligible at \Nmg $\ge 10^{15} \unitNHI$ since the ratio of the path lengths of the lines is $\sim 1$ (see Fig.~\ref{fig:path_ratio_smooth}). 
Thus, \Rmg does not strongly depend on the dust optical depth.

\subsubsection{Dependence on \vexp}
The right top panel of Fig.~\ref{fig:spec_smooth_flat} shows the \mgii spectra for three \vexp = 100, 200, and 500 \kms at \Nmg = $10^{14.5}$ \unitNHI.
When \vexp increases, the widths of the emission and absorption become broader.
A higher \vexp causes scattering in a wider range of incident continuum radiation. 
Accordingly, \EWemis as well as \EWabs increase with increasing \vexp, which can be seen in the left and center panels of Fig.~\ref{fig:EW_smooth}, respectively.
Since both K and H lines show the same behavior, \Rmg does not strongly depend on \vexp (also because here we focus on the \vexp < 700 \kms case) as shown in the right panel of Fig.~\ref{fig:EW_smooth}.

\subsubsection{Dependence on \fc}\label{sec:EW_fc}
Lastly, the right bottom panel of Fig.~\ref{fig:spec_smooth_flat} displays
\mgii spectra for various degrees of clumpiness, characterized by \fc at \Nmg = $10^{14.5} \unitNHI$. 
The trend of spectra varying \fc is similar to that found in the previous section for the Gaussian source.

As seen before, the spectrum of the clumpy medium shown in Fig.~\ref{fig:spec_smooth_flat} becomes similar to that of the smooth medium with increasing \fc.
At \fc = 1, the scattering is negligible, and the escaping spectrum is flat like the incident radiation.
At \fc = 10, even though the absorption feature is weaker than that of the smooth medium, the doublet ratio of emission features is still $\sim 1$.
The amount of photons undergoing scattering decreases with decreasing \fc.

The \mgii doublet ratio varies through scattering by the fast-moving medium (\vexp > 700 \kms) as shown in Fig.~\ref{fig:ratio_smooth}. 
In this case, \Rmg converges to the intrinsic ratio 2 with decreasing \fc (see Fig.~\ref{fig:ratio_clumpy}) because the number of photons moving from the K/H line to the H/K line by the strong outflow/inflow decreases.
However, since \Rmg in the clumpy medium is mainly determined by the ratio of scattered K and H photons which hinge on \Nmg, \Rmg is not sensitive for \fc and still depends on the total column density \Nmg.

\subsection{Stellar continuum with in situ \mgii emission}\label{sec:ratio_emission_continuum}

\begin{figure*}
	\includegraphics[width=175mm]{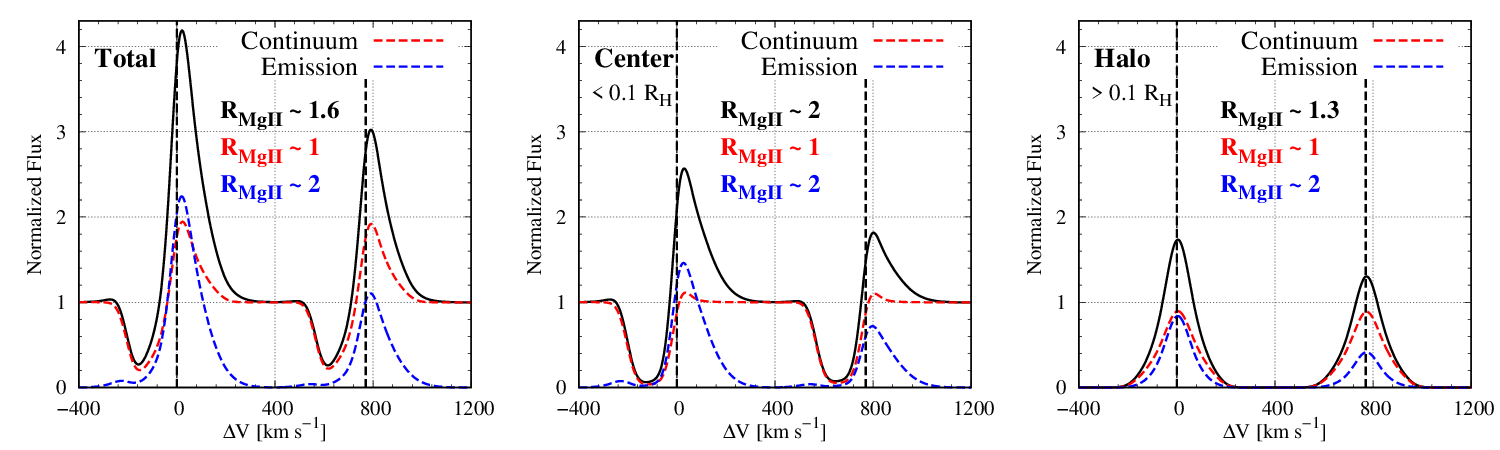}
    \caption{The total integrated spectrum (left), central spectrum escaping at projected radii $R_p$ < 0.1 $\RH$ (center), and halo spectrum ($R_p > 0.1 \RH$; right panel). The incident radiation consists of a flat continuum and Gaussian emission with $\sigma_{\rm emit} = 100 \kms$ and equivalent width EW = 5\,\AA. \vexp, \Nmg, and \fc are fixed at 200 \kms, $10^{15} \unitNHI$, and 100, respectively. 
    Spectra are shown as normalized flux with the flat continuum being unity.
    The black solid line represents the total escaping spectrum of the incident radiation, whereas the red and blue dashed lines show the escaping spectrum of only the flat continuum and the Gaussian emission, respectively.
    The values of \Rmg indicated in each panel mark the doublet ratio of each emission part in the corresponding color. \Rmg of the total and center spectra with the continuum is the ratio of \mgii K and H emission features over the continuum, which is identical to the emission EW ratio of \mgii in Eq.~\eqref{eq:doublet_ratio_continuum}. \Rmg of the spectra without the continuum is the double ratio of the emission case in Eq.~\eqref{eq:doublet_ratio_gaussian}.
    }
    \label{fig:spec_con_emis}
\end{figure*}

\begin{figure}
	\includegraphics[width=\columnwidth]{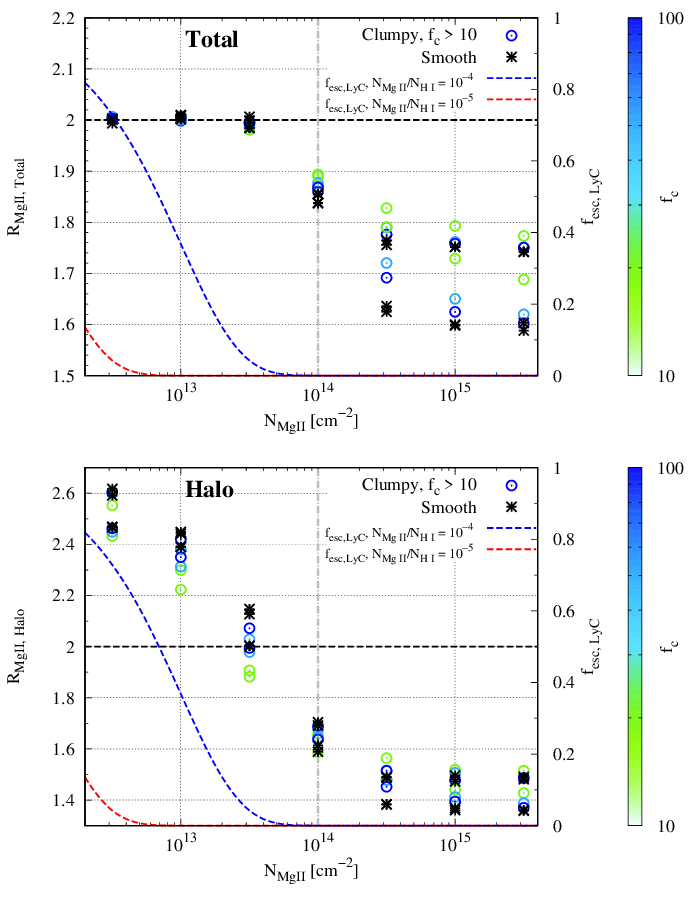}
    \caption{\Rmg as a function of \Nmg (=\fc\Nmgcl for clumpy medium) considering the flat continuum and in situ emission with the intrinsic equivalent width \EWint= 5\AA\ as the incident radiation like the spectrum in Fig.~\ref{fig:spec_con_emis}.
    The star marks and open circles represent the smooth and clumpy media for \fc > 10. The colors of the circles indicate the values of covering factor \fc, according to the color bar on the right side.
    The top and bottom panels are for the doublet ratio of the total integrated spectrum \Rmgtotal and halo spectrum \Rmghalo, respectively.
    In the top panel, because of the existence of the continuum in the total spectrum (see the left panel of Fig.~\ref{fig:spec_con_emis}), \Rmgtotal is the doublet ratio of \mgii emission feature, which is the ratio of \EWemis the K and H photons over the continuum in Eq.~\eqref{eq:doublet_ratio_continuum}.
    In the bottom panel, Eq.~\eqref{eq:doublet_ratio_gaussian} for the doublet ratio of the emission case is adopted to calculate \Rmghalo because of the halo spectrum without the continuum (see the right panel of Fig.~\ref{fig:spec_con_emis}. The blue and red dashed lines are the profile of the LyC escape fraction \fesclyc as a function of \Nmg assuming the \mgii fraction \fmghi = \Nmg/\NHI $\sim 10^{-4}$ and $10^{-5}$, respectively.
    Generally, \Rmgtotal and \Rmghalo decrease with increasing \Nmg.
    }
    \label{fig:ratio_con_emis}
\end{figure}

\begin{figure}
	\includegraphics[width=\columnwidth]{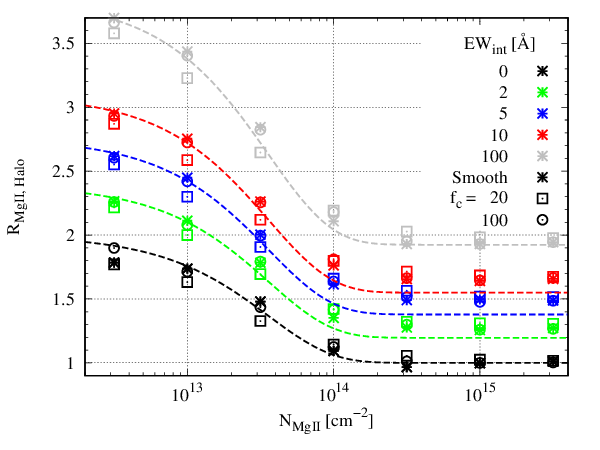}
    \caption{\Rmghalo for various intrinsic equivalent width \EWint as a function of \Nmg at \vexp = 100\kms. The colors represent \EWint = 0 (black), 2 (green), 5 (blue), 10 (red), and 100 (gray). The dashed lines show the relation between \Rmghalo and \Nmg noted in Eq.~\eqref{eq:flux_ratio}. The shapes of points present the smooth medium (star) and the clumpy media with \fc = 20 (square) and 100 (circle) The blue points correspond to \Rmghalo at \vexp = 100 \kms in the bottom panel of Fig.~\ref{fig:ratio_con_emis}. 
    \vexp is fixed at 100 \kms.
    }
    \label{fig:ratio_EW}
\end{figure}


\mgii emission is observed alongside continuum radiation with measured equivalent widths (EW) $\sim 2-10$ \AA\ \citep{henry18,xu23}. It is commonly also more spatially extended than UV continuum 
\citep{leclercq22,dutta23,guo23b}.
\cite{chisholm20} reported that the \mgii doublet ratio in the central region, including the source galaxy, is higher than in the region surrounding the galaxy. They found the doublet ratio of inside and outside are $\sim 2$ and $1.3$, respectively.

In the previous sections, we identified two mechanisms to change the \mgii doublet ratio,
(1) scattering in strong outflow/inflow and (2) scattering of the flat continuum.
In \S~\ref{sec:ratio_smooth}, we showed that the doublet ratio varies when the expansion speed \vexp is higher than the velocity separation of K and H lines, $\sim 750 \kms$.
Furthermore, in \S~\ref{sec:spec_flat}, we noted that considering the flat continuum as the incident radiation, the doublet ratio decreases with increasing \Nmg; \Rmg $\sim 1$ at \Nmg > $10^{14} \unitNHI$.

In this section, we combine the two emission mechanisms introduced in the previous sections and consider the incident radiation, including a flat continuum and in situ Gaussian emission, to investigate the \mgii spectra and the behavior of the \mgii doublet ratio under more realistic conditions.
Hence, hereafter, we define that the doublet ratio \Rmg is \Rmgemis in Eq.~\eqref{eq:doublet_ratio_gaussian} (\Rmgcon in Eq.~\eqref{eq:doublet_ratio_continuum}) if the spectrum is without (with) a continuum.
Furthermore, we separate in this section the \mgii spectra originating from the central and halo regions as opposed to simply focusing on the integrated spectrum as we have in the previous sections.

\subsubsection{Central versus Halo \mgii emission}
Fig.~\ref{fig:spec_con_emis} shows the simulated \mgii spectra of the source consisting of a flat continuum and an in situ Gaussian emission with the width $\sigsrc = 100\kms$ and the intrinsic equivalent width \EWint = 5 \AA. \Nmg and \vexp are fixed at $10^{15} \unitNHI$ and 200\kms, respectively.
The spectrum shown is smoothed with a Gaussian profile with a width of $30\kms$, corresponding to the spectral resolution of $R \sim 10^4$ in order to be more easily compared to observations.

The left panel of Fig.~\ref{fig:spec_con_emis} shows the total integrated spectrum. The center and right spectra are `central' and `halo' spectra composed of escaping photons at the projected radius $r_p < 0.1\RH$ and $> 0.1\RH$, respectively.
The red and blue dashed lines represent the spectrum of the continuum and emission to visualize the contribution of the two types of sources independently.

The right panel of Fig.~\ref{fig:spec_con_emis} shows that the halo spectrum is an emission feature without continuum radiation because
the halo spectrum is composed of only scattered photons.
For most photons, the medium is optically thin, and they directly escape as the `central spectrum'. This is not the case for the photons in the range from the systemic velocity to $\Delta V = -\vexp$, which are scattered and make up the `halo spectrum'.
Hence, in the center panel of Fig.~\ref{fig:spec_con_emis}, the central spectrum has stronger absorption and weaker emission features than the total spectrum in the left panel due to the absence of scattered photons. In addition, the central spectrum from the continuum does have significant emission features and looks like the incident radiation with the absorption feature by scattering medium in the line of sight.

The contribution of the stellar continuum and in situ emission determines the \mgii doublet ratio \Rmg.
Fig.~\ref{fig:spec_con_emis} shows that the doublet ratio \Rmg is always $\sim 1$ and $2$ from continuum and Gaussian emission, respectively -- both in the central and halo spectra.
However, the observable doublet ratio \Rmg differs between the central and the halo region.
The central panel of Fig.~\ref{fig:spec_con_emis} shows the central spectrum with \Rmg $\sim$ 2.
The emission feature from the continuum (red dashed line) is much weaker than that from the in situ emission (blue dashed line) because most initial photons in the blueward are spatially diffused via scattering and classified as the halo spectrum.
Thus, \Rmg $\sim$ 2 as the in situ emission dominates the emission features of the center spectrum.
In the right panel of Fig.~\ref{fig:spec_con_emis}, the strengths of the K line emissions from the continuum (\Rmg $\sim 2$) and in situ emission (\Rmg $\sim 1$) are comparable in the halo spectrum. 
Hence, \Rmg is $\sim 1.3$ and smaller than in the central spectrum.
The left panel of Fig.~\ref{fig:spec_con_emis} shows that \Rmg of the total spectrum is $\sim 1.6$. Both directly escaping photons of the in situ emission with $\Rmg \sim 2$ (central spectrum) and scattering photons with $\Rmg \sim 1.3$ (halo spectrum) contribute to the formation of the \mgii emission features above the continuum.

In summary, the combination of continuum and Gaussian in situ \mgii emission can lead to doublet ratios $\Rmg < 2$ also with moderate ($<700\kms$) velocities if $\Nmg > 10^{14} \unitNHI$. This is because the absorption of the continuum part is saturated, resulting in $\Rmg\sim 1$ of the scattered emission, whereas, for the Gaussian component, the ratio remains close to its fiducial value $\Rmg\sim 2$. The (weighted) averaging of these two mechanisms leads thus to $\Rmg<2$ with a lower value -- due to the larger contribution from the continuum -- found in the halo spectrum. This is a plausible mechanism to explain the low doublet ratios observed, and we will investigate this more systematically below.

\subsubsection{\mgii doublet ratio  \& \Nmg}\label{sec:ratio_Nmg}
In order to study the dependence of \Rmg on \Nmg with the joined intrinsic spectrum, Fig.~\ref{fig:ratio_con_emis} shows \Rmg as a function of \Nmg in the slow outflow and inflow ($0 \kms <|\vexp| < 500 \kms$) regime. 
The star marks and open circles represent the smooth medium and the clumpy medium with \fc= 20, 50, and 100.
In the top panel for the total integrated spectrum, at \Nmg < $10^{14} \unitNHI$, the doublet ratio of the total spectrum \Rmgtotal is $\sim 2$ regardless of other parameters.
In this low \Nmg regime, the in situ emission mainly contributes to the formation of \Rmgtotal.
In other words, the in situ emission dominates the emission features of the escaping spectrum, and the effect of scattering is relatively negligible in the total spectrum.

At high column densities (\Nmg > $10^{14} \unitNHI$), the doublet ratio becomes \Rmgtotal < 2 since the scattering of the flat continuum is sufficient to make the emission features with \Rmgtotal $\sim$ 1 as discussed in the previous section and shown in the right panel of Fig.~\ref{fig:EW_smooth}.

In the bottom panel of Fig.~\ref{fig:ratio_con_emis}, the profile of the halo doublet ratio \Rmghalo shows a clear trend comparing that of \Rmgtotal; \Rmghalo declines with increasing \Nmg. 
In the low \Nmg regime, \Rmghalo is > 2 because the halo spectrum is composed of scattered photons.
The ratio of scattering rate of the K and H lines at $\Nmg < 10^{14} \unitNHI$ is higher than unity as shown in the right panel of Fig.~\ref{fig:EW_smooth} since the optical depth of the K line \tauK $\sim 2\tauH$.
Therefore, when the in situ emission with the intrinsic doublet ratio 2 undergoes scattering,
the doublet ratio of scattered photons is > 2. We will discuss this further below.

For large \Nmg, however, the contribution of scattered photons of the flat continuum leads to \Rmg $< 2$ -- as seen in the example shown in Fig.~\ref{fig:EW_smooth} above.
As a result, the \mgii doublet ratios of both the total and halo spectra become less than 2 at \Nmg $> 10^{14} \unitNHI$ as shown by Fig.~\ref{fig:ratio_con_emis}.

Fig.~\ref{fig:ratio_EW} shows the halo doublet ratio for various intrinsic equivalent widths of the in situ emission \EWint at $\vexp = 100 \kms.$
The overall \Rmghalo increases with increasing \EWint and does not strongly depend on \fc.
In the high \Nmg regime, \Rmghalo is $\sim$ 1 for \EWint = 0 and 2 for \EWint = 100 \AA, i.e., fully determined by whether the spectrum is dominated by the in situ emission or the flat continuum.\\

From the above findings, one can construct a simple analytical model to predict the doublet ratio given an intrinsic equivalent width and \mgii column density.
As the halo is composed of scattered photons, the ratio of scattered K and H photons determines \Rmghalo.
In the optically thin regime (\tauzero $<$ 1), 
the ratio of scattered K and H line photons is $\sim 2$ since the scattering fraction is $1-\exp(-\tauzero)\approx\tauzero$ and $\tau_{\rm 0,K} = 2\tau_{\rm 0,H}$. Here, \tauK and \tauH are the optical depth near the K and H transition, respectively\footnote{Note that throughout this work, we refer to $\tau$ as the optical depth of the K line unless specified otherwise.}.
In the optically thick regime (\tauzero $>$ 1), however,
the continuum absorption is saturated, and most K and H photons undergo scattering.
The scattering fraction in the medium with optical depth $\tau$ is given by
\begin{equation}
    f_{\rm scat} = 1 - e^{-\tau}.
\end{equation}
Therefore, the ratio of the K and H lines' scattering fractions given by
\begin{equation}\label{eq:ratio_EW}
    {f_{\rm scat,K} \over f_{\rm scat,H}} = { {1 - e^{-\tauK}} \over {1 - e^{-\tauH}}} = 1 + e^{-\tauH}
\end{equation}
where in the last step we used $\tauK \approx 2 \tauH$.

To estimate \Rmghalo analytically using the scattering fractions above,
one requires the doublet ratio near the K and H lines of the intrinsic radiation.
The scattering geometry for our simulation has a Hubble-like outflow where the outflow velocity is proportional to the distance from the central source.
This -- and the fact that \mgii scattering in the Lorentzian part is negligible -- implies that photons in the velocity range from $-\vexp$ to 0 \kms undergo scattering.
In this velocity range, 
the intrinsic doublet ratio is given by 
\begin{equation}\label{eq:intrinsic_ratio}
   R_{\rm MgII,int} = { {\int^{\lambdaK}_{\lambdaK - \Delta \lambdaK} F_i(\lambda) d \lambda } \over {\int^{\lambdaH}_{\lambdaH- \Delta \lambdaH} F_i(\lambda) d \lambda }  },
\end{equation}
where $F_i(\lambda)$ is the intrinsic radiation and $\Delta \lambda_{\rm K, H}=\vexp \lambda_{\rm K,H} / c$.
This implies that for our fiducial choice of $v_{\rm exp}=100\kms$, $R_{\rm MgII,int}$ is 1, 1.38, 1.55, and 2 for $\EWint = 0$~\AA\ (a flat continuum), 5 \AA, 10 \AA, and $\infty$ (\mgii emission without a continuum), respectively.

The intrinsic radiation with \EWint normalized by the continuum flux $F_c$ is given by
\begin{equation}\label{eq:incident_radiation}
  F_i(\lambda)/F_{c} = 1 + {\EWint \over 3}\left[ 2{\rm G}(\lambda, \lambdaK, \sigsrc) + {\rm G}(\lambda, \lambdaH,\sigsrc) \right],
\end{equation}
where ${\rm G}(x,\mu,\sigma)=\exp(-(x-\mu)^2 /2 \sigma^2)/\sqrt{2 \pi \sigma^2}$ is a Gaussian profile with mean $\mu$ and variance $\sigma^2$.

Therefore, adopting Eq.~\eqref{eq:intrinsic_ratio} and \eqref{eq:incident_radiation}, we can write the intrinsic doublet ratio 
\begin{equation}\label{eq:ratio_int}
   R_{\rm MgII,int} ={ { 6{\Delta \lambdaK + 2\EWint{\rm erf}({\vexp /\sqrt{2} \sigsrc }) } } \over { {6\Delta \lambdaH + \EWint{\rm erf}({\vexp /\sqrt{2} \sigsrc }) } } }.
\end{equation}
Thus, \Rmg in the halo is given by
\begin{equation}\label{eq:flux_ratio}
    \Rmghalo =  R_{\rm MgII,int}  {f_{\rm scat,K} \over f_{\rm scat,H}} =  R_{\rm MgII,int} \left( 1 + e^{-\overline{\tau_{\rm H}}} \right).
\end{equation}
Here, $\overline{\tau_{\rm H}}$ is the average optical depth of the K line from $-\vexp$ to 0 \kms.
Using $\sigma_\nu$ in Eq.~\eqref{eq:cross_section} and assuming the homogeneous medium with uniform number density and a Hubble-like outflow,
$\overline{\tau_{\rm H}}$ is given by
\begin{equation}\label{eq:ratio_tau}
    \overline{\tau_{\rm H}} = \left( \Nmg \over {3.6 \times 10^{13}\unitNHI}\right) \left( \vexp \over {100 \kms} \right)^{-1}.
\end{equation}
Fig.~\ref{fig:ratio_EW} shows that the halo doublet ratio \Rmghalo can be well described by Eq.~\eqref{eq:flux_ratio}. Only for $\Nmg \gtrsim 10^{14} \unitNHI$ small differences can be found due to multiple scatterings.

Note that the dashed lines in Fig.~\ref{fig:ratio_con_emis} represent
LyC escape fraction \fesclyc as a function of \Nmg assuming \mgii fractions $ = 10^{-4}$ and $10^{-5}$.
Specifically, the figure shows that \Rmgtotal < 2 at \fesclyc $\sim 0$ (top panel) and \Rmghalo decreases with decreasing \fesclyc (bottom panel).
This illustrates that the \mgii doublet ratio is a sensitive proxy for \fesclyc. 
Hence, the \mgii doublet ratio can be a powerful tool to indicate LyC leakage and estimate the LyC escape fraction.
In \S~\ref{sec:mgii_lyC}, we will discuss this in detail and test further \Rmghalo as a tracer of LyC escape.

\section{Discussion}\label{sec:discussion}

\subsection{\mgii \& \lya escape fractions}\label{sec:fesc_mgii_lya}

\begin{figure}
	\includegraphics[width=\columnwidth]{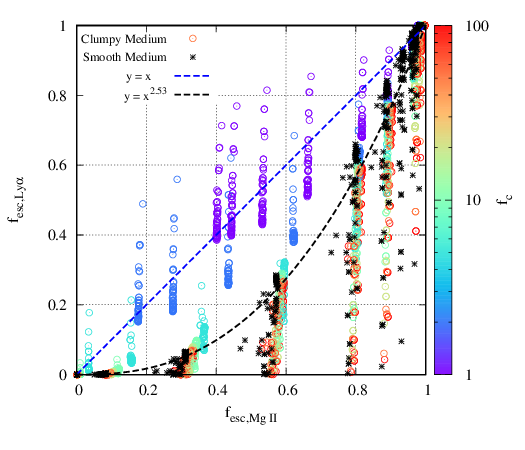}
    \caption{Escape fractions of \mgii and \lya, \fescmgii and \fesclya, of the clumpy medium (color open circles) and the smooth medium (black star marks). The colors of the circles are for various \fc following the color bar on the right side. The blue and black dashed lines represent the function of \fesclya = \fescmgii and $\fescmgii^{2.53}$, respectively. Recall that 2.53 is the \taudabs ratio of \lya and \mgii.
    At \fc < 10, \fesclya is comparable to \fescmgii because the surface scattering effect for \lya is stronger than for \mgii due to \NHI $\gg$ \Nmg.
    }
    \label{fig:fesc_compare}
\end{figure}
The escape fraction of radiation is generally given by $f_{\rm esc}=\exp(-\taudabs)$ where $\taudabs = \int {\rm d}s\,\sigma_{\rm d,abs} n_{\rm d}$; $\sigma_{\rm d,abs}$ and $n_{\rm d}$ are the dust absorption cross section and dust number density, respectively. Importantly, however, the integral is carried out along the trajectory of the photon and thus is highly non-trivial for resonant line photons.
\footnote{Note that the number of scatterings does not enter directly into $f_{\rm esc}$, i.e., it is not the case that photons that scatter more necessarily have a lower $f_{\rm esc}$ as often assumed.}

Specifically, comparing the escape fractions of \mgii and \lya, we can see three effects at play:
\begin{enumerate}
    
\item the dust absorption cross section of \lya is higher than for \mgii, generally resulting in lower escape fractions. In particular, $\sigma_{\rm d,abs}^{\rm Ly\alpha}\approx 2.53 \sigma_{\rm d,abs}^{\mgii}$ in the Milky Way and even larger differences in low metallicity environments (the ratio is, e.g., $3.8$ and $9.3$ in the LMC and SMC dust models, respectively; \citealp{draine03a,draine03b}). Note that this effect alone can affect the escape fraction by orders of magnitude (recall that, e.g., $\exp(-9)\approx 10^{-4}$), for instance, at high-$z$. We illustrate this point in Fig.~\ref{fig:f_esc_appendix} and \ref{fig:f_esc_compare_appendix} of Appendix~\ref{sec:dust_model}, where we compare escape fractions of \mgii and \lya using the MW, LMC, and SMC dust models assuming a monochromatic source.

\item in addition, another effect leading to a decrease in \lya escape fraction is, of course, that the path length $l_{\rm path}$ of \lya is generally longer than that of \mgii because \NHI $\gg$ \Nmg. It is important to note, though, that while for moderately optically thin systems, photons escape via a `single flight' \citep{osterbrock62}, thus, $l_{\rm path}\sim R$, i.e., the extent of the system, for extremely optically thick media ($\tauzero \gtrsim 10^5$; i.e., when \lya photons escape via excursion) $l_{\rm path}$ is only increased by a factor of $x_{\rm peak}\approx 6.5 (\NHI / 10^{19} \unitNHI)^{1/3}$ for $T=10^4\,$K \citep{adams72,neufeld90}. Hence, while the number of scatterings (which, for \lya in this regime, is $\sim \tau_0$) is increased by orders of magnitude, the increase in total path length -- and thus dust absorption optical depth -- is typically only a factor of few. We show this explicitly in Fig.~\ref{fig:path_ratio_smooth} and \S~\ref{sec:fesc_mono_smooth}, where we compare $l_{\rm path}$ of \mgii K and H lines measured from the simulations.

\item Lastly, in a multiphase medium \lya can undergo surface scatterings off of cold and dense regions, thus preventing the exposure to dust grains \citep{neufeld91,hansen06,gronke17}.
We explicitly show that this effect exists in \mgii RT in Fig.~\ref{fig:fesc_fc} and \S~\ref{sec:fesc_clumpy}.
The effect of scattering becomes strong when the optical depth of the cold region increases.
Therefore, as this effect would be weaker for \mgii photons due to $\NHI \gg \Nmg$, it would increase the \lya escape fraction with respect to \mgii. 

\end{enumerate}

We show an overview of escape fractions computed in a range of smooth and clumpy models in Fig.~\ref{fig:fesc_compare} (with the color coding corresponding to different values of \fc). 
In this figure, all three effects described above can be seen at play. Overall, the \lya escape fraction is lower than that of \mgii; especially the smooth media simulations (indicated with black star marks) cluster around the $\fesclya = \fescmgii^{2.53}$ curve, i.e., can be explained by the higher dust optical depth for \lya (point \textit{(i)} above). However, a substantial part of smooth and clumpy runs with $\fc \gtrsim 10$ show an even lower \lya escape fraction -- due to a longer escape trajectory (point \textit{(ii)} above). Only a fraction of runs with $\fc \lesssim 10$ shows larger \lya than \mgii escape fractions since the effect of surface scattering in \lya RT is stronger than that of \mgii due to $\NHI \gg \Nmg$ (point \textit{(iii)} above).
Here, a higher column density simultaneously causes a longer path length and stronger surface scattering effect, which can decrease and increase \fesc, respectively.
Thus, the clumpiness (i.e., \fc) is crucial to understanding the relation between \fesc of \lya and \mgii.

Note that interestingly, some observational studies report that they found \fescmgii comparable or even smaller than \fesclya \citep{henry18,xu23}.
Given the above discussion, this could be evidence of surface scatterings at play. Note that this can only be achieved if the medium is clumpy with only a few clumps per line-of-sight and the velocity dispersion is not too large \citep{laursen13,gronke18}. In principle, the ratio of escape fractions can give us interesting insights into the geometry of the scattering medium.

\subsection{Mg II as a proxy for LyC leakage}\label{sec:mgii_lyC}

\begin{figure}
	\includegraphics[width=80mm]{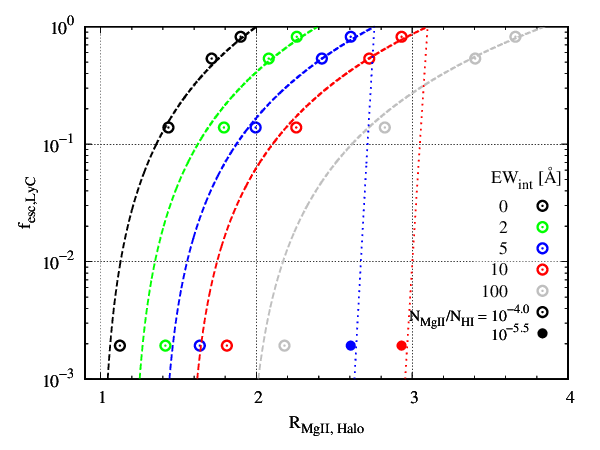}
    \caption{The LyC escape fraction \fesclyc as a function of the halo doublet ratio \Rmghalo for various intrinsic equivalent widths of the in situ emission $\rm EW_{int}$ = 0 (the flat continuum), 2, 5, 10, and 100 \AA\, at \vexp = 100 \kms.
    Open and filled circles represent the \mgii fraction \Nmg/\NHI = $10^{-4}$ and $10^{-5.5}$, respectively.
    The color dashed (dotted) lines are the analytic solution in Eq.~\eqref{eq:lyc_mgii} assuming \Nmg/\NHI = $10^{-4}$ ($10^{-5.5}$).
    In order to get \fesclyc from \Rmghalo, the intrinsic EW and the \mgii fraction are crucial parameters.
    }
    \label{fig:ratio_LyC}
\end{figure}

\begin{figure*}
	\includegraphics[width=175mm]{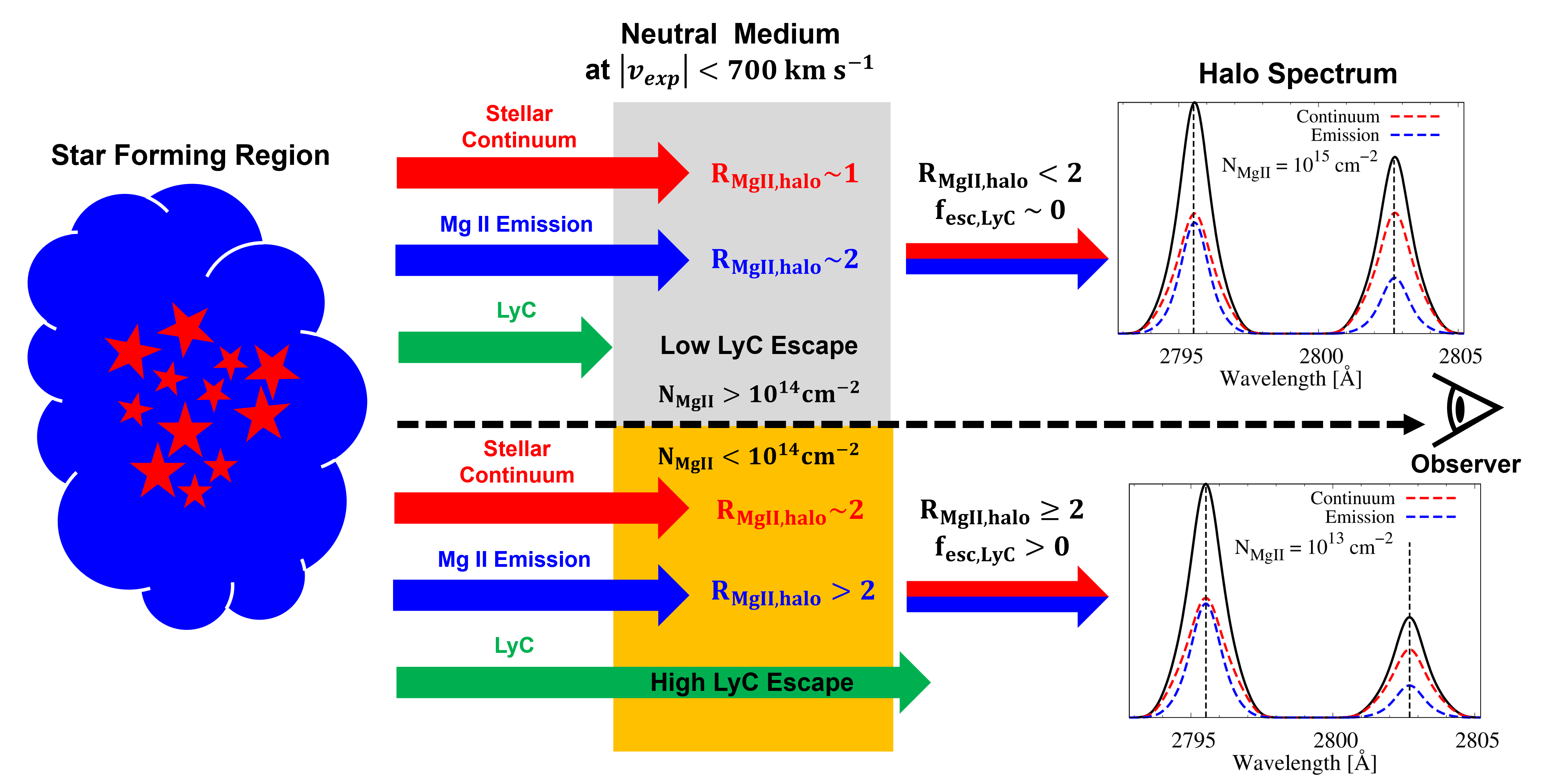}
    \caption{Schematic illustration to demonstrate the halo doublet ratio of \mgii lines \Rmghalo as a tracer of LyC escape. A small doublet ratio \Rmg < 2 can represent the low LyC escape fraction as shown in Fig.~\ref{fig:ratio_LyC}.
    On the left side, a star-forming region emits \mgii in situ emission (blue), stellar continuum (red), and LyC (green). Intrinsically, the doublet ratio is two for recombination or collisional excitation.
    The gray and orange neutral media divided by the black dashed line represent \Nmg > $10^{14} \unitNHI$ and $< 10^{14} \unitNHI$, respectively, with moderate ($|\vexp|<700 \kms$) kinematics.
    The upper and bottom right panels are the halo spectra at \Nmg = $10^{15}$ and $10^{13} \unitNHI$, respectively, with the black solid lines indicating the total spectrum and the red (blue) dashed lines just the contribution of the stellar continuum (the in situ emission).
    The spectra in the upper panel are identical to the halo spectra in the right panel of Fig.~\ref{fig:spec_con_emis}.
    In the high \Nmg regime (> $10^{14} \unitNHI$), the doublet ratios of the halo spectra from the in situ emission and stellar continuum are $\sim 2$ and $\sim 1$ (also see Fig.~\ref{fig:spec_con_emis} \& \S~\ref{sec:ratio_emission_continuum}).
    In the low \Nmg regime, \Rmghalo is higher than 2, as discussed in \S~\ref{sec:ratio_emission_continuum} and shown in Fig.~\ref{fig:ratio_EW}.
    }
    \label{fig:mgii_lyc}
\end{figure*}
The escape of LyC radiation in order to reionize the Universe is one of the main outstanding puzzles at high-$z$ \citep[see reviews][]{2010Natur.468...49R,madau14,eldridge22}. Thus far, only a few individual LyC leakers have been found \citep{shapley16,izotov16,izotov18_obs,vanzella16,vanzella18, revira-thorsen17,rivera-thorsen19,kim23}, and progress comes from survey observation \citep{steidel01,steidel18,flury22_a} and proxy observables. The latter is particularly needed since, already at intermediate redshift, the IGM becomes opaque to ionizing radiation making direct detections of LyC hard to impossible.

Since -- as \lya -- the \mgii doublet is susceptible to cold, neutral $\sim 10^4\,$K gas,
it is a natural proxy for probing ionizing photon escape. 
This connection has been explored in work by \citet{henry18,chisholm20,izotov22,xu23} who suggest that a doublet ratio of $\Rmg < 2$ represents low LyC escape fractions. 
Their argument is based on the fact that the optical depth of the K line is two times greater than that of the H line, leading to $\Rmg < 2$ for larger optical depths, whereas, for negligible amounts of scattering, one would observe the fiducial value of $\Rmg \sim 2$.

Later, \cite{katz22} tested this relationship between \Rmg and LyC escape fraction using cosmological hydrodynamical simulation post-processed with \mgii radiative transfer.
They concluded that the \mgii doublet ratio could be used as an indicator of LyC leakers for galaxies in the optically thin regime.

However, as pointed out above, there are other important physical mechanisms -- mainly strong ($|v|\gtrsim 700\kms$) kinematics (\S~\ref{sec:ratio_smooth}), dust (\S~\ref{sec:ratio_mono}),    and scattered stellar continuum (\S~\ref{sec:flat_continuum}) -- that influence strongly the emergent doublet ratio, and thus the inferred LyC escape fraction.

To recap: we showed in Fig.~\ref{fig:ratio_smooth} \& \S~\ref{sec:ratio_smooth} how in- or outflows close to or exceeding the doublet separation can in- or decrease \Rmg, respectively. Furthermore, for intermediate optical depths ($0.5\lesssim \tau \lesssim 10$), the path lengths of the H and K lines are different, leading to different dust attenuation and thus a decreased \Rmg (cf. \S~\ref{sec:ratio_mono}). And lastly, we showed in \S~\ref{sec:flat_continuum} that scattered stellar continuum can also lower the doublet ratio.

Because of the restricted range of parameters, the former two effects can influence \Rmg; we focus here on the latter one. However, we do want to point out that also the other effects can play a role and lead to a potentially wrongly inferred \fesclyc. We discuss the effect of strong kinematics on the doublet ratio further in \S~\ref{sec:ratio_outflow}. 

In \S~\ref{sec:flat_continuum}, we showed that the doublet ratio of from a pure flat continuum ($\EWint = 0$) is generally $<2$, even approaching $\sim 1$ in the limit where both K and H lines are very optically thick (\tauzero > 10). Generally, this reduction of \Rmg is also present with $\EWint>0$ as the emergent spectrum is a composition of the flat continuum ($\Rmg <2$) as well as the Gaussian part ($\Rmg\sim 2$).
Note that this effect is generally at play for $\tauzero \gtrsim 10$, i.e., column densitites $\Nmg\gtrsim 10^{14}\unitNHI$. 
Consequently, at this high \Nmg, where the LyC escape \fesclyc $\approx 0$ (cf. Fig.~\ref{fig:ratio_con_emis}), \Rmg < 2 due to the contribution of the continuum pumping.
Albeit for somewhat different reasons, it is, therefore, consistent with the picture of the pioneering observational work quoted above, which suggested $\Rmg$ as a proxy for LyC leakage.
\\

On top of the previously explored $\Rmg<2$ proxy for LyC escape, in \S~\ref{sec:ratio_Nmg}, we highlight an additional measure to constrain the column density and, thus, ionizing photon escape: the doublet ratio in the halo. As this observable only consists of scattered radiation, it follows a similar trend for $\Nmg\gtrsim 10^{14}\unitNHI$, i.e., there $\Rmgtotal \sim \Rmghalo <2$. However, for smaller column densities -- that is, in the LyC escape relevant regime -- the integrated doublet ratio flattens to $\sim 2$ whereas the halo part shows $\Rmghalo>2$ (cf. Fig.~\ref{fig:ratio_EW}). This is because, in this regime, the lines are not saturated (i.e., \tauzero < 1).

These results suggest a novel `smoking gun' signature for ionizing photon escape: if $\Rmghalo>2$, the column densities are low (\Nmg$\lesssim 10^{13.5}\unitNHI$) implying $\fesclyc\gtrsim 0.2$ (for \Nmg/\NHI$\sim 10^{-4}$; cf. Fig.~\ref{fig:ratio_con_emis} which shows a clear $\Nmg-\Rmghalo$ correlation). Naturally, the exact numbers do depend on, e.g., the intrinsic equivalent width of the line (cf. Fig.~\ref{fig:ratio_EW}).

We show the relation between \Rmghalo and \fesclyc explicitly in Fig.~\ref{fig:ratio_LyC} for a suite of our smooth and clumpy models. Here, we show \fesclyc as a function of the halo \Rmg for various \EWint = 0 (flat continuum), 2, 5, 10, and 100 \AA\ (cf. Eq.~\eqref{eq:incident_radiation}). We also show how the correlation changes when considering different \mgii abundances. It is clear from Fig.~\ref{fig:ratio_EW} that independent of on (non) multiphase structure, the radiative transfer models follow a clear $\Rmghalo-\fesclyc$ trend -- which is shifted to higher \Rmg for larger \EWint.

From the equations derived in \S~\ref{sec:ratio_Nmg} for the doublet ratio, we can write down a simple formulation for this trend. Rewriting Eq.~\eqref{eq:flux_ratio} to give \Nmg as a function of \Rmghalo, and adopting $\fesclyc=\exp(-\tau_{\rm LyC})\approx \exp[-\NHI / ({1.6 \times 10^{17}}\unitNHI)]$, one can obtain
\begin{equation}\label{eq:lyc_mgii}
\fesclyc \approx  \exp \left[ -2.2 \frac{\Nmg}{3.6\times 10^{13}\unitNHI} \left({\fmghi \over 10^{-4}}\right)^{-1}      \right].
\end{equation}
Here, $\Nmg$ is the \mgii column density, which can be analytically estimated using the halo doublet ratio (cf. Eq.~\eqref{eq:flux_ratio}) and is then by
\begin{equation}
{\Nmg \over {3.6 \times 10^{13}\unitNHI} }= -\ln \left( {\Rmghalo \over R_{\rm MgII, int}} - 1 \right)  \left( \vexp \over {100 \kms}\right)
\end{equation}
where $R_{\rm MgII, int}\approx 2$ is the fiducial doublet ratio (cf. \S~\ref{sec:source}).
Eq.~\eqref{eq:lyc_mgii} reproduces our radiative transfer simulations well (shown as dashed lines in Fig.~\ref{fig:ratio_LyC} and can be used to convert observed halo doublet ratios to ionizing escape fractions. This can, in principle, also be applied for high redshifts (where the direct detection of ionizing photons is impossible). However, note that for very low magnesium abundances \fmghi $< 10^{-4.5}$, \mgii scattering is negligible due to small \Nmg ($< 10^{13} \unitNHI$) at the canonical \NHI for LyC (1.6 $\times 10^{17} \unitNHI$) as shown in Fig.~\ref{fig:ratio_LyC}.\\

A visual summary of the discussed relation and our new proxy of ionizing photon escape \Rmghalo is shown in Fig.~\ref{fig:mgii_lyc}. There we show schematically the process leading to low or high LyC escape and the corresponding \mgii halo spectrum.

\subsection{Connection to other resonant lines}

\begin{table*}
    \caption{Atomic data of \lya and metal resonance doublets. $\lambdaK$ and $\lambdaH$ are atomic line center wavelength of K and H lines (second and third columns). $f_{\rm K}$ and $f_{\rm H}$ are oscillation strengths of K and H lines (fourth and fifth columns). $\Delta V_{\rm K,H}$ is the velocity separation of the K and H lines (sixth column). $E_{\rm ion}$ is the ionization energy of the atoms (seventh column). A is the ionization energy to create atoms, e.g., the Mg~I ionization energy in the case of Mg~II. (eighth column). ${n / n_{\rm H}}$ is the fraction of the atom in the solar metalicity (ninth column).}
\resizebox{\linewidth}{!}
{\begin{tabular}{|c|r|r|r|r|c|c|c|c|}
\hline
Atom &  $\lambdaK$$^a$ & $\lambdaH$$^a$ & $f_{\rm K}$ &  $f_{\rm H}$ & $\Delta V_{\rm K,H}$$^b$ & ${E_{\rm ion}}^c$ & ${E^c_{\rm ion}}^d$ & $\log (n/n_{\rm H})$ \cr 
\hline
\lya & 1215.668 & 1215.674 & 0.278 &  0.139 & 1.3 & 13.6 & $-$  & $-$ \cr
Mg~II & 2796.4 	 & 2803.5  & 0.608 &  0.303 & 760 & 15.0 & 7.64 & $-4.46^*$ \cr
C~IV & 1548.2 & 1550.8 & 0.190 &  0.0952 & 500 & 64.5 & 47.9 & $-3.61^{\dagger}$ \cr
O~VI & 1031.9 & 1037.6 & 0.133 &  0.0660 & 1660 & 138 & 114 & $-3.31^{\dagger}$ \cr
N~V & 1238.9 & 1242.8 & 0.157 &  0.0782 & 970 & 97.9 & 77.5 & $-4.07^{\dagger}$ \cr
Si~IV & 1393.8 & 1402.8 & 0.513 &  0.255 & 1926 & 45.1 & 33.5 & $-4.46^{*}$ \cr
\hline
 \end{tabular}}
 \footnotesize{a: vacuum wavelength in \AA, b: velocity separation of K and H lines $\Delta V_{\rm K,H} \sim (\lambdaK - \lambdaH)c/\lambdaK$ in \kms, c, d: ionization energy in eV, *: \cite{holweger01}, $\dagger$: \cite{grevesse98}}\\
 \label{tab:atomic}
\end{table*}

In this study, we focused on the resonant \mgii and \lya lines. However, also other resonant lines are commonly studied in astrophysics. For instance, 
\citet{hayes16} reported the detection of an extended O~VI nebula with a size larger than the \lya counterpart. They speculate in their work that this could be due to the scattering of O~VI photons in the hot CGM at the temperature of several $10^5\, \rm K$.
Furthermore, \cite{berg19} noted that the scattering with C~IV causes the profile of C~IV emission to be broader than non-resonant He~II emission.
Those studies show that metal doublets are affected by radiative transfer processes and thus carry additional information about the gas properties within them.

The behavior and physical properties of scattered metal doublets are similar to \lya because of the atomic physics of \mgii, \civ, \ovi, \nv, and \siiv atoms, which have one electron in the outer orbit. 
We described the atomic physics of \mgii in detail in \S~\ref{sec:cross_section}, and analogous arguments apply for the other metal doublets.
For instance, 
\civ$\lambda\lambda$ 1549, 1551, \ovi$\lambda\lambda$ 1032, 1038,  \nv$\lambda\lambda$ 1239, 1243, and \siiv$\lambda\lambda$ 1393, 1402 have the same atomic structure composed of two transitions, K ($S_{1/2}-P_{3/2}$) and H ($S_{1/2}-P_{1/2}$). 
Furthermore, due to their resonant nature, the doublets of highly ionized atoms can undergo scatterings before escaping the system.
It is also important to note that the ratio of all the oscillator strengths is $\sim 2$, implying that the intrinsic doublet ratio is 2 when the emission lines originate from the collisional excitation and recombination. This canonical ratio also implies the same difference in scattering optical depths as described here for \mgii and the associated physical processes discussed.

The main differences between other resonant lines lie in the energy level as well as in the velocity separation between the K and the H line. In Table~\ref{tab:atomic}, we list a selection of \lya and metal resonance doublets commonly used in astrophysics. One can see that the ionization energy ranges from 13.6 eV to 138 eV, i.e., probing temperatures from several $10^3\, \rm K$ to several $10^5\, \rm K$. This implies that multiwavelength access to these emission lines can constrain the multiphase nature of the CGM -- as often already done in absorption line studies (e.g., \citealp{fox07,fox09,crighton14,rubin15,mas-ribas17}; or review by \citealp{tumlinson17}).

It is also evident from Table~\ref{tab:atomic} that the velocity separation of the K and H lines varies between 1.3\kms for \lya and $1657\kms$ for \ovi. As we discussed already in \S~\ref{sec:cross_section}, this separation for \lya is too small to be observed. However, for the other lines, this means that our conclusions drawn in \S~\ref{sec:smooth_gaussian} have to be rescaled from $\sim 760\kms$ for \mgii to the respective doublet separation $\Delta V_{K,H}$, which can yield interesting constraints on galactic flows.
However, higher column density ($> 10^{14} \unitNHI$) is required to utilize the doublet ratio of other lines (\ovi, \civ, and \nv) to indicate the strong outflow/inflow,
since the oscillation strengths of the three doublets are $3-4$ times less than \mgii.
We will discuss other metal lines as a tracer of the galactic flows in the next section.


\subsection{Metal doublet ratios as a probe of strong outflows}\label{sec:ratio_outflow}

Galactic winds are a cornerstone in our theory of galaxy formation and evolution, but their energetics and extents are still debated \citep[for reviews see, e.g.,][]{veilluex05,veilluex20,rupke18}.

As we show in \S~\ref{sec:ratio_smooth} for the example of \mgii, a strong change in the doublet ratio can be due to strong kinematics, i.e., larger than the doublet separation $\Delta V_{\rm K,H}$; specifically outflows leading to $\Rmg<2$ and inflows to $\Rmg>2$.
Although, in this work, we also discuss other mechanisms to alter the doublet ratio, such as scattering of the continuum photons, which can also decrease \Rmg, it is important to note that for all these processes, the doublet ratio is always higher than unity (see, e.g., Figures~\ref{fig:ratio_smooth} and \ref{fig:ratio_clumpy}). Thus, a doublet ratio of less than unity is clear evidence of the strong outflows with velocities at least approaching $\Delta V_{\rm K,H}$.

Table~\ref{tab:atomic} summarizes the atomic data of \mgii and the three doublets of highly ionized atoms, \civ, \ovi, \nv, and \siiv with columns showing the velocity separation of the K \& H lines, ionization energy, and the fraction of atom in the solar metalicity. 
Those three doublets have high ionization energies, $> 60\, \rm eV$, and thus are more relevant for probing for a hot medium with $T > 10^5\, \rm K$.

 The emission of \civ doublet is a potentially good tracer of hot wind \& CGM with $T \sim 10^{5}\, \rm K$ because the \civ doublet has the smallest separation ($\sim 500 \kms$) and strongest oscillation strength ($\sim 0.19$) of three doublets of highly ionized atoms in Table~\ref{tab:atomic}.
In Fig.~\ref{fig:ratio_smooth}, the \mgii doublet ratio significantly varies when the outflow/inflow velocity is higher than the velocity separation and \Nmg $\gtrsim 10^{14} \unitNHI$.
Adopting the atomic data in Table~\ref{tab:atomic}, 
the \civ ratio becomes less than 2 when the outflow velocity > 500 \kms and a C~IV column density $\gtrsim 3 \times 10^{14} \unitNHI$.
The literature, \citet{kafatos85,michalitsianos88,michalitsianos92}, reported that the ratio of C~IV$\lambda$ 1549 and $\lambda$1551 is less than 1 in the {IUE} \citep[International Ultraviolet Explorer;][]{iue_obs} spectra of symbiotic stars because of the strong outflow > 600 \kms.
This \civ ratio is too small to be explained by the scattering of the continuum.
Consequently, these previous studies suggest the potential of the \civ doublet ratio as the indicator of the hot wind > 500 \kms.

\subsection{Spatially varying \mgii emission}


\begin{figure*}
	\includegraphics[width=\textwidth]{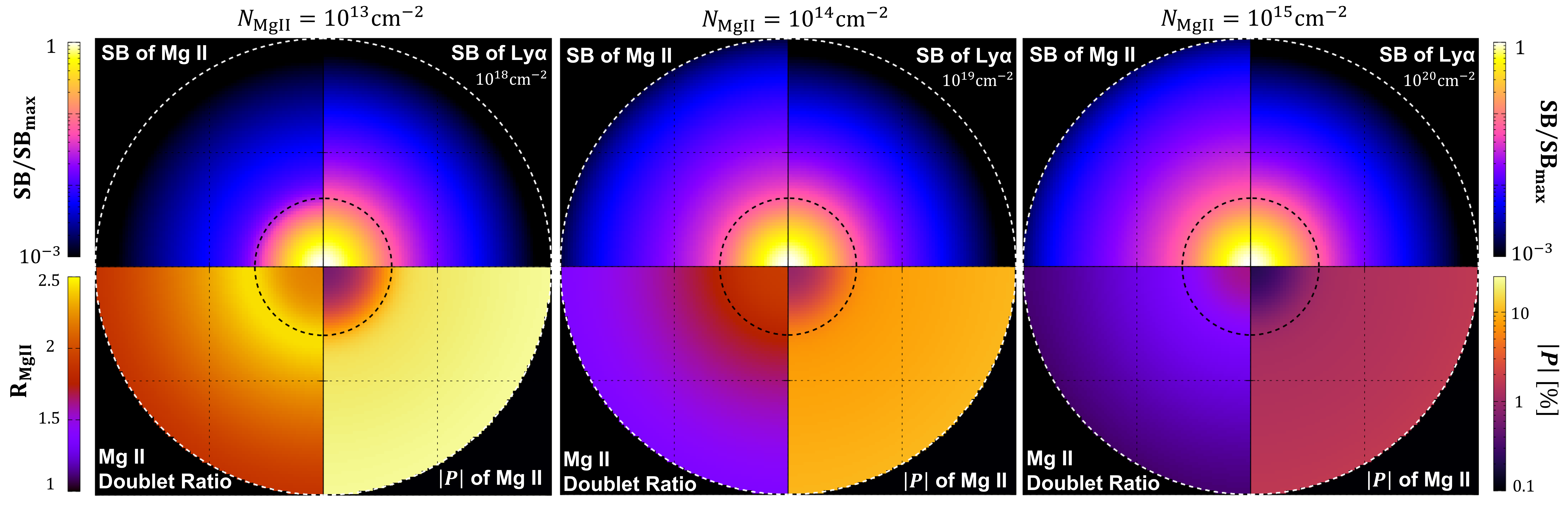}
    \caption{Projected images for three \Nmg = $10^{13}\unitNHI$ (left), $10^{14}\unitNHI$ (center), and $10^{15} \unitNHI$ (right panel).
    \vexp, \fc, and \taudabs are fixed at 200 \kms, 100, and 0, respectively.
    The incident radiation consists of the flat continuum and the in situ emission with \sigsrc = 100 \kms and $\rm EW_{int}$~=~5~\AA. 
    Each panel is composed of four patches of the projected images, \mgii surface brightness (top left), \lya surface brightness (top right), a doublet ratio \Rmg (bottom left), and a degree of polarization of \mgii (bottom right).
    \Nmg/\NHI = $10^{-5}$ is assumed to get \lya surface brightness.
    The seeing effect with the radius $0.1\RH = 10\, \rm kpc$ is adopted as an angular scale at $z\approx 3$ is $\sim$ 7.855 kpc arcsec$^{-1}$ \citep{wright06}. The black and white dashed lines represent the projected radius $0.2\RH$ and \RH, respectively.
    The inner region of the black dashed circle in the top patches is strongly affected by the seeing; typically, two times the radius for seeing (20 kpc $\sim$ 2 arcsec) but strongly dependent on the spatial resolution of the instrument and redshift.
    The bright core and the unpolarized feature in this inner region mainly originate from directly escaping photons at the central source. 
    }
    \label{fig:image}
\end{figure*}
\begin{figure}
	\includegraphics[width=\columnwidth]{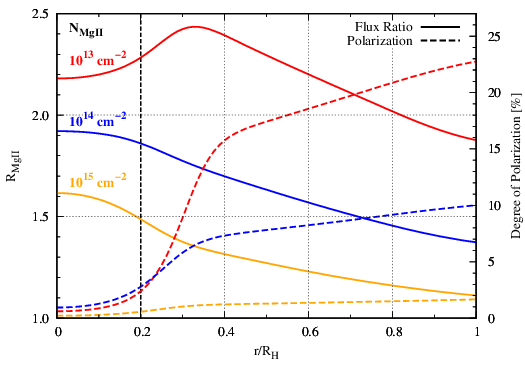}
    \caption{The doublet ratio \Rmg (solid) and the degree of polarization $|P|$ (dashed line) as a function of the projected radius $r/\RH$. 
    The profiles are extracted from the projected images in Fig.~\ref{fig:image}.
    Each color represent \Nmg $10^{13}$ (red), $10^{14}$ (blue) and $10^{15}\unitNHI$ (orange).
    }
    \label{fig:radi}
\end{figure}

As mentioned in the introduction, the spatially extended emission halo obtained by IFS such as MUSE, KCWI, and HETDEX allows us to investigate the spatial variation of \lya \citep{claeyssens19,leclercq20,guo23}  and \mgii \citep{chisholm20,burchett21,dutta23,guo23b}.
Additionally, polarized \lya emission has been observed in spatially extended \lya nebulae \citep{hayes11,you17,kim20}, providing compelling evidence of scattering effects \citep{dijkstra08,eide18,chang23}. Given the similarity in atomic physics between \mgii and \lya, it follows that \mgii emission is also subject to polarization through scattering.
Using our radiative transfer code, we can study how the spectra, surface brightness, and even polarization pattern change in space.

Fig.~\ref{fig:image} shows as an example the projected images of \mgii and \lya for various column densities, \NHI \& \Nmg.
Note that here, we adopt a Gaussian point spread function with width $\sigma=0.1\RH$ leading to an extended bright core in the central region. 
As expected, the SB profile of \mgii becomes more extended with increasing \Nmg because of the increased scattering rate. 
The overall \Rmg decreases with increasing \Nmg because of the contribution of the continuum as discussed in \S~\ref{sec:ratio_emission_continuum} (cf.Fig.~\ref{fig:ratio_con_emis}).
The degree of polarization $|P|$ of \mgii also decreases with increasing \Nmg since the multiple scattering process decreases the strength of polarization \citep{chandrasekhar60,chang17,seon22,chang23}.

Fig.~\ref{fig:radi} shows the profiles of \Rmg and $|P|$ as a function of the projected radius $r_p/\RH$ for three \Nmg = $10^{13}$, $10^{14}$, and $10^{15} \unitNHI$. 
When $\NHI=10^{13}\unitNHI$, \Rmg increases until $\sim$ 0.3$r/\RH$ and the peak value of \Rmg is higher than 2.
In this low \NHI, the scattering of the in situ emission causes the high doublet ratio $\Rmg > 2$ near $0.2 r/\RH$ as shown in the right panel of Fig.~\ref{fig:ratio_con_emis}.
At $\NHI = 10^{14}$ and $10^{15} \unitNHI$, \Rmg decreases with increasing $r/\RH$ .
Since the outflow velocity in our sphere geometry is proportional to the radius,
the outer \mgii halo mainly causes the scattering of the flat continuum.
Thus, \Rmg in the outer halo is smaller than in the inner region.

Fig.~\ref{fig:radi} shows the dependence on \Nmg of $|P|$, which increases radially outward.
In the inner region with  $r/\RH< 0.2$, $|P|$ is less than 5 \% because seeing effects spread unpolarized photons directly escaping from the central source.
However, the outer halo is strongly polarized over 10\% at $\Nmg < 10^{15} \unitNHI$.
Thus, if \mgii emission mainly comes from scattering, \mgii is strongly polarized (>10\%).
In the case of high \Nmg=$10^{15} \unitNHI$, if it is not strongly polarized, 
the evidence of the scattering is imprinted on the spectral profile.

In \S~\ref{sec:mgii_lyC}, we discussed \Rmghalo as an indicator of LyC escape 
since the halo spectrum is composed of \mgii photon undergoing scattering.
However, in cases where intrinsic \mgii emission is spatially extended, distinguishing scattered \mgii photons becomes challenging from an observational perspective. 
Consequently, the polarization of \mgii can serve as a valuable tool for estimating the halo doublet ratio of \mgii, \Rmghalo. 
Therefore, the measurement of the doublet ratio in polarized \mgii nebulae can serve as an analogous quantity to \Rmghalo.

\subsection{Probing the  `clumpiness' of the CGM through resonant lines}\label{sec:f_crit_clumpy}

Many radiative transfer studies focus on the effect of column density and kinematics on emergent properties. However, also the `clumpiness' of the structure plays a crucial role in shaping them -- and given most astrophysical systems (such as the ISM and CGM) are multiphase, one cannot neglect this property in our studies.

The specific parameter governing radiative transfer through a clumpy medium is \fc, the mean number of clumps per line-of-sight \citep{hansen06}. Previous studies have focused on \lya radiative transfer through a clumpy medium and have found that, in general, the observable information (e.g., spectrum, surface brightness, escape fraction, polarization) is different from those in the smooth medium \citep[e.g.,][]{neufeld91,hansen06,laursen13,duval14,gronke16,chang23}. However, if \fc is greater than some critical number of clumps \fccrit (which is a function of the column density and the kinematics), 
a `phase transition' occurs, and the emergent observables of a clumpy medium are the same as the ones of a homogeneous medium \citep{gronke17,chang23}.

In \S~\ref{sec:f_crit}, we extend these previous studies and investigate the radiative transfer of \mgii in a clumpy, multiphase medium. We find that akin to \lya, a critical number of clumps exists, above which the escaping \mgii spectrum is close to the homogeneous one. We also provide an analytic estimate for this \fccrit in Eq.~\eqref{eq:fit_mgii} (as well as update the \lya ones; Eq.~\eqref{eq:fit_lya}) and show that it reproduces the simulated results reasonably well.

This provides a potentially interesting probe to study the clumpiness of the CGM. As most \lya spectra are well fit using homogeneous models (see, e.g., \citealp{verhamme2008,karman2017,gronke2017}), this can be interpreted as $\fc>\fccrit$ in these systems \citep{gronke17}. Such a clumpy CGM is supported by other observables such as quasar absorption line studies \citep[e.g.,][]{crighton14,lan17} -- as well as by theoretical models showing that cold clumps naturally fragment in a hot environment \citep{mccourt18,liang20,gronke22}.

However, importantly, all these probes can only provide lower limits on the `clumpiness' or the mean number of clumps, i.e., the smallest scale or characteristic size of this cold gas is unknown.
As we show in \S~\ref{sec:f_crit} (cf. Fig.~\ref{fig:fc_crit}), \mgii usually has a different \fccrit compared to \lya. This implies that, in principle, one could use a combination of these two lines to provide both an upper and lower bound to \fc and thus constrain the `clumpiness' of the CGM. This could be, for instance, possible for a static medium with \NHI = $10^{20.5} \unitNHI$ and \Nmg = $10^{15} \unitNHI$ for which the critical covering factors are $\fc \sim 20$ and $\sim 100$ for \lya and \mgii, respectively (see Fig.~\ref{fig:fc_map}). Thus, if for such a medium a double peaked \lya profile, but a single peaked \mgii profile is observed, the \fc can be constrained to lay between these two values. Naturally, this prediction is still quite theoretical and would have to be further specified using more complex kinematics and gas structures.

\section{Conclusion}\label{sec:conclusion}

Resonant lines provide us with an important window into the low surface brightness regime and, thus, allow us to study high redshift objects as well as the dim outskirts of galaxies. As these fields are the focus of many future instruments, telescopes, and surveys -- observationally, progress is rapid, and we now know already, e.g., much more about the galactic ecosystem than a decade ago. To interpret this flood of data, it is therefore important that this development is accompanied by theoretical progress.

This work extended our Monte Carlo radiative transfer code in \cite{chang23} to include \mgii doublet.
Since the atomic structures of \mgii \& \lya with their H ($S_{1/2} - P_{1/2}$) and K ($S_{1/2} - P_{3/2}$) transitions, are identical, we could implement \mgii scattering based on the atomic physics of \lya (\S~\ref{sec:RT}).

Our simulation code `{\tt RT-scat}' supports arbitrary 3D gas configurations and kinematics as input. It supports both a Cartesian grid as well as a multiphase `clumpy' medium with many ($\gtrsim 10^9$) clumps. 
As for the latter, a `brute force' method would be memory intensive; we implement `subgrid' clumps that can exist within the cells.
The code furthermore supports the calculation of the polarization behavior (adopting the numerical method in \citealp{seon22}) and dust scattering and extinction using the MW as well as the SMC and LMC dust models in \cite{draine03a,draine03b} (see Appendix~\ref{sec:dust_model}).
Notably, after the submission of this paper, \cite{seon23} appeared, which also studies \mgii radiative transfer. The author focuses on spherical and cylindrical geometries with a uniform density. Hence, the two studies are highly complementary.

With this code, we study the joined \mgii and \lya escape in both homogeneous (\S~\ref{sec:smooth}) and clumpy (\S~\ref{sec:clumpy}) media. Our main focus is the emergent \mgii lines. In particular, we systematically study \textit{(i)} the \mgii doublet ratio in various environments, \textit{(ii)} the general spectral shape, and \textit{(iii)} the \mgii escape fraction. Apart from the scattering geometry, we also vary the intrinsic source between a monochromatic light, a Gaussian emission, and a flat continuum to study how the observables' dependence on it.

Our main conclusions are the following:
\begin{itemize}
\item Because typically \hi column densities in neutral media are much greater than \mgii column densities, the emergent \mgii spectra differ from \lya spectra, even though the atomic physics of the two lines are similar.
As \NHI \& \Nmg increases in the outflow, the spectral peak of \lya gets redshifted while the peak of \mgii is always near the systematic velocity (Fig.~\ref{fig:smooth_spec_NH}). 
In the dusty medium, the scattered red wings of \lya and \mgii show opposite effects, i.e., are weakened and strengthened, respectively (Fig.~\ref{fig:smooth_spec_tauD}).

\item  In general, the \mgii escape fraction \fescmgii is higher than the \lya escape fraction \fesclya because $\NHI \gg \Nmg$ and the dust optical depth of \lya is at least $\sim$ 2.53 times that of \mgii (see \S~\ref{sec:smooth_gaussian} and Fig.~\ref{fig:fesc_smooth} as well as the discussion in \S~\ref{sec:fesc_mgii_lya}).
In addition, the higher column density causes a longer path length, which induces more dust extinction.
However, in a moderately clumpy medium, when an incident photon meets an optically thick clump, the photon suffers scattering on the surface of the clump with a shorter path length. It is surface scatterings that increase escape fractions (see \S~\ref{sec:clumpy_gaussian} and Fig.~\ref{fig:fesc_clumpy}). 
The effect of these surface scatterings is stronger for \lya than in \mgii photons -- again due to $\NHI \gg \Nmg$.
In summary, when the covering factor \fc < 10 and \NHI > $10^{20} \unitNHI$ (\Nmg > $10^{14.5} \unitNHI$), \fesclya $\sim$ \fescmgii due to the surface scattering (see \S~\ref{sec:fesc_mgii_lya} and Fig.~\ref{fig:fesc_compare}). In conclusion, the \lya and \mgii escape fractions are correlated in a non-trivial fashion (cf. \S~\ref{sec:fesc_mgii_lya} for discussion on \fesc).

\item In a clumpy medium, a `critical covering factor' \fccrit above which the radiative transfer of multiphase matches the one through a homogeneous medium also exists for \mgii. However, generally \fccrit for \mgii and \lya do not match (see \S~\ref{sec:f_crit} and Fig.~\ref{fig:fc_crit}).
We provide analytic solutions of \fccrit as a function of \vexp, and column densities in Eq.~\eqref{eq:fit_lya} and \eqref{eq:fit_mgii} for \lya and \mgii, respectively. The differing \fccrit values can be used to probe the clumpiness of the CGM, which we discuss in \S~\ref{sec:f_crit_clumpy}.

\item The doublet ratio of \mgii, \Rmg, can be a tracer for strong in- or outflows.
The ratio can can be de- or increased from its intrinsic doublet ratio of $\sim 2$ due to outflowing and inflowing media, respectively, if $|\vexp| > 700 \kms$ and $\Nmg > 10^{14} \unitNHI$ (see Fig.~\ref{fig:smooth_ratio}).
This is because of the fact that if the kinematic velocity of the scattering medium is larger than the velocity separation of \mgii doublet, $\sim 750 \kms$,
K/H photons are scattered by the H/K transition of the outflowing/inflowing \mgii atom (see Fig.~\ref{fig:smooth_spec_vexp}).
However, this trend weakens at the clumpy medium with low \fc (see Fig.~\ref{fig:ratio_clumpy}). While we also discuss other effects that alter \Rmg (see points below), strong outflows are the only ones that can lead to $\Rmg < 1$. This can be used as a complementary probe for the energetics of galactic winds (cf. \S~\ref{sec:ratio_outflow}).

\item A particularly interesting case is the \mgii line formation due to scattered stellar continuum photons. In this scenario, \Rmg is lowered from its intrinsic value of $2$, and \Nmg determines \Rmg  (see \S~\ref{sec:flat_continuum}). 
Assuming a flat continuum, the \mgii emission appears by scattering of \mgii leading to a P-Cygni-like profile (Fig.~\ref{fig:spec_smooth_flat}); the strength of the emission depends on \Nmg.
In the low \Nmg regime (< $10^{14} \unitNHI$), 
the scattering rate of photons near the K transition is less than unity and two times higher than near the H transition.
In the high \Nmg regime (> $10^{14} \unitNHI$), 
both scattering rates of photons near the K and H transition are close to unity.
Thus, \Rmg decreases with increasing \Nmg (Fig.~\ref{fig:EW_smooth}).

\item We also study the most general and realistic emission case and combine in-situ Gaussian \mgii emission and a stellar continuum contribution. Also, in this scenario, the doublet ratio \Rmg on \Nmg -- as well as the intrinsic equivalent width of the in situ emission \EWint (see \S~\ref{sec:ratio_emission_continuum} and Fig.~\ref{fig:ratio_con_emis}).
Here, we also propose a more sensitive measure of the gas properties, the doublet ratio of the halo $\Rmghalo$.
Since the halo photons are only composed of scattered photons, $\Rmghalo$ is a less `contaminated' probe and can be used to constrain, e.g., the cold gas column density.
This is particularly interesting for the constraint of ionizing photons (cf. discussion in \S~\ref{sec:mgii_lyC}). We find that the halo double ratio $\Rmghalo \gtrsim 2$ is a clear proxy for ionizing photon escape.
For this purpose, we provide an analytic fit relating the column density (\S~\ref{sec:ratio_emission_continuum}, Eq.~\ref{eq:flux_ratio}) as well as the ionizing escape fraction (Eq.~\ref{eq:lyc_mgii}) with \Rmghalo.
We show that those solutions match our simulated results well. 
\end{itemize}

We expect that other metal resonance doublets in the UV regime, \ovi$\lambda\lambda$ 1032, 1038, \nv$\lambda\lambda$ 1239, 1243, and \civ$\lambda\lambda$ 1549, 1551, can show similar behavior and trends as discussed here for \mgii. They can, therefore, provide interesting constraints on the warm CGM with $T \gtrsim 10^5\, \rm K$.
We plan to investigate the scattering processes of other metal doublets in future work.
Furthermore, we hope to adopt our simulated results shown here to model and analyze the observations of spatially extended \mgii emission.

\section*{Acknowledgements}
The authors thank the anonymous referee for a detailed report, which significantly improved the quality of the work. The authors also thank Hee-Won Lee and Kwang-Il Seon for their insightful comments that greatly contributed to the completion of this work.
MG thanks the Max Planck Society for support through the Max Planck Research Group.
\section*{Data Availability}
Data related to this work will be shared on reasonable request to the corresponding author.



\bibliographystyle{mnras}
\bibliography{reference} 



    
\appendix

\section{Escape Fraction by Different Dust Models}\label{sec:dust_model}

\begin{figure*}
	\includegraphics[width=\textwidth]{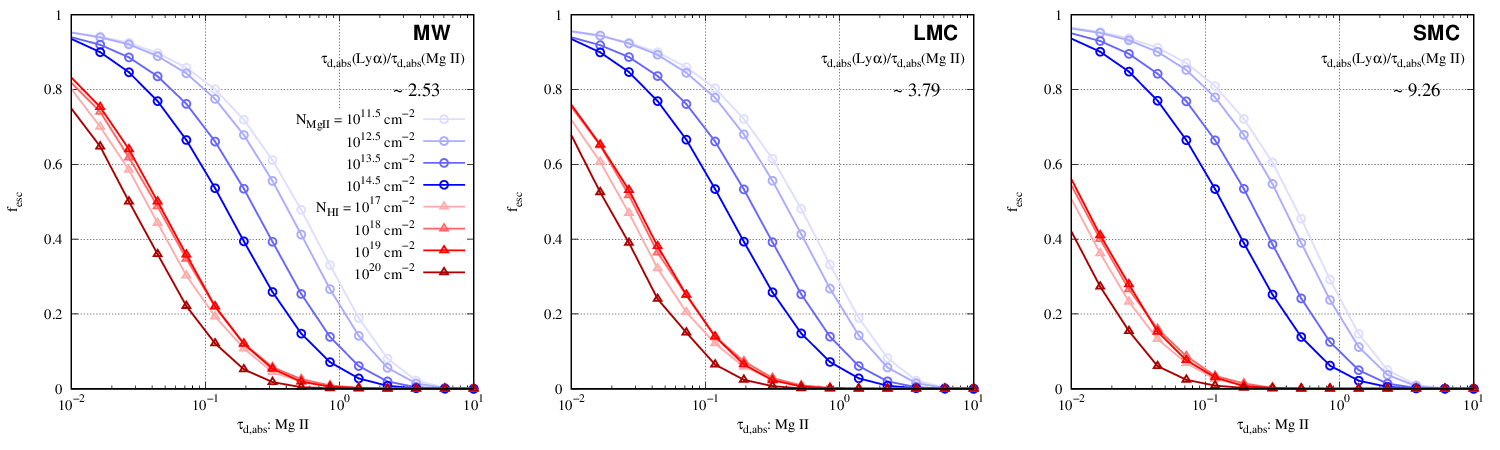}
    \caption{Escape fractions \fesc of \mgii (blue) and \lya (red) for three types of dust models: MW (left), LMC (center), and SMC (right).
    The model composed of a central source and homogeneous medium considers the monochromatic source (\sigsrc = 0 \kms), the static medium (\vexp = 0\kms), and the thermal speed as the random motion of scattering medium  (\sigr = \vth).
    The $x$-axis is the dust absorption optical depth of \mgii near 2800 \AA. 
    The shade of color represents various \NHI \& \Nmg assuming $\Nmg / \NHI = 10^{-5.5}$.
    }
    \label{fig:f_esc_appendix}
\end{figure*}

Fig.~\ref{fig:f_esc_appendix} shows the escape fractions of \mgii and \lya, denoted as \fescmgii and \fesclya, respectively, as a function of the dust absorption optical depth \taudabs of \mgii. Three different dust models are considered: Milky Way (MW), Large Magellanic Cloud (LMC), and Small Magellanic Cloud (SMC) dust models, as described in \cite{draine03a,draine03b}. The ratio of \taudabs between \lya and \mgii, $\taudabs(\rm Ly\alpha) / \taudabs(\rm MgII)$, is 2.53, 3.79, and 9.26 for the MW, LMC, and SMC dust models, respectively. \fesclya decreases with increasing $\taudabs(\rm Ly\alpha) / \taudabs(\rm MgII)$.

\begin{figure}
	\includegraphics[width=\columnwidth]{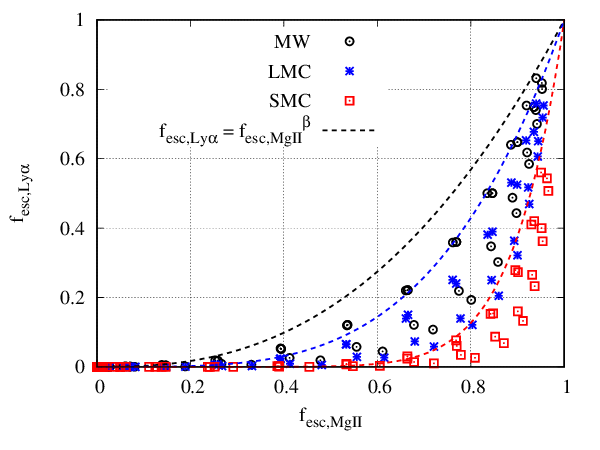}
    \caption{\fesc of \mgii and \lya, \fescmgii and \fesclya in Fig.~\ref{fig:f_esc_appendix}.
    The colors represent the dust models: MW (black), LMC (blue), and SMC (red).
    The dashed lines represent $\fesclya = \fescmgii^\beta$, where $\beta$ is the ratio of dust absorption optical depth of \lya and \mgii, $\taudabs(\rm Ly\alpha) / \taudabs(\rm MgII)$, shown in Fig.~\ref{fig:f_esc_appendix}.
    $\beta$ is 2.53, 3.79, and 9.26 for the MW, LMC, and SMC dust models, respectively.
    }
    \label{fig:f_esc_compare_appendix}
\end{figure}

Fig.~\ref{fig:f_esc_compare_appendix} shows a comparison between \fesclya and \fescmgii, assuming $\Nmg / \NHI = 10^{-5.5}$. 
Since the path length of \lya is typically larger than that of \mgii due to the significantly higher column density of \hi compared to \mgii (i.e., $\NHI \gg \Nmg$), \fesclya is generally smaller than $\fescmgii^{\beta}$, where $\beta$ is $\taudabs(\rm Ly\alpha) / \taudabs(\rm MgII)$ for each dust model.
However, it is important to note that in a clumpy medium, the trend can be changed, as discussed in Fig.~\ref{fig:fesc_compare} and \S~\ref{sec:fesc_mgii_lya}.

\bsp	
\label{lastpage}
\end{document}